\documentclass[12pt,letterpaper]{article}

\usepackage[left=1in,right=1in,top=1.5in,bottom=1.5in]{geometry}
\usepackage{setspace}

\usepackage{amsmath,amssymb,amsthm,mathtools,bm}

\usepackage{graphicx}
\usepackage{booktabs}
\usepackage{threeparttable}
\usepackage{longtable}
\usepackage{array}
\usepackage{multirow}
\usepackage{caption}
\usepackage{subcaption}
\usepackage{float}
\usepackage{placeins}
\usepackage{makecell}

\usepackage{enumitem}
\usepackage{xcolor}
\usepackage{appendix}
\usepackage{pdflscape}
\usepackage{csquotes}
\usepackage{bbm}

\usepackage{chngcntr}
\counterwithin{figure}{section}
\counterwithin{table}{section}

\usepackage{tikz}
\usetikzlibrary{positioning,arrows.meta}

\usepackage[authoryear]{natbib}

\usepackage{xurl}
\usepackage{etoolbox}

\AtBeginEnvironment{thebibliography}{%
	\setlength{\emergencystretch}{3em}%
}

\usepackage[colorlinks=true,linkcolor=blue,citecolor=blue,urlcolor=blue]{hyperref}
\usepackage[nameinlink,capitalize]{cleveref}

\AtBeginDocument{
	\setlength{\abovedisplayskip}{7pt plus 2pt minus 2pt}
	\setlength{\belowdisplayskip}{7pt plus 2pt minus 2pt}
	\setlength{\abovedisplayshortskip}{4pt plus 2pt minus 1pt}
	\setlength{\belowdisplayshortskip}{4pt plus 2pt minus 1pt}
}

\theoremstyle{definition}

\title{The Year-End Toll:\\Frictions Embedded in Option-Implied Rates}
\author{Useong Shin\thanks{
		Sogang Business School, Sogang University (Seoul, Korea).\\
		ORCID: \href{https://orcid.org/0009-0003-0197-9003}{0009-0003-0197-9003}\\
		Email: \texttt{useong@sogang.ac.kr}
}}
\date{\today}

\begin{document}
	
	\maketitle
	
	\pagenumbering{arabic}
	\setcounter{page}{1}
	
	\begin{abstract}
		Option-implied rates are often treated as frictionless because completed boxes deliver riskless payoffs. I show that this interpretation requires option- and benchmark-side implementation wedges to offset. Using SPX and RUT options from 2012--2025, I find a 2--3 bp unannualized increase in the option funding basis when maturity first crosses December 31. The effect behaves as a fixed price wedge, strengthens in the mid-2010s, survives alternative contracts and benchmarks, appears independently in government-bond CIP, and is reproduced in an independently constructed option panel. Put--call parity can identify discount rates precisely without establishing their economic purity.
	\end{abstract}
	
	\begin{flushleft}
		\textbf{\small JEL:} G12; G13; G21; G28\\
		\textbf{\small Keywords:} option funding basis; option-implied interest rates; put--call parity; intermediary balance-sheet constraints; year-end effects; safe-asset convenience yield
	\end{flushleft}
	
	\paragraph{Acknowledgments}
	I am grateful to Michele Azzone (Politecnico di Milano) for generously sharing research data, for guidance on implementing the implied-discount-factor pipeline, and for detailed feedback on earlier drafts. I also thank Young Ho Eom (Yonsei University) for helpful suggestions that led to additional holiday-based robustness checks and improvements to the empirical design. All remaining errors are my own.
	
	\clearpage
	\hypersetup{pageanchor=false}
	\thispagestyle{empty}
	
	\begin{center}
		{\large\bfseries Conflict-of-Interest Disclosure Statement}
	\end{center}
	
	\vspace{1.5em}
	
	\noindent\textbf{Useong Shin}
	
	\medskip
	
	\noindent This research received no specific financial support from any public, commercial, or not-for-profit entity. I have no other conflicts of interest to disclose.
	
	\vspace{2em}
	
	\noindent\textbf{Use of Artificial Intelligence}
	
	\medskip
	
	\noindent During the preparation of this manuscript, the author used ChatGPT (OpenAI; GPT-5.6 Sol) and Claude (Anthropic; Claude Opus 5.0), accessed from July to August 2026, to assist with debugging MATLAB research code, empirical analysis and interpretation, drafting and editing text, and correcting \LaTeX{} syntax. The author did not use these tools to generate research data. All research code was reviewed and executed by the author, and all empirical results and AI-assisted outputs were independently checked and validated by the author. The author takes full responsibility for the content of the manuscript.
	
	\clearpage
	\hypersetup{pageanchor=true}
	\setcounter{page}{2}
	
	\section{Introduction}
	\label{sec:intro}
	
	Option-implied discount rates extracted from index options provide an alternative measure of the risk-free rate without directly relying on safe-asset prices. Put--call parity makes same-maturity call--put spreads linear in strike, allowing the discount factor to be recovered from the slope \citep{AB21,BDG22}. Recent work interprets the resulting spread between option-implied and safe-asset rates as a convenience yield \citep{BDG22,DVT26}. This interpretation relies on the fixed payoff of a completed box, which provides no safe-asset convenience services. Yet payoff certainty does not imply frictionless implementation: maintaining a box may require cash, collateral, margin, internal limits, or intermediary balance-sheet capacity \citep{CBOEBox24}.
	
	I study this distinction using the option funding basis,
	\begin{equation}
		\operatorname{OFB}_{i,t}^{b}(T)
		=
		\frac{10^{4}}{\tau_{t,T}}
		\log\!\left(
		\frac{D_{t}^{b}(T)}
		{\widehat B_{i,t}(T)}
		\right),
		\label{eq:intro_ofb}
	\end{equation}
	where \(\widehat B_{i,t}(T)\) is the option-implied discount factor and \(D_{t}^{b}(T)\) is a maturity-matched external benchmark discount factor. With Treasuries as the benchmark, the OFB is algebraically identical to the option-implied-rate minus Treasury-rate spread used as a convenience-yield measure. The question is therefore not whether the spread is measured precisely, but whether its entire economic content can be attributed to safe-asset convenience.
	
	The central result is a discontinuity when maturity first crosses a future December 31. Among maturities priced on the same trade date, the unannualized spread \(\tau_{t,T}\operatorname{OFB}_{i,t}^{b}(T)\) rises by about \(2\)--\(3\) bp in both SPX and RUT, with nearly identical estimates under DGS and DTB (\Cref{fig:ofbrd,tab:monthly_local_yearend_rd}). Regular weeklies reproduce the positive jump on a much denser maturity grid (\Cref{fig:rd_weekly}). Functional-form tests favor a fixed price wedge rather than an annualized flow premium, so the effect scales approximately as \(\kappa/\tau\): a representative \(2.5\) bp charge corresponds to about \(10\) annualized bp at three months (\Cref{tab:yearend_functional_form}).
	
	The full maturity cross section yields the same conclusion. Contracts spanning year-end carry an additional unannualized price difference of about \(2.9\)--\(3.2\) bp after controlling for nonlinear maturity structure, quoted liquidity, the NFCI, and the same-day spread between 90-day AA asset-backed commercial paper and three-month Treasury bills (\Cref{tab:yearend_regression_spx,tab:yearend_regression_rut}). The OFB covaries positively with this benchmark-relative funding proxy, while the year-end coefficient remains strongly positive. The first crossing is precisely estimated, whereas thinner long-maturity support prevents sharp inference about later crossings (\Cref{tab:additional_yearend_crossings}). The boundary effect also survives Treasury and OIS benchmarks, alternative calendar and maturity controls, dividend adjustments, and symmetric quarter-end placebo tests (\Cref{tab:option_benchmark_jump_decomposition,fig:ois_common_rd,tab:ois_common_results,tab:quarterly_expiration_confound}).
	
	The year-end charge is strongly time varying. Annual estimates from the full-panel and strict same-date designs move closely together, and the option evidence places a substantial strengthening of the boundary effect in the mid-2010s (\Cref{fig:yearly_method_comparison,fig:yearly_regime_break}). This pattern is reproduced in an economically distinct market. Government-bond CIP exhibits both a positive year-end boundary charge and an independently estimated regime shift in essentially the same transition window (\Cref{tab:cip_yearend_regime,tab:cip_regime_break}). An independently constructed option-implied-rate panel from \citet{DVT26} also produces a positive U.S.\ crossing pattern, with three- and six-month magnitudes consistent with the fixed-price scaling found in my data (\Cref{subsec:dvt_rep}). These external results make explanations based solely on my option-processing pipeline, fixed option-expiration rules, or generic calendar seasonality substantially less plausible.
	
	The common regime timing is consistent with a higher shadow price of year-end balance-sheet capacity during the post-crisis regulatory transition. The estimated option transition window overlaps the phase-in of the U.S.\ G-SIB surcharge and contains the implementation of the enhanced supplementary leverage ratio, while government-bond CIP independently locates a similar change (\Cref{subsec:yearly_regime_shift,subsec:cip_regime_shift}). I do not interpret this timing as identifying the causal effect of either regulation. The narrower conclusion is that independently measured prices of year-end balance-sheet exposure became materially larger during the same period in which balance-sheet regulation of major intermediaries tightened.
	
	The year-end toll is therefore an existence proof rather than a complete decomposition of option-side frictions. December 31 provides unusually sharp identification, whereas funding, collateral, netting, inventory, and balance-sheet costs may vary smoothly and leave no comparable cutoff. Even perfectly removing the identified component would not establish that the residual option-implied rate is frictionless.
	
	The paper makes three contributions. First, it identifies a reporting-boundary discontinuity in the relative price of option-implied and external benchmark discount factors, shows that it behaves as a fixed price wedge, and documents a mid-2010s strengthening independently echoed in government-bond CIP. Second, it shows that precise put--call parity and a riskless box payoff do not imply a zero option-side implementation wedge; option-implied rates covary with funding conditions and can embed balance-sheet-sensitive pricing. Third, it makes explicit the relative-purity condition required to interpret the entire option-implied-rate minus safe-asset-rate spread as a convenience yield: option- and benchmark-side implementation wedges must offset. Put--call parity identifies the discount rate precisely; it does not identify its economic purity.
	
	\section{Related Literature}
	\label{sec:lit}
	
	\paragraph{Option-implied discount rates and convenience yields.}
	
	Recovering implied interest rates from option prices and box spreads has a long history \citep{BG86,RR89}. More recent work extracts maturity-specific discount factors from the linear relation between same-maturity call--put spreads and strike \citep{AB21}. \citet{BDG22} use the resulting option-implied rate to measure Treasury convenience yields, and \citet{DVT26} extend the approach internationally to separate safe-asset convenience yields from covered-interest-parity deviations.
	
	These papers do not require option markets themselves to be free of intermediary frictions. In particular, \citet{DVT26} emphasize that box rates may reflect the funding conditions of institutions active in derivatives markets while remaining free of convenience services specific to safe assets. My contribution is therefore narrower than arguing that prior option-implied rates are statistically misspecified. I instead make explicit the additional relative-purity condition required for the option--safe-asset spread to identify convenience alone and test whether economically meaningful option-side implementation wedges enter that spread.
	
	This distinction connects to the broader literature on safe-asset convenience yields, which attributes low safe yields to liquidity, money-like services, collateral value, and demand for safety \citep{KVJ12,Nagel16,DIS18}. Treasury relative prices can themselves vary with market functioning and intermediary constraints \citep{HNS22,DHL23}, while implied borrowing rates in S\&P 500 derivatives can reflect funding illiquidity faced by option market makers \citep{GJS24}. The empirical object studied here can therefore contain both safe-asset convenience and relative implementation wedges.
	
	\paragraph{Derivative funding, limits to arbitrage, and market segmentation.}
	
	Derivative prices can depend on funding, collateral, margin, settlement, and netting arrangements even when terminal cash flows are closely related \citep{ADS19}. Intermediary balance-sheet capacity is reflected in relative-value prices \citep{FLVN20}, consistent with the broader limits-to-arbitrage literature in which capital and funding constraints impede convergence despite sharp pricing relations \citep{SV97,GV02,BP09,GP11,HK13,HMV23}. Dealer capacity and end-user demand are likewise reflected in option prices \citep{GPP09}.
	
	Most directly, \citet{SSW25} show that funding rates associated with box, CIP, and equity spot--futures arbitrage are segmented rather than unified by a single universal risk-free rate. I extend this perspective along a different dimension by asking whether the relative price implied by an option position changes when the position must remain outstanding across a future reporting boundary. The implication is not that the entire option--benchmark spread is an implementation cost, but that payoff equivalence alone does not establish its economic purity.
	
	\paragraph{Reporting boundaries and external relative-value evidence.}
	
	Year-end interest-rate effects have long been associated with settlement demand, liquidity preferences, and preferred habitat \citep{Ogden87,GW97,GW05,KSW08,Kotomin13,BW18}. More recent work shows that regulatory reporting affects quantities and prices in balance-sheet-intensive markets around quarter ends and year ends \citep{Munyan15,BMPW22,BBGW24}. My design differs in timing: rather than comparing prices observed at year-end, I compare positions whose maturities fall on opposite sides of a future December 31. The resulting discontinuity measures the ex ante price of requiring a position to survive the reporting date.
	
	CIP deviations provide a natural external comparison because they also respond to funding demand, dealer capacity, and regulatory constraints \citep{DTV18,CDW21,COZ21}. Government-bond CIP further combines international-arbitrage frictions with bond-specific convenience services \citep{DKS25}. Government-bond CIP exhibits both a positive year-end boundary effect and a mid-2010s strengthening closely aligned with the option-market regime shift (\Cref{subsec:cip_boundary_effect,subsec:cip_regime_shift}). An independently constructed international option panel provides a separate construction check (\Cref{subsec:dvt_rep}). These comparisons distinguish the paper from conventional year-end seasonality tests: the relevant object is a time-varying price of future reporting-boundary exposure across distinct relative-value markets.
	
	\section{Economic Decomposition of the Option Funding Basis}
	\label{sec:theory}
	
	A riskless terminal payoff and a frictionless discount rate are distinct concepts. A completed box pays a fixed amount at maturity, yet constructing and maintaining it may require cash, collateral, margin, internal limits, and intermediary balance-sheet capacity. This section formalizes how such implementation wedges enter the option funding basis (OFB) and motivates the empirical tests.
	
	\subsection{Option-implied rates and relative pricing wedges}
	\label{subsec:theory_relative_wedge}
	
	For option market \(i\), trade date \(t\), maturity \(T\), and benchmark \(b\), define
	\begin{equation}
		r_{i,t}^{\mathrm{opt}}(T)
		=
		-\frac{10^{4}}{\tau_{t,T}}
		\log \widehat B_{i,t}(T),
		\qquad
		r_{t}^{b}(T)
		=
		-\frac{10^{4}}{\tau_{t,T}}
		\log D_{t}^{b}(T).
		\label{eq:theory_option_benchmark_rates}
	\end{equation}
	
	The OFB is their difference:
	\begin{equation}
		\operatorname{OFB}_{i,t}^{b}(T)
		=
		\frac{10^{4}}{\tau_{t,T}}
		\log\!\left(
		\frac{D_{t}^{b}(T)}
		{\widehat B_{i,t}(T)}
		\right)
		=
		r_{i,t}^{\mathrm{opt}}(T)
		-
		r_{t}^{b}(T).
		\label{eq:theory_ofb}
	\end{equation}
	
	With Treasuries as benchmark \(b\), the OFB in \Cref{eq:theory_ofb} is algebraically identical to the option-implied-rate minus Treasury-rate spread. This identity, however, does not determine the economic components of the spread.
	
	Let \(r_{t}^{0}(T)\) denote a common reference rate stripped of convenience services and implementation wedges. Write
	\begin{align}
		r_{i,t}^{\mathrm{opt}}(T)
		&=
		r_{t}^{0}(T)
		+
		\phi_{i,t}^{\mathrm{opt}}(T),
		\label{eq:theory_option_rate_decomposition}\\
		r_{t}^{b}(T)
		&=
		r_{t}^{0}(T)
		-
		CY_{t}^{b}(T)
		+
		\phi_{t}^{b}(T),
		\label{eq:theory_benchmark_rate_decomposition}
	\end{align}
	where \(CY_{t}^{b}(T)\) is the benchmark convenience yield and \(\phi_{i,t}^{\mathrm{opt}}(T)\) and \(\phi_{t}^{b}(T)\) are option- and benchmark-side implementation wedges. The resulting decomposition is
	\begin{equation}
		\operatorname{OFB}_{i,t}^{b}(T)
		=
		CY_{t}^{b}(T)
		+
		\phi_{i,t}^{\mathrm{opt}}(T)
		-
		\phi_{t}^{b}(T).
		\label{eq:theory_ofb_decomposition}
	\end{equation}
	
	Thus, the OFB contains the benchmark convenience yield plus a relative implementation wedge. Attributing the entire spread to convenience requires
	\begin{equation}
		\phi_{i,t}^{\mathrm{opt}}(T)
		=
		\phi_{t}^{b}(T).
		\label{eq:theory_relative_purity}
	\end{equation}
	
	The equality in \Cref{eq:theory_relative_purity} is a relative-purity condition, not a requirement that both wedges equal zero. Put--call parity and the fixed payoff of a completed box impose neither restriction: they identify the discount factor but do not require the trades sustaining its price to use no capital or the option- and benchmark-side wedges to cancel.
	
	\paragraph{Riskless payoffs and funding conditions.}
	
	A completed long box with strikes \(K_{1}<K_{2}\) pays \(K_{2}-K_{1}\) at maturity regardless of the underlying price. Payoff certainty removes terminal price risk but not the resources required to finance or maintain the position.
	
	This fixed terminal claim makes asset-backed short-term funding conditions a natural reduced-form comparison for the option-side wedge. I therefore use the daily spread between the 90-day AA asset-backed commercial-paper rate and the three-month Treasury-bill rate, denoted \(\operatorname{ABCP90\!-\!DTB3}_{t}\). Subtracting the matched Treasury-bill rate removes a common short safe-rate component and yields a market measure of the relative price of asset-backed short-term funding.
	
	The proxy does not imply that box traders literally issue ABCP or borrow at the observed ABCP rate. Its coefficient is therefore not a structural funding-cost pass-through parameter. The reduced-form prediction is only that, if financing and balance-sheet conditions enter \(\phi_{i,t}^{\mathrm{opt}}(T)\), the OFB should covary positively with \(\operatorname{ABCP90\!-\!DTB3}_{t}\) after controlling for maturity, quoted liquidity, and broader financial conditions.
	
	Funding covariation and the year-end discontinuity identify different dimensions of the relative wedge. Funding covariation uses daily time-series variation, whereas the year-end discontinuity compares maturities observed on the same trade date. Persistence of the discontinuity across alternative benchmarks further tests whether benchmark-side pricing alone can explain the effect.
	
	\subsection{A fixed price wedge at the reporting boundary}
	\label{subsec:theory_reporting_boundary}
	
	Let \(T_{t}^{\mathrm{YE}}\) denote December 31 of the trade year and define
	\begin{equation}
		D_{t,T}^{\mathrm{YC}}
		=
		\mathbbm 1
		\left\{
		T>T_{t}^{\mathrm{YE}}
		\right\}.
		\label{eq:theory_year_crossing}
	\end{equation}
	
	A contract maturing exactly on December 31 is classified as noncrossing. Thus, \(D_{t,T}^{\mathrm{YC}}\) indicates whether the position must remain outstanding beyond the reporting boundary.
	
	If first exposure to that boundary creates a fixed price wedge, write
	\begin{equation}
		\phi_{i,t}^{\mathrm{opt}}(T)
		=
		\overline\phi_{i,t}^{\mathrm{opt}}(T)
		+
		\kappa_{i,t}
		\frac{D_{t,T}^{\mathrm{YC}}}{\tau_{t,T}},
		\label{eq:theory_fixed_boundary_wedge}
	\end{equation}
	where \(\overline\phi_{i,t}^{\mathrm{opt}}(T)\) is continuous through the boundary and \(\kappa_{i,t}\) is an unannualized boundary price wedge. The factor \(1/\tau_{t,T}\) converts a fixed price wedge into an annualized rate difference; it does not imply that the underlying price cost declines with maturity.
	
	As an alternative, an annualized flow premium would imply
	\begin{equation}
		\phi_{i,t}^{\mathrm{opt}}(T)
		=
		\overline\phi_{i,t}^{\mathrm{opt}}(T)
		+
		\lambda_{i,t}
		D_{t,T}^{\mathrm{YC}}.
		\label{eq:theory_flow_boundary_wedge}
	\end{equation}
	
	Under \Cref{eq:theory_fixed_boundary_wedge}, the unannualized effect is \(\kappa_{i,t}\). Under \Cref{eq:theory_flow_boundary_wedge}, it is \(\tau_{t,T}\lambda_{i,t}\) and therefore increases with contract duration. The horse race between \(D_{t,T}^{\mathrm{YC}}/\tau_{t,T}\) and \(D_{t,T}^{\mathrm{YC}}\) in \Cref{eq:data_functional_form} distinguishes these two functional forms.
	
	For the strict same-date local design, define the unannualized outcome as
	\begin{equation}
		G_{i,t}^{b}(T)
		=
		\tau_{t,T}
		\operatorname{OFB}_{i,t}^{b}(T).
		\label{eq:theory_unannualized_outcome}
	\end{equation}
	
	By \Cref{eq:theory_ofb}, this outcome can equivalently be written as
	\begin{equation}
		G_{i,t}^{b}(T)
		=
		10^{4}
		\log\!\left(
		\frac{D_{t}^{b}(T)}
		{\widehat B_{i,t}(T)}
		\right),
		\label{eq:theory_unannualized_log_wedge}
	\end{equation}
	so \(G_{i,t}^{b}(T)\) is the unannualized log-price difference between the option-implied and benchmark discount factors.
	
	Combining \Cref{eq:theory_ofb_decomposition,eq:theory_fixed_boundary_wedge,eq:theory_unannualized_outcome} gives
	\begin{equation}
		G_{i,t}^{b}(T)
		=
		\tau_{t,T}
		\left[
		CY_{t}^{b}(T)
		+
		\overline\phi_{i,t}^{\mathrm{opt}}(T)
		-
		\phi_{t}^{b}(T)
		\right]
		+
		\kappa_{i,t}
		D_{t,T}^{\mathrm{YC}}.
		\label{eq:theory_unannualized_decomposition}
	\end{equation}
	
	If \(\tau_{t,T}\) and the bracketed component in \Cref{eq:theory_unannualized_decomposition} are continuous at \(T=T_{t}^{\mathrm{YE}}\), the discontinuity in \(G_{i,t}^{b}(T)\) equals \(\kappa_{i,t}\):
	\begin{equation}
		\lim_{T\downarrow T_{t}^{\mathrm{YE}}}
		G_{i,t}^{b}(T)
		-
		\lim_{T\uparrow T_{t}^{\mathrm{YE}}}
		G_{i,t}^{b}(T)
		=
		\kappa_{i,t}.
		\label{eq:theory_boundary_jump}
	\end{equation}
	
	The transformation removes the mechanical \(1/\tau_{t,T}\) amplification of annualized OFB and gives the local coefficient a direct price interpretation. The result in \Cref{eq:theory_boundary_jump}, however, does not by itself establish an option-side source: identification requires \(CY_{t}^{b}(T)\) and \(\phi_{t}^{b}(T)\) to remain continuous at the cutoff. The benchmark decomposition and alternative-curve tests examine this restriction.
	
	The theoretical \(\kappa_{i,t}\) may vary across markets and dates. Pooled estimates therefore recover design-specific average boundary effects, while annual estimates characterize their time variation.
	
	\paragraph{First and subsequent boundaries.}
	
	Long-dated contracts can span multiple December 31 boundaries. The specification described in \Cref{app:additional_yearend_crossings} and reported in \Cref{tab:additional_yearend_crossings} therefore separates the first, second, and third crossings. The first is estimated precisely, whereas later crossings have much thinner support and cannot distinguish zero from economically meaningful positive effects.
	
	\subsection{Empirical predictions}
	\label{subsec:theory_empirical_predictions}
	
	First, the OFB should rise with \(\Delta_{t}^{\mathrm{ABCP-TB}}\) if asset-backed funding and balance-sheet conditions enter the option-side wedge. This is a reduced-form sign prediction, not a structural pass-through restriction.
	
	Second, a distinct positive discontinuity may remain when maturity first crosses December 31. The funding coefficient uses daily time-series variation, whereas the local discontinuity is identified from cross-maturity variation within the same trade date.
	
	Third, a fixed boundary charge predicts a positive coefficient on \(D_{t,T}^{\mathrm{YC}}/\tau_{t,T}\) with little incremental contribution from \(D_{t,T}^{\mathrm{YC}}\). The first crossing should also be positive, while the theory places no zero restriction on subsequent crossings (\Cref{app:additional_yearend_crossings,tab:additional_yearend_crossings}).
	
	Fourth, if the discontinuity is not generated by a particular benchmark's curve construction or year-end turn, OFBs constructed using DGS, DTB, and a non-Treasury benchmark should exhibit similar jumps on identical option cells.
	
	Together, these tests do not fully decompose the OFB or identify a unique funding, collateral, margin, or regulatory mechanism. They instead test whether the OFB contains a funding-related component and a separate reporting-boundary wedge. Either finding constrains the relative-purity condition in \Cref{eq:theory_relative_purity} required to treat the option-implied rate as a frictionless primitive.
	
	\section{Data and Empirical Design}
	\label{sec:data}
	
	Standard monthly options form the baseline sample for full-maturity and local year-end tests, while regular weeklies provide an independent local sample with denser maturity support. The two contract groups are never pooled. Detailed quote processing, contract selection, benchmark construction, maturity matching, and sample audits are reported in \Cref{app:pipeline_construction}; synthetic-forward fit and precision are reported in \Cref{app:ab_fit}.
	
	Throughout, \(i\) indexes the option market, \(t\) trade date, \(s\) trading minute, \(T\) maturity, \(K\) strike, and \(b\) benchmark. Time to maturity, \(\tau_{t,T}\), uses ACT/365. All regressions are estimated separately by market and benchmark.
	
	\subsection{Option-implied discount factors and sample construction}
	\label{subsec:data_option_discount}
	
	The option data are one-minute NBBO quotes for European-style SPX, RUT, SPXW, and RUTW index options, with contract and settlement conventions following Cboe specifications \citep{CBOESPX26,CBOERUT26}. Standard monthlies cover June 2012--December 2025 and \(30\)--\(1{,}095\) days to maturity. Regular weeklies cover SPXW over the same period and RUTW from January 2016, with \(7\)--\(365\) days to maturity. Exchange-calendar, expiration, settlement, early-close, quote, and matching rules are detailed in \Cref{app:option_pipeline}.
	
	For each market--minute--maturity cell, I estimate
	\begin{equation}
		C_{i,t,s}(K,T)-P_{i,t,s}(K,T)
		=
		a_{i,t,s}(T)
		+
		\xi_{i,t,s}(T)K
		+
		u_{i,t,s}(K,T),
		\label{eq:data_ab_regression}
	\end{equation}
	so the common strike slope identifies
	\begin{equation}
		\widehat B_{i,t,s}^{\mathrm{AB}}(T)
		=
		-\widehat\xi_{i,t,s}(T).
		\label{eq:data_ab_discount}
	\end{equation}
	
	The baseline \(\widehat B_{i,t}(T)\) is the intraday median of quality-screened estimates within each market--trade-date--maturity cell. The quoted-liquidity measure \(\operatorname{BA}_{i,t}(T)\) is aggregated at the same level. Estimation screens and aggregation are detailed in \Cref{app:option_discount_estimators,app:liquidity_intraday_aggregation}; adjacent-strike Box estimates provide a measurement check in \Cref{app:full_panel_sample_measurement,tab:full_panel_sample_measurement_robustness}.
	
	\subsection{Benchmark discount curves and analysis panels}
	\label{subsec:data_benchmarks}
	
	The primary benchmark is a DGS Treasury discount curve supporting maturities through three years. DTB provides a distinct Treasury-bill curve through one year. Both are constructed independently of option prices, matched without future information, and used only within observed curve support; details are in \Cref{app:benchmark_curve_construction}.
	
	Full-panel regressions use \(30\)--\(1{,}095\) days for DGS and \(30\)--\(365\) days for DTB. Direct DGS--DTB comparisons and the monthly local design use exactly matched cells within \(30\)--\(365\) days, while the weekly local design uses exact common cells within \(7\)--\(365\) days. The OFB is defined in \Cref{eq:theory_ofb}; a positive value means that the option-implied rate exceeds the matched benchmark rate.
	
	\begin{table}[!t]
		\centering
		\singlespacing
		\caption{Descriptive Statistics for the Standard-Monthly Option Funding Basis}
		\label{tab:descriptives}
		\begin{threeparttable}
			\footnotesize
			\setlength{\tabcolsep}{5pt}
			\renewcommand{\arraystretch}{1.02}
			\begin{tabular*}{0.96\textwidth}{@{\extracolsep{\fill}}llrrrrrr@{}}
				\toprule
				Benchmark
				& Market
				& Observations
				& Mean
				& Std. dev.
				& p5
				& Median
				& p95 \\
				\midrule
				DGS 3Y full
				& SPX
				& 41,404
				& 34.00
				& 16.49
				& 12.88
				& 33.25
				& 58.38 \\
				DGS 3Y full
				& RUT
				& 29,244
				& 38.06
				& 18.09
				& 12.48
				& 37.23
				& 67.30 \\
				DTB 1Y full
				& SPX
				& 28,000
				& 37.35
				& 16.83
				& 15.67
				& 36.31
				& 62.46 \\
				DTB 1Y full
				& RUT
				& 19,038
				& 40.28
				& 18.37
				& 16.34
				& 39.29
				& 68.39 \\
				\bottomrule
			\end{tabular*}
		\end{threeparttable}
		
		\vspace{0.4em}
		
		\begin{minipage}{0.96\textwidth}
			\footnotesize
			\textit{Notes:} The table reports descriptive statistics, in basis points, for the annualized AB-based OFB. The observation unit is a market--trade-date--maturity cell. DGS uses its natural \(30\)--\(1{,}095\)-day support and DTB its \(30\)--\(365\)-day support. Statistics precede availability restrictions from the daily funding control.
		\end{minipage}
	\end{table}
	
	\begin{table}[!t]
		\centering
		\singlespacing
		\caption{Descriptive Statistics for the Regular-Weekly Option Funding Basis}
		\label{tab:weekly_descriptives}
		\begin{threeparttable}
			\footnotesize
			\setlength{\tabcolsep}{5pt}
			\renewcommand{\arraystretch}{1.02}
			\begin{tabular*}{0.96\textwidth}{@{\extracolsep{\fill}}llrrrrrr@{}}
				\toprule
				Benchmark
				& Market
				& Observations
				& Mean
				& Std. dev.
				& p5
				& Median
				& p95 \\
				\midrule
				DGS matched
				& SPXW
				& 39,179
				& 49.97
				& 39.65
				& 11.47
				& 43.88
				& 110.80 \\
				DGS matched
				& RUTW
				& 19,327
				& 55.15
				& 108.56
				& \(-22.01\)
				& 45.76
				& 184.07 \\
				DTB matched
				& SPXW
				& 39,179
				& 50.66
				& 39.62
				& 12.53
				& 44.13
				& 112.24 \\
				DTB matched
				& RUTW
				& 19,327
				& 56.18
				& 108.61
				& \(-19.14\)
				& 46.32
				& 185.49 \\
				\bottomrule
			\end{tabular*}
		\end{threeparttable}
		
		\vspace{0.4em}
		
		\begin{minipage}{0.96\textwidth}
			\footnotesize
			\textit{Notes:} The table reports descriptive statistics, in basis points, for the annualized AB-based OFB in the regular-weekly sample. The observation unit is a market--trade-date--maturity cell. DGS and DTB use exactly matched \(7\)--\(365\)-day cells within each market, so benchmark rows share identical option-side observations. Statistics precede the fourth-quarter, local-bandwidth, and same-date-support restrictions.
		\end{minipage}
	\end{table}
	
	As shown in \Cref{tab:descriptives,tab:weekly_descriptives}, OFB levels are higher and more dispersed for the RUT family, especially among regular weeklies, while DGS and DTB yield similar distributions within each contract group.
	
	\subsection{Full-maturity cross-sectional regressions around year-end}
	\label{subsec:data_full_panel}
	
	The full-panel analysis uses standard monthlies to test whether the OFB covaries with asset-backed funding conditions and whether a distinct year-end wedge remains after controlling for smooth maturity structure.
	
	The funding proxy is the same-day spread between the 90-day AA ABCP rate and the three-month Treasury-bill rate:
	\begin{equation}
		\Delta_{t}^{\mathrm{ABCP-TB}}
		=
		100
		\left[
		\operatorname{ABCP90}_{t}
		-
		\operatorname{DTB3}_{t}
		\right].
		\label{eq:data_abcp_tb}
	\end{equation}
	
	The baseline requires both rates on the option trade date. A robustness specification carries only a previously observed completed spread for at most three calendar days and never uses future information. Data sources and matching rules are detailed in \Cref{app:data_sources,app:full_panel_sample_measurement}.
	
	Residual maturity is centered and standardized within each estimation sample and enters as a cubic polynomial. The common controls are
	\begin{equation}
		\begin{aligned}
			\mathbf x_{i,t,T}^{b\prime}\boldsymbol\theta
			={}&
			\alpha
			+
			\mu_{\operatorname{year}(t)}
			+
			\sum_{\ell=1}^{3}
			\delta_{\ell}
			\widetilde{\tau}_{t,T}^{\,\ell}
			+
			\gamma
			\frac{\operatorname{BA}_{i,t}(T)}{\tau_{t,T}}
			\\
			&+
			\eta\operatorname{NFCI}_{t}
			+
			\beta_{\mathrm{ABCP-TB}}
			\Delta_{t}^{\mathrm{ABCP-TB}}.
		\end{aligned}
		\label{eq:data_full_panel_controls}
	\end{equation}
	
	Calendar-year fixed effects absorb low-frequency common variation; the remaining controls capture maturity structure, quoted liquidity, broad financial conditions, and asset-backed funding conditions. \Cref{subsec:theory_empirical_predictions} predicts \(\beta_{\mathrm{ABCP-TB}}>0\).
	
	Using \(D_{t,T}^{\mathrm{YC}}\) from \Cref{eq:theory_year_crossing}, I estimate
	\begin{align}
		\mathcal M_{1}:\qquad
		\operatorname{OFB}_{i,t}^{b}(T)
		&=
		\mathbf x_{i,t,T}^{b\prime}\boldsymbol\theta^{(1)}
		+
		\varepsilon_{i,t,T}^{b,(1)},
		\label{eq:data_full_panel_m1}
		\\
		\mathcal M_{2}:\qquad
		\operatorname{OFB}_{i,t}^{b}(T)
		&=
		\mathbf x_{i,t,T}^{b\prime}\boldsymbol\theta^{(2)}
		+
		\kappa
		\frac{D_{t,T}^{\mathrm{YC}}}{\tau_{t,T}}
		+
		\varepsilon_{i,t,T}^{b,(2)}.
		\label{eq:data_full_panel_m2}
	\end{align}
	
	The two models use identical observations. The funding coefficient measures reduced-form conditional covariation, while \(\kappa\) measures the average unannualized boundary wedge. First and subsequent crossings are separated in \Cref{app:additional_yearend_crossings,tab:additional_yearend_crossings}.
	
	Inference uses Newey--West HAC(21) on trade-date scores \citep{NW87} and two-sided calendar-year score sign flips for the main boundary coefficients \citep{CRS17}. Complete-year samples, common support, limited prior-day funding carry, Box estimates, alternative HAC bandwidths, and leave-one-year-out diagnostics appear in \Cref{app:full_panel_robustness}.
	
	\subsection{Strict same-date local design at the year-end boundary}
	\label{subsec:data_local_design}
	
	Define maturity distance from December 31 of the trade year as
	\begin{equation}
		h_{t,T}
		=
		\operatorname{days}
		\left(
		T-T_{t}^{\mathrm{YE}}
		\right).
		\label{eq:data_running_variable}
	\end{equation}
	
	Crossing maturities have \(h_{t,T}>0\). Contracts with \(h_{t,T}=0\) are noncrossing by definition but excluded from the baseline local sample. The outcome is
	\begin{equation}
		G_{i,t}^{b}(T)
		=
		\tau_{t,T}
		\operatorname{OFB}_{i,t}^{b}(T)
		=
		10^{4}
		\log\!\left(
		\frac{D_{t}^{b}(T)}
		{\widehat B_{i,t}(T)}
		\right).
		\label{eq:data_local_outcome}
	\end{equation}
	
	The monthly design uses exact DGS--DTB common cells within \(30\)--\(365\) days and the weekly design exact common cells within \(7\)--\(365\) days. Both require fourth-quarter trade dates, \(0<\lvert h_{t,T}\rvert<90\), at most one year-end crossing, and maturities on both sides of the cutoff on every retained trade date.
	
	I estimate
	\begin{align}
		G_{i,t}^{b}(T)
		&=
		\alpha_t
		+
		\kappa
		\mathbbm 1
		\left\{
		h_{t,T}>0
		\right\}
		+
		\beta_{-}
		h_{t,T}
		\mathbbm 1
		\left\{
		h_{t,T}<0
		\right\}
		\notag\\
		&\quad
		+
		\beta_{+}
		h_{t,T}
		\mathbbm 1
		\left\{
		h_{t,T}>0
		\right\}
		+
		\gamma
		\operatorname{BA}_{i,t}(T)
		+
		\varepsilon_{i,t,T}^{b,\mathrm{L}},
		\label{eq:data_local_yearend}
	\end{align}
	with triangular weights \(w_{t,T}=1-\lvert h_{t,T}\rvert/90\).
	
	Trade-date fixed effects absorb all date-level conditions, including \(\operatorname{ABCP90\!-\!DTB3}_{t}\), so \(\kappa\) is identified entirely from cross-maturity variation within a trade date. Because maturities lie on discrete exchange-specified dates, I interpret the design as a strict same-date boundary comparison rather than a nonparametric continuous-running-variable limit. Inference uses HAC(21) trade-date scores and calendar-year episode sign flips \citep{LL10,NW87,CRS17}.
	
	\begin{table}[!t]
		\centering
		\singlespacing
		\caption{Local Maturity Support around Year-End: Standard Monthlies and Regular Weeklies}
		\label{tab:local_support_comparison}
		\begin{threeparttable}
			\footnotesize
			\setlength{\tabcolsep}{5.5pt}
			\renewcommand{\arraystretch}{1.05}
			\begin{tabular*}{0.96\textwidth}{@{\extracolsep{\fill}}lrrrr@{}}
				\toprule
				&
				\multicolumn{2}{c}{Standard monthly}
				&
				\multicolumn{2}{c}{Regular weekly} \\
				\cmidrule(lr){2-3}
				\cmidrule(lr){4-5}
				&
				SPX
				&
				RUT
				&
				SPXW
				&
				RUTW \\
				\midrule
				Local-sample observations
				& 1,993
				& 1,548
				& 7,256
				& 2,974 \\
				Two-sided-support trade dates
				& 479
				& 479
				& 694
				& 464 \\
				Year-end episodes
				& 14
				& 14
				& 14
				& 10 \\
				Maturities per trade date, median
				& 4
				& 3
				& 11
				& 6 \\
				Maturities per trade date, p25--p75
				& 4--4
				& 3--3
				& 9--12
				& 6--7 \\
				Nearest left-side distance, median
				& 12
				& 12
				& 4
				& 4 \\
				Nearest right-side distance, median
				& 18
				& 18
				& 6.5
				& 29 \\
				Worst-side distance, p95
				& 21
				& 79
				& 21
				& 88 \\
				\bottomrule
			\end{tabular*}
		\end{threeparttable}
		
		\vspace{0.4em}
		
		\begin{minipage}{0.96\textwidth}
			\footnotesize
			\textit{Notes:} The table compares maturity support under the baseline strict same-date local design on DGS--DTB matched cells. Monthly options use \(30\)--\(365\) days to maturity and regular weeklies \(7\)--\(365\) days; all samples require fourth-quarter trade dates, \(0<\lvert h_{t,T}\rvert<90\), at most one crossing, and both sides of the cutoff on each retained trade date. Nearest-side distances are measured in calendar days; the worst-side distance is the larger of the nearest distances on the two sides. SPX, RUT, and SPXW contain 14 year-end episodes from 2012--2025, and RUTW 10 from 2016--2025.
		\end{minipage}
	\end{table}
	
	\Cref{tab:local_support_comparison} shows that regular weeklies substantially increase maturity density, especially for SPXW. RUTW also improves local support but remains less balanced on the right side of the cutoff. Further diagnostics appear in \Cref{app:local_yearend_design,tab:local_rd_robustness}.
	
	\subsection{Functional form of the boundary effect}
	\label{subsec:data_boundary_functional_form}
	
	To distinguish the fixed wedge in \Cref{eq:theory_fixed_boundary_wedge} from the flow premium in \Cref{eq:theory_flow_boundary_wedge}, I use standard monthlies with at most one crossing and both crossing and noncrossing maturities on each retained trade date:
	\begin{align}
		\operatorname{OFB}_{i,t}^{b}(T)
		&=
		\alpha_t
		+
		\sum_{\ell=1}^{3}
		\delta_{\ell}
		\widetilde{\tau}_{t,T}^{\,\ell}
		+
		\pi
		D_{t,T}^{\mathrm{YC}}
		+
		\kappa
		\frac{D_{t,T}^{\mathrm{YC}}}{\tau_{t,T}}
		\notag\\
		&\quad
		+
		\gamma
		\frac{\operatorname{BA}_{i,t}(T)}{\tau_{t,T}}
		+
		\varepsilon_{i,t,T}^{b,\mathrm{F}}.
		\label{eq:data_functional_form}
	\end{align}
	
	Trade-date fixed effects absorb all date-level conditions. Hence \(\pi\) measures an annualized flow component and \(\kappa\) a fixed unannualized wedge; \(\pi=0\) with \(\kappa>0\) favors the fixed-price specification. Repeated crossings are examined in \Cref{app:additional_yearend_crossings,tab:additional_yearend_crossings}.
	
	\FloatBarrier
	
	\section{The Year-End Boundary Jump}
	\label{sec:jump}
	
	This section tests whether the OFB covaries with asset-backed short-term funding conditions and whether a distinct price change remains when maturity crosses December 31. The full-panel regressions estimate the conditional loading on the same-day \(\operatorname{ABCP90\!-\!DTB3}_{t}\) spread together with the average year-end boundary effect across the maturity structure. The strict same-date local design instead compares maturities on opposite sides of the cutoff within the same trade date, absorbing all date-level funding and financial conditions and identifying the boundary effect from cross-maturity variation alone.
	
	\subsection{Identification from full-panel regressions}
	\label{subsec:jump_regression}
	
	\subsubsection{Regression results}
	\label{subsub:regressions}
	
	I estimate \Cref{eq:data_full_panel_m1,eq:data_full_panel_m2} on the same-day sample with an exact \(\operatorname{ABCP90\!-\!DTB3}\) observation. Model \(\mathcal M_{1}\) estimates the conditional relation between asset-backed funding conditions and the OFB, while \(\mathcal M_{2}\) adds only \(D_{t,T}^{\mathrm{YC}}/\tau_{t,T}\) on identical observations. Thus, \(\beta_{\mathrm{ABCP-TB}}\) captures funding-related covariation and \(\kappa\) the additional unannualized year-end price wedge.
	
	\begin{table}[!t]
		\centering
		\singlespacing
		\caption{Full-Panel Regressions with the Year-End Boundary Term: SPX}
		\label{tab:yearend_regression_spx}
		\begin{threeparttable}
			\footnotesize
			\setlength{\tabcolsep}{2.8pt}
			\renewcommand{\arraystretch}{1.08}
			\begin{tabular*}{0.96\textwidth}{@{\extracolsep{\fill}}lrrrrrrrr@{}}
				\toprule
				&
				\multicolumn{4}{c}{DGS 3Y}
				&
				\multicolumn{4}{c}{DTB 1Y} \\
				\cmidrule(lr){2-5}
				\cmidrule(lr){6-9}
				&
				\multicolumn{2}{c}{\(\mathcal M_{1}\)}
				& \multicolumn{2}{c}{\(\mathcal M_{2}\)}
				& \multicolumn{2}{c}{\(\mathcal M_{1}\)}
				& \multicolumn{2}{c}{\(\mathcal M_{2}\)} \\
				\cmidrule(lr){2-3}
				\cmidrule(lr){4-5}
				\cmidrule(lr){6-7}
				\cmidrule(lr){8-9}
				&
				Coeff.
				& HAC \(t\)
				& Coeff.
				& HAC \(t\)
				& Coeff.
				& HAC \(t\)
				& Coeff.
				& HAC \(t\) \\
				\midrule
				
				\(D^{\mathrm{YC}}/\tau\)
				&
				&
				& \(2.938^{***}\)
				& (6.468)
				&
				&
				& \(2.869^{***}\)
				& (6.747) \\
				
				Exact \(p\)
				&
				&
				& \multicolumn{2}{c}{[0.0011]}
				&
				&
				& \multicolumn{2}{c}{[0.0011]} \\
				
				\addlinespace[0.3em]
				
				\(\Delta_{t}^{\mathrm{ABCP-TB}}\)
				& \(0.290^{***}\)
				& (4.243)
				& \(0.266^{***}\)
				& (4.616)
				& \(0.376^{***}\)
				& (4.773)
				& \(0.341^{***}\)
				& (5.344) \\
				
				\midrule
				
				Observations
				& \multicolumn{2}{c}{40,855}
				& \multicolumn{2}{c}{40,855}
				& \multicolumn{2}{c}{27,626}
				& \multicolumn{2}{c}{27,626} \\
				
				Trade dates
				& \multicolumn{2}{c}{3,371}
				& \multicolumn{2}{c}{3,371}
				& \multicolumn{2}{c}{3,371}
				& \multicolumn{2}{c}{3,371} \\
				
				Adjusted \(R^{2}\)
				& \multicolumn{2}{c}{0.362}
				& \multicolumn{2}{c}{0.405}
				& \multicolumn{2}{c}{0.340}
				& \multicolumn{2}{c}{0.398} \\
				
				\(\Delta\) adjusted \(R^{2}\)
				&
				&
				& \multicolumn{2}{c}{0.043}
				&
				&
				& \multicolumn{2}{c}{0.058} \\
				
				\bottomrule
			\end{tabular*}
		\end{threeparttable}
		
		\vspace{0.4em}
		
		\begin{minipage}{0.96\textwidth}
			\footnotesize
			\textit{Notes:} The dependent variable is the annualized OFB. DGS and DTB cover \(30\)--\(1{,}095\) and \(30\)--\(365\) days, respectively. All regressions include an intercept, calendar-year fixed effects, a cubic maturity polynomial, \(\operatorname{BA}/\tau\), the NFCI, and the same-day \(\Delta_{t}^{\mathrm{ABCP-TB}}=100(\operatorname{ABCP90}_{t}-\operatorname{DTB3}_{t})\). Model \(\mathcal M_{2}\) adds only \(D_{t,T}^{\mathrm{YC}}/\tau_{t,T}\) to \(\mathcal M_{1}\) on the identical sample. Parentheses report Newey--West HAC(21) \(t\)-statistics from trade-date scores; brackets report two-sided calendar-year score sign-flip \(p\)-values for the boundary term. \(^{*}\), \(^{**}\), and \(^{***}\) denote two-sided HAC significance at the \(10\%\), \(5\%\), and \(1\%\) levels.
		\end{minipage}
	\end{table}
	
	\begin{table}[!t]
		\centering
		\singlespacing
		\caption{Full-Panel Regressions with the Year-End Boundary Term: RUT}
		\label{tab:yearend_regression_rut}
		\begin{threeparttable}
			\footnotesize
			\setlength{\tabcolsep}{2.8pt}
			\renewcommand{\arraystretch}{1.08}
			\begin{tabular*}{0.96\textwidth}{@{\extracolsep{\fill}}lrrrrrrrr@{}}
				\toprule
				&
				\multicolumn{4}{c}{DGS 3Y}
				&
				\multicolumn{4}{c}{DTB 1Y} \\
				\cmidrule(lr){2-5}
				\cmidrule(lr){6-9}
				&
				\multicolumn{2}{c}{\(\mathcal M_{1}\)}
				& \multicolumn{2}{c}{\(\mathcal M_{2}\)}
				& \multicolumn{2}{c}{\(\mathcal M_{1}\)}
				& \multicolumn{2}{c}{\(\mathcal M_{2}\)} \\
				\cmidrule(lr){2-3}
				\cmidrule(lr){4-5}
				\cmidrule(lr){6-7}
				\cmidrule(lr){8-9}
				&
				Coeff.
				& HAC \(t\)
				& Coeff.
				& HAC \(t\)
				& Coeff.
				& HAC \(t\)
				& Coeff.
				& HAC \(t\) \\
				\midrule
				
				\(D^{\mathrm{YC}}/\tau\)
				&
				&
				& \(3.199^{***}\)
				& (5.302)
				&
				&
				& \(3.165^{***}\)
				& (5.773) \\
				
				Exact \(p\)
				&
				&
				& \multicolumn{2}{c}{[0.0005]}
				&
				&
				& \multicolumn{2}{c}{[0.0005]} \\
				
				\addlinespace[0.3em]
				
				\(\Delta_{t}^{\mathrm{ABCP-TB}}\)
				& \(0.250^{***}\)
				& (4.026)
				& \(0.221^{***}\)
				& (4.476)
				& \(0.338^{***}\)
				& (4.506)
				& \(0.295^{***}\)
				& (5.232) \\
				
				\midrule
				
				Observations
				& \multicolumn{2}{c}{28,841}
				& \multicolumn{2}{c}{28,841}
				& \multicolumn{2}{c}{18,767}
				& \multicolumn{2}{c}{18,767} \\
				
				Trade dates
				& \multicolumn{2}{c}{3,370}
				& \multicolumn{2}{c}{3,370}
				& \multicolumn{2}{c}{3,370}
				& \multicolumn{2}{c}{3,370} \\
				
				Adjusted \(R^{2}\)
				& \multicolumn{2}{c}{0.267}
				& \multicolumn{2}{c}{0.317}
				& \multicolumn{2}{c}{0.237}
				& \multicolumn{2}{c}{0.310} \\
				
				\(\Delta\) adjusted \(R^{2}\)
				&
				&
				& \multicolumn{2}{c}{0.051}
				&
				&
				& \multicolumn{2}{c}{0.074} \\
				
				\bottomrule
			\end{tabular*}
		\end{threeparttable}
		
		\vspace{0.4em}
		
		\begin{minipage}{0.96\textwidth}
			\footnotesize
			\textit{Notes:} The dependent variable is the annualized OFB. DGS and DTB cover \(30\)--\(1{,}095\) and \(30\)--\(365\) days, respectively. All regressions include an intercept, calendar-year fixed effects, a cubic maturity polynomial, \(\operatorname{BA}/\tau\), the NFCI, and the same-day \(\Delta_{t}^{\mathrm{ABCP-TB}}=100(\operatorname{ABCP90}_{t}-\operatorname{DTB3}_{t})\). Model \(\mathcal M_{2}\) adds only \(D_{t,T}^{\mathrm{YC}}/\tau_{t,T}\) to \(\mathcal M_{1}\) on the identical sample. Parentheses report Newey--West HAC(21) \(t\)-statistics from trade-date scores; brackets report two-sided calendar-year score sign-flip \(p\)-values for the boundary term. \(^{*}\), \(^{**}\), and \(^{***}\) denote two-sided HAC significance at the \(10\%\), \(5\%\), and \(1\%\) levels.
		\end{minipage}
	\end{table}
	
	Across \Cref{tab:yearend_regression_spx,tab:yearend_regression_rut}, the year-end coefficient ranges from \(2.869\) to \(3.199\) bp, with HAC(21) \(t\)-statistics of \(5.30\)--\(6.75\) and exact \(p\)-values no larger than \(0.0011\). DGS and DTB produce closely similar estimates, including on exactly matched cells in \Cref{tab:full_panel_sample_measurement_robustness}.
	
	The same-day \(\operatorname{ABCP90\!-\!DTB3}\) loading is positive in every specification. In \(\mathcal M_{1}\), it ranges from \(0.250\) to \(0.376\), and remains \(0.221\)--\(0.341\) after adding the boundary term. Thus, a 10 bp wider asset-backed funding spread is associated with about \(2.2\)--\(3.4\) bp higher annualized OFB. Meanwhile, adding \(D_{t,T}^{\mathrm{YC}}/\tau_{t,T}\) raises adjusted \(R^{2}\) by \(4.3\)--\(7.4\) percentage points. The data therefore contain both funding-related covariation and a distinct reporting-boundary component.
	
	\subsubsection{Functional form of the jump}
	\label{subsub:functional_form}
	
	A fixed price wedge loads on \(D_{t,T}^{\mathrm{YC}}/\tau_{t,T}\), whereas a maturity-invariant annualized flow premium loads on \(D_{t,T}^{\mathrm{YC}}\). \Cref{eq:data_functional_form} includes both terms in the same trade-date fixed-effect regression.
	
	\begin{table}[!t]
		\centering
		\singlespacing
		\caption{Fixed Boundary Wedge versus Annualized Flow Premium}
		\label{tab:yearend_functional_form}
		\begin{threeparttable}
			\footnotesize
			\setlength{\tabcolsep}{5.5pt}
			\renewcommand{\arraystretch}{1.03}
			\begin{tabular*}{0.96\textwidth}{@{\extracolsep{\fill}}lrrrr@{}}
				\toprule
				&
				\multicolumn{2}{c}{DGS 3Y}
				&
				\multicolumn{2}{c}{DTB 1Y} \\
				\cmidrule(lr){2-3}
				\cmidrule(lr){4-5}
				&
				SPX
				&
				RUT
				&
				SPX
				&
				RUT \\
				\midrule
				Fixed boundary wedge: \(D^{\mathrm{YC}}/\tau\)
				& 2.409
				& 2.926
				& 3.119
				& 2.881 \\
				HAC(21) \(t\)-statistic
				& 3.033
				& 3.571
				& 3.880
				& 3.412 \\
				Exact \(p\)-value
				& 0.0110
				& 0.0045
				& 0.0018
				& 0.0046 \\
				\addlinespace[0.3em]
				Annualized flow premium: \(D^{\mathrm{YC}}\)
				& 0.932
				& 0.006
				& \(-0.665\)
				& 0.391 \\
				HAC(21) \(t\)-statistic
				& 0.596
				& 0.004
				& \(-0.399\)
				& 0.199 \\
				Exact \(p\)-value
				& 0.6384
				& 0.9977
				& 0.7980
				& 0.8330 \\
				Observations
				& 31,300
				& 21,653
				& 23,742
				& 14,312 \\
				\bottomrule
			\end{tabular*}
		\end{threeparttable}
		
		\vspace{0.4em}
		
		\begin{minipage}{0.96\textwidth}
			\footnotesize
			\textit{Notes:} The table estimates \Cref{eq:data_functional_form} using contracts with at most one crossing and trade dates containing maturities on both sides of year-end. All regressions include trade-date fixed effects, a cubic maturity polynomial, and \(\operatorname{BA}/\tau\). HAC statistics use trade-date scores; exact \(p\)-values use two-sided calendar-year score sign flips.
		\end{minipage}
	\end{table}
	
	\Cref{tab:yearend_functional_form} favors a fixed price wedge. The coefficient on \(D_{t,T}^{\mathrm{YC}}/\tau_{t,T}\) is \(2.409\)--\(3.119\) bp and significant in all four specifications, while the coefficient on \(D_{t,T}^{\mathrm{YC}}\) changes sign and is never significant. The long-maturity specification in \Cref{app:additional_yearend_crossings,tab:additional_yearend_crossings} likewise identifies a positive first crossing, but later crossings are too imprecise to establish whether additional boundaries carry further charges.
	
	\subsection{Identification from the strict same-date local design}
	\label{subsec:jump_RD}
	
	\subsubsection{Standard monthly options}
	\label{subsub:RD_monthly}
	
	I apply the strict same-date local design in \Cref{eq:data_local_yearend} to standard monthly options. The sample is restricted to option cells observed under both DGS and DTB from 2012 through 2025 and requires fourth-quarter trade dates, \(0<\lvert h_{t,T}\rvert<90\), at most one year-end crossing, and maturities on both sides of the cutoff on the same trade date. Estimation uses triangular weights, trade-date fixed effects, separate local-linear maturity slopes on the two sides of the boundary, and the maturity-level median bid--ask control.
	
	\begin{table}[!t]
		\centering
		\singlespacing
		\caption{Same-Date Local Year-End Discontinuity in Standard Monthly Options}
		\label{tab:monthly_local_yearend_rd}
		\begin{threeparttable}
			\footnotesize
			\setlength{\tabcolsep}{4.5pt}
			\renewcommand{\arraystretch}{1.02}
			\begin{tabular*}{0.96\textwidth}{@{\extracolsep{\fill}}llrrrrrr@{}}
				\toprule
				Market
				& Benchmark
				& Jump
				& HAC(21) \(t\)
				& Exact \(p\)
				& Observations
				& \makecell{Two-sided\\trade dates}
				& Episodes \\
				\midrule
				SPX
				& DGS
				& 2.678
				& 5.264
				& 0.0004
				& 1,993
				& 479
				& 14 \\
				SPX
				& DTB
				& 2.650
				& 5.276
				& 0.0004
				& 1,993
				& 479
				& 14 \\
				RUT
				& DGS
				& 2.844
				& 4.921
				& 0.0012
				& 1,548
				& 479
				& 14 \\
				RUT
				& DTB
				& 2.776
				& 4.904
				& 0.0012
				& 1,548
				& 479
				& 14 \\
				\bottomrule
			\end{tabular*}
		\end{threeparttable}
		
		\vspace{0.4em}
		
		\begin{minipage}{0.96\textwidth}
			\footnotesize
			\textit{Notes:} The dependent variable is the unannualized AB option funding basis \(G_{i,t}^{b}(T)=\tau_{t,T}\operatorname{OFB}_{i,t}^{b}(T)\), measured in basis points. The sample consists of standard-monthly cells observed identically under DGS and DTB from 2012 through 2025. All regressions require fourth-quarter trade dates, \(0<\lvert h_{t,T}\rvert<90\), at most one year-end crossing, and maturities on both sides of the cutoff on the same trade date. Estimation uses triangular weights, weighted trade-date fixed effects, separate local-linear maturity slopes on the two sides of the boundary, and the maturity-level median bid--ask spread. HAC(21) \(t\)-statistics are computed from trade-date scores. Exact \(p\)-values are from two-sided sign-flip tests that reverse the scores of the 14 calendar-year episodes under the null.
		\end{minipage}
	\end{table}
	
	\begin{figure}[!t]
		\centering
		\includegraphics[width=\textwidth]{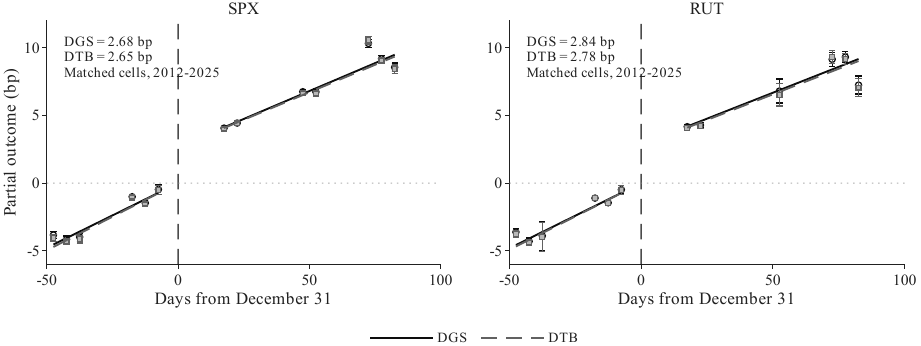}
		\caption{Same-Date Local Year-End Discontinuity in Standard Monthly Options}
		\label{fig:ofbrd}
		\begin{minipage}{0.96\textwidth}
			\footnotesize
			\textit{Notes:} The horizontal axis is the calendar-day distance \(h_{t,T}\) from December 31 of the trade year to option maturity. The vertical axis is the unannualized OFB after partialling out trade-date fixed effects, separate local-linear maturity slopes on the two sides of the boundary, and the maturity-level median bid--ask spread. Points and bars report 5-day-bin means and trade-date-level standard errors, and the lines are local-linear fits on the two sides of the cutoff using triangular weights. The sample and regression design are the same as in \Cref{tab:monthly_local_yearend_rd}. Values reported inside the figure are the estimated unannualized boundary jumps.
		\end{minipage}
	\end{figure}
	
	\Cref{fig:ofbrd,tab:monthly_local_yearend_rd} show a positive discontinuity in the unannualized OFB when maturity crosses December 31 in both markets. The SPX jump is \(2.678\) bp under DGS and \(2.650\) bp under DTB, while the corresponding RUT estimates are \(2.844\) and \(2.776\) bp. HAC(21) \(t\)-statistics range from \(4.90\) to \(5.28\), and all exact \(p\)-values are at most \(0.0012\).
	
	Identification comes from 14 year-end episodes and 479 trade dates with maturities on both sides of the cutoff. The final samples contain \(1{,}993\) market--trade-date--maturity cells for SPX and \(1{,}548\) for RUT. The estimates therefore exploit cross-maturity price differences within the same trade date rather than differences in financial conditions across year-end dates.
	
	The DGS--DTB difference is only \(0.028\) bp for SPX and \(0.068\) bp for RUT. Thus, two Treasury curves constructed from different source data and procedures produce nearly identical discontinuities on exactly the same option cells, weakening explanations based on a particular Treasury bootstrap or interpolation procedure.
	
	Because the outcome \(G_{i,t}^{b}(T)\) removes the \(1/\tau_{t,T}\) amplification induced by annualization, the \(2.65\)--\(2.84\) bp jump cannot be attributed mechanically to shorter maturity. Instead, it is consistent with a fixed unannualized price wedge that appears when the contract begins to span the reporting boundary.
	
	\subsubsection{Regular weekly options}
	\label{subsub:RD_weekly}
	
	Regular weekly options provide an independent replication sample with substantially denser maturity support around December 31. Standard monthlies have a relatively sparse local maturity grid, as documented in \Cref{tab:local_support_comparison}, so the weekly sample provides a direct test of whether the positive discontinuity depends on the fixed standard-monthly expiration structure.
	
	The sample consists of the SPXW and RUTW regular weeklies defined in \Cref{subsec:data_option_discount}, using only trade-date--maturity cells exactly shared by DGS and DTB within each market. Daily, end-of-month, standard-monthly, and other irregular expirations are excluded. I then apply the strict same-date local design in \Cref{subsec:data_local_design} separately to each market.
	
	\begin{figure}[!t]
		\centering
		\includegraphics[width=\textwidth]{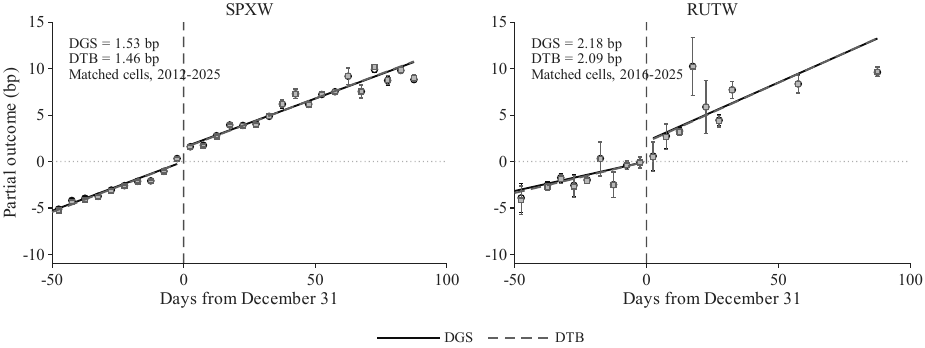}
		\caption{Same-Date Local Year-End Discontinuity in Regular Weekly Options}
		\label{fig:rd_weekly}
		
		\vspace{0.4em}
		
		\begin{minipage}{0.96\textwidth}
			\footnotesize
			\textit{Notes:} The horizontal axis is the calendar-day distance \(h_{t,T}\) from December 31 of the trade year to option maturity. The vertical axis is the unannualized OFB after partialling out trade-date fixed effects, separate local-linear maturity slopes on the two sides of the boundary, and the maturity-level median bid--ask spread. Points and bars report 5-day-bin means and trade-date-level standard errors, and the lines are local-linear fits on the two sides of the cutoff using triangular weights. The sample and regression design are the same as in \Cref{tab:weekly_local_yearend_rd}. Values reported inside the figure are the estimated unannualized boundary jumps.
		\end{minipage}
	\end{figure}
	
	\Cref{fig:rd_weekly,tab:weekly_local_yearend_rd} show positive boundary jumps in the weekly samples as well. For SPXW, the estimates are \(1.530\) bp under DGS and \(1.456\) bp under DTB; for RUTW, they are \(2.183\) and \(2.093\) bp. On identical option cells, the DGS--DTB differences are only \(0.074\) bp for SPXW and \(0.091\) bp for RUTW.
	
	For SPXW, the HAC(21) \(t\)-statistics are \(4.892\) and \(4.661\), with exact \(p\)-values of \(0.0101\) and \(0.0129\). Identification comes from 14 episodes, 694 trade dates with two-sided support, and \(7{,}256\) option cells. The median nearest maturity lies only 4 days below and 6.5 days above the cutoff (\Cref{tab:local_support_comparison}). SPXW therefore provides a direct test against the possibility that the monthly discontinuity is mechanically generated by sparse standard-monthly maturity support.
	
	\begin{table}[!t]
		\centering
		\singlespacing
		\caption{Same-Date Local Year-End Discontinuity in Regular Weekly Options}
		\label{tab:weekly_local_yearend_rd}
		\begin{threeparttable}
			\footnotesize
			\setlength{\tabcolsep}{4.5pt}
			\renewcommand{\arraystretch}{1.02}
			\begin{tabular*}{0.96\textwidth}{@{\extracolsep{\fill}}llrrrrrr@{}}
				\toprule
				Market
				& Benchmark
				& Jump
				& HAC(21) \(t\)
				& Exact \(p\)
				& Observations
				& \makecell{Two-sided\\trade dates}
				& Episodes \\
				\midrule
				SPXW
				& DGS
				& 1.530
				& 4.892
				& 0.0101
				& 7,256
				& 694
				& 14 \\
				SPXW
				& DTB
				& 1.456
				& 4.661
				& 0.0129
				& 7,256
				& 694
				& 14 \\
				RUTW
				& DGS
				& 2.183
				& 2.682
				& 0.0957
				& 2,974
				& 464
				& 10 \\
				RUTW
				& DTB
				& 2.093
				& 2.570
				& 0.1211
				& 2,974
				& 464
				& 10 \\
				\bottomrule
			\end{tabular*}
		\end{threeparttable}
		
		\vspace{0.4em}
		
		\begin{minipage}{0.96\textwidth}
			\footnotesize
			\textit{Notes:} The dependent variable is the unannualized AB option funding basis \(G_{i,t}^{b}(T)=\tau_{t,T}\operatorname{OFB}_{i,t}^{b}(T)\), measured in basis points. The sample consists of regular-weekly cells observed identically under DGS and DTB within each market. The SPXW sample covers 2012--2025 and the RUTW sample 2016--2025, yielding 14 and 10 year-end episodes, respectively. All regressions require fourth-quarter trade dates, \(0<\lvert h_{t,T}\rvert<90\), at most one year-end crossing, and maturities on both sides of the cutoff on the same trade date. Estimation uses triangular weights, weighted trade-date fixed effects, separate local-linear maturity slopes on the two sides of the boundary, and the maturity-level median bid--ask spread, following the same design as \Cref{tab:monthly_local_yearend_rd}. HAC(21) \(t\)-statistics are computed from trade-date scores, and exact \(p\)-values are from two-sided sign-flip tests that reverse calendar-year episode scores under the null.
		\end{minipage}
	\end{table}
	
	The RUTW jumps are likewise positive at roughly \(2.1\)--\(2.2\) bp under both benchmarks, with HAC(21) \(t\)-statistics of \(2.682\) and \(2.570\). The corresponding exact \(p\)-values are \(0.0957\) and \(0.1211\), reflecting the lower power of exact inference with only 10 episodes beginning in 2016 and relatively unbalanced right-side maturity support. I therefore treat RUTW as evidence that the sign and economic magnitude replicate in a separate contract group rather than as an equally strong statistical replication.
	
	The role of the weekly sample is not to produce a larger coefficient than the monthly sample. Its contribution is that the positive discontinuity survives in SPXW despite substantially denser maturity support around the cutoff. This makes it difficult to attribute the baseline evidence in \Cref{fig:ofbrd,tab:monthly_local_yearend_rd} to the fixed expiration calendar or sparse cutoff support of standard monthly options.
	
	\subsection{Economic Magnitude and Relation to Measured Convenience Yields}
	\label{subsec:jump_economic_magnitude}
	
	The boundary coefficients are measured in unannualized price units, not as \(2\)--\(3\) bp annualized interest-rate effects. Because the functional-form evidence in \Cref{tab:yearend_functional_form} favors a fixed price wedge, an unannualized boundary charge \(\kappa\) contributes approximately
	\begin{equation}
		\Delta \operatorname{OFB}^{\mathrm{YC}}(\tau)
		=
		\frac{\kappa}{\tau}
		\label{eq:jump_economic_magnitude}
	\end{equation}
	to annualized OFB. Taking \(2.5\) bp as a representative pooled magnitude, \Cref{eq:jump_economic_magnitude} implies about \(5\) annualized bp at six months, \(10\) bp at three months, and \(30\) bp at one month. A price wedge of only a few unannualized basis points can therefore produce a double-digit distortion in short-maturity option-implied rates.
	
	This magnitude matters because the affected object is the same option-implied-rate minus Treasury-rate spread interpreted as a convenience yield in \citet{BDG22,DVT26}. Those studies report U.S.\ convenience yields measured in tens of annualized basis points, including at short maturities. The comparison is not one-for-one: the year-end coefficient is an incremental charge conditional on boundary exposure, whereas reported convenience yields summarize broader samples and states. The relevant point is instead that an identified option-side component of the same measured spread can itself reach economically meaningful annualized magnitudes.
	
	The pooled coefficient also masks substantial time variation. Annual estimates in \Cref{fig:yearly_method_comparison} become much larger in high-effect years, so the reporting-boundary component is neither a constant additive bias nor a fixed fraction of the measured spread. Together with its discontinuous activation and \(1/\tau\) maturity scaling, this makes the wedge difficult to remove with a simple correction. Moreover, the design identifies only the component exposed by an observable reporting boundary; it does not rule out other implementation wedges that vary smoothly or lack a similarly sharp source of identification. The year-end toll should therefore be viewed as an economically material identified violation of relative purity, not as an exhaustive measure of option-side frictions.
	
	\section{Time Variation and a Mid-2010s Regime Shift}
	\label{sec:yearly_boundary_charge}
	
	\subsection{Yearly Variation in the Year-End Boundary Charge}
	\label{subsec:yearly_variation}
	
	The year-end boundary charge varies substantially over time. I compare two DGS-based monthly estimates over the 13 complete calendar years from 2013 through 2025. The first re-estimates the \(D_{t,T}^{\mathrm{YC}}/\tau_{t,T}\) coefficient from \Cref{subsub:regressions} separately within each year using the same-day \(\operatorname{ABCP90\!-\!DTB3}\) specification. The second applies the strict same-date local design in \Cref{subsub:RD_monthly} separately to each December 31 episode. The two methods use different maturity ranges and identifying variation, so their levels need not coincide.
	
	\begin{figure}[!t]
		\centering
		\includegraphics[width=\textwidth]{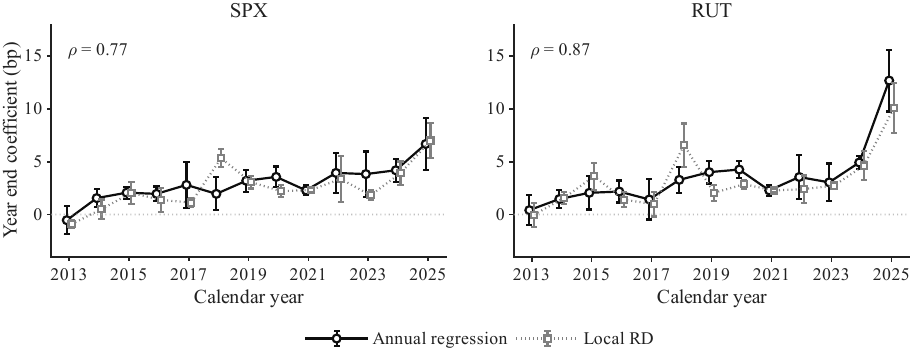}
		\caption{Year-by-Year Boundary Charges: Full-Panel Regression and Local Discontinuity}
		\label{fig:yearly_method_comparison}
		
		\vspace{0.4em}
		
		\begin{minipage}{0.96\textwidth}
			\footnotesize
			\textit{Notes:} The left panel reports SPX and the right panel RUT for 2013--2025. The black solid line plots the annual \(D_{t,T}^{\mathrm{YC}}/\tau_{t,T}\) coefficient from the DGS-based full-panel regression on the DGS--DTB one-year common support. The gray dashed line plots the strict same-date local year-end estimate from \Cref{subsub:RD_monthly}. The annual regression includes a centered cubic maturity polynomial, \(\operatorname{BA}/\tau\), the NFCI, and same-day \(\operatorname{ABCP90\!-\!DTB3}\). Both coefficients are in unannualized basis points. Vertical bars report Newey--West HAC(21) \(95\%\) confidence intervals based on trade-date scores.
		\end{minipage}
	\end{figure}
	
	The two series comove strongly in \Cref{fig:yearly_method_comparison}, with correlations of \(0.77\) for SPX and \(0.87\) for RUT. Both methods produce relatively small effects in much of the early sample and larger positive coefficients later. Their agreement despite different maturity ranges, controls, and identifying variation indicates that the annual heterogeneity is not specific to one estimation method.
	
	The correspondence is not point-for-point. In particular, the local estimates rise more sharply around 2018 than the full-panel estimates. Because the two designs use different samples, maturity ranges, and regression weights, I do not assign a structural interpretation to these annual differences. The relevant result is the strong comovement in the broader time variation of the boundary charge.
	
	\subsection{A Mid-2010s Regime Shift}
	\label{subsec:yearly_regime_shift}
	
	The annual local estimates suggest a shift from a relatively weak early-sample regime to a substantially larger late-sample boundary charge. I examine this pattern using the DGS local-RD series for monthly SPX, monthly RUT, and regular-weekly SPXW, which share year-end episodes from 2012 through 2025. RUTW begins only in 2016 and is therefore excluded from the common break search.
	
	For each admissible first-post year \(c\), I compare the equally weighted mean annual discontinuity before \(c\) with that from \(c\) onward. The common criterion sums the three series-specific Wald statistics. Inference re-runs the full break search under synchronized calendar-year sign flips, while candidate-date inversion characterizes uncertainty about the selected break year \citep{CRS17}.
	
	\begin{figure}[!t]
		\centering
		\includegraphics[width=\textwidth]{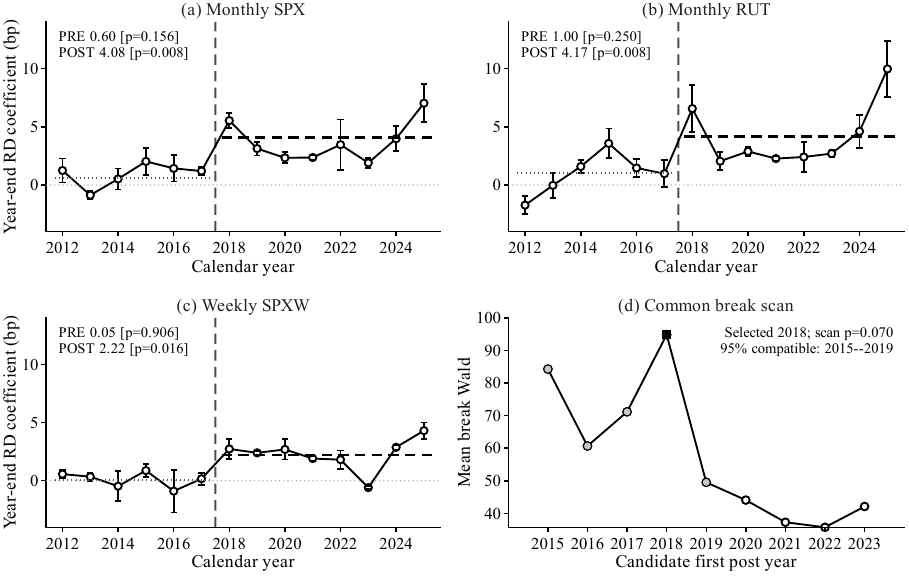}
		\caption{Annual Year-End Boundary Charges and the Common Regime-Break Search}
		\label{fig:yearly_regime_break}
		
		\vspace{0.4em}
		
		\begin{minipage}{0.96\textwidth}
			\footnotesize
			\textit{Notes:} Panels (a)--(c) report annual strict same-date local year-end coefficients under the DGS benchmark for monthly SPX, monthly RUT, and regular-weekly SPXW. Coefficients are in unannualized basis points, and vertical bars report HAC(21) \(95\%\) confidence intervals based on trade-date scores. The vertical dashed line separates PRE and POST periods at the data-selected common first-post year, 2018; horizontal dashed lines report the corresponding pooled coefficients. Bracketed \(p\)-values are two-sided calendar-year episode sign-flip \(p\)-values. Panel (d) reports the mean of the three series-specific break Wald statistics across admissible candidate first-post years. The scan-adjusted \(p\)-value accounts for break-date search, and the \(95\%\) compatibility set is obtained by inverting candidate-date tests conditional on a one-break specification.
		\end{minipage}
	\end{figure}
	
	The common criterion in \Cref{fig:yearly_regime_break} is maximized at 2018. The equally weighted post-minus-pre increase across the three series is \(2.724\) bp and is positive in every series. The candidate-specific exact \(p\)-value is \(0.0043\), while the scan-adjusted \(p\)-value is \(0.070\). The \(90\%\) compatibility set is 2016--2018 and the \(95\%\) set is 2015--2019. The evidence therefore supports a mid-2010s regime shift more strongly than a uniquely identified 2018 breakpoint.
	
	The economic contrast across regimes is considerably sharper than the break date itself. At the selected 2018 split, the pooled DGS discontinuity rises from \(0.60\) to \(4.08\) bp for monthly SPX, from \(1.00\) to \(4.17\) bp for monthly RUT, and from \(0.05\) to \(2.22\) bp for SPXW. PRE-period exact \(p\)-values are \(0.156\), \(0.250\), and \(0.906\), compared with POST-period values of \(0.008\), \(0.008\), and \(0.016\). The same qualitative pattern---a larger POST coefficient, an insignificant PRE coefficient, and a significant positive POST coefficient---holds for every candidate first-post year from 2015 through 2019 in the \(95\%\) compatibility set.
	
	The result is also robust to nuisance specification and individual episodes. Allowing the bid--ask loading to vary by calendar year shifts the common criterion maximum to 2015 but preserves the weak-PRE and strong-POST pattern. Leave-one-episode-out searches select 2018 in 12 of 14 folds; omitting 2017 moves the selected date to 2015, while omitting 2018 moves it to 2019. Excluding either endpoint year, 2012 or 2025, leaves 2018 as the selected break. Thus, the regime interpretation does not depend on one extreme year or on the exact nuisance specification.
	
	The timing overlaps with the post-crisis tightening of U.S.\ bank balance-sheet regulation. The U.S.\ G-SIB capital surcharge was finalized in 2015, phased in beginning in 2016, and reached full implementation in 2019, while the enhanced supplementary leverage ratio became effective in 2018 \citep{FedGSIB15,FedESLR14}. The G-SIB framework is also directly tied to year-end systemic-indicator reporting and associated balance-sheet adjustments \citep{BMPW22}. The statistically compatible 2015--2019 transition window therefore overlaps closely with the period in which year-end balance-sheet constraints on major U.S.\ intermediaries tightened.
	
	I do not interpret this timing as identifying either regulation causally. Several post-crisis capital and leverage requirements changed over overlapping dates, and the exact breakpoint is not sharply estimated. The narrower conclusion is that the year-end boundary charge became materially larger during the same regulatory transition in which year-end balance-sheet exposure became more costly.
	
	This timing pattern also appears outside the option market. In \Cref{subsec:cip_regime_shift}, government-bond CIP independently selects 2018 as the first post-regime year under three aggregation methods, with scan-adjusted \(p\)-values of \(0.010\)--\(0.024\) and a \(95\%\) compatibility set of 2014--2019. The close overlap with the OFB transition window, despite different contracts, prices, and market participants, supports interpreting the mid-2010s change as a broader shift in the pricing of year-end balance-sheet exposure rather than an option-market-specific phenomenon.
	
	\section{Alternative Explanations and Robustness Tests}
	\label{sec:alt}
	
	The strict same-date design in \Cref{fig:ofbrd,tab:monthly_local_yearend_rd} absorbs interest rates and financial conditions common to all maturities priced on a given trade date, but several alternative explanations remain. I consider three broad classes: benchmark and funding-market effects, calendar-boundary mechanics, and measurement or specification artifacts. Unless otherwise noted, local tests retain the baseline strict same-date design and unannualized price units, while full-panel tests retain the baseline sample and controls and change only the feature under examination.
	
	\subsection{Option- and Benchmark-Side Decomposition}
	\label{subsec:alt_option_benchmark_decomposition}
	
	Because the OFB is the difference between option-implied and benchmark rates, its local discontinuity can be decomposed exactly. In unannualized units,
	\begin{align}
		G_{i,t}^{\mathrm{opt}}(T)
		&=
		\tau_{t,T}r_{i,t}^{\mathrm{opt}}(T)
		=
		-10^{4}\log \widehat B_{i,t}(T),
		\notag\\
		G_{t}^{\mathrm{bench},b}(T)
		&=
		\tau_{t,T}r_{t}^{b}(T)
		=
		-10^{4}\log D_{t}^{b}(T),
		\notag\\
		G_{i,t}^{\mathrm{OFB},b}(T)
		&=
		\tau_{t,T}\operatorname{OFB}_{i,t}^{b}(T)
		=
		G_{i,t}^{\mathrm{opt}}(T)
		-
		G_{t}^{\mathrm{bench},b}(T).
		\label{eq:alt_option_benchmark_components}
	\end{align}
	
	Here, \(G_{i,t}^{\mathrm{OFB},b}(T)\) is the local outcome in \Cref{eq:data_local_outcome}. Estimating \Cref{eq:data_local_yearend} on the three components using identical option cells and regressors therefore gives
	\begin{equation}
		\widehat\kappa_{i}^{\mathrm{OFB},b}
		=
		\widehat\kappa_{i}^{\mathrm{opt}}
		-
		\widehat\kappa_{i}^{\mathrm{bench},b}.
		\label{eq:alt_option_benchmark_jump_identity}
	\end{equation}
	
	This accounting decomposition does not separately identify structural option-side and benchmark-side wedges. It instead tests whether the OFB discontinuity can be attributed to a stable positive jump in the benchmark discount curve.
	
	\begin{table}[!t]
		\centering
		\singlespacing
		\caption{Option-Side and Benchmark-Side Decomposition of the Local Year-End Discontinuity}
		\label{tab:option_benchmark_jump_decomposition}
		\begin{threeparttable}
			\footnotesize
			\setlength{\tabcolsep}{5.0pt}
			\renewcommand{\arraystretch}{1.08}
			\begin{tabular*}{0.96\textwidth}{@{\extracolsep{\fill}}llrrrr@{}}
				\toprule
				Outcome
				& Statistic
				& SPX--DGS
				& SPX--DTB
				& RUT--DGS
				& RUT--DTB \\
				\midrule
				
				\multirow{3}{*}{Option-implied component}
				& Jump
				& 3.859
				& 3.859
				& 1.995
				& 1.995 \\
				& HAC(21) \(t\)
				& 1.357
				& 1.357
				& 0.958
				& 0.958 \\
				& Exact \(p\)
				& 0.2843
				& 0.2843
				& 0.1737
				& 0.1737 \\
				
				\addlinespace[0.4em]
				
				\multirow{3}{*}{Benchmark component}
				& Jump
				& 1.182
				& 1.210
				& \(-0.848\)
				& \(-0.780\) \\
				& HAC(21) \(t\)
				& 0.452
				& 0.462
				& \(-0.443\)
				& \(-0.409\) \\
				& Exact \(p\)
				& 0.7610
				& 0.7534
				& 0.4626
				& 0.4991 \\
				
				\addlinespace[0.4em]
				
				\multirow{3}{*}{OFB}
				& Jump
				& 2.678
				& 2.650
				& 2.844
				& 2.776 \\
				& HAC(21) \(t\)
				& 5.264
				& 5.276
				& 4.921
				& 4.904 \\
				& Exact \(p\)
				& 0.0004
				& 0.0004
				& 0.0012
				& 0.0012 \\
				
				\bottomrule
			\end{tabular*}
		\end{threeparttable}
		
		\vspace{0.4em}
		
		\begin{minipage}{0.96\textwidth}
			\footnotesize
			\textit{Notes:} The table estimates \Cref{eq:data_local_yearend} on the three components in \Cref{eq:alt_option_benchmark_components}, using the same observations, controls, weights, and inference as \Cref{tab:monthly_local_yearend_rd}. Jumps are in unannualized basis points. HAC(21) \(t\)-statistics use trade-date scores, and exact \(p\)-values use two-sided sign flips of 14 year-end episode score blocks. DGS and DTB use identical option cells within each market, so their option-implied components coincide. \Cref{eq:alt_option_benchmark_jump_identity} holds within each column.
		\end{minipage}
	\end{table}
	
	\Cref{tab:option_benchmark_jump_decomposition} shows that the option-implied component jumps positively in both markets, although neither estimate is precise enough to identify that component separately. The benchmark component is positive for SPX but negative for RUT and is imprecisely estimated in every specification. In contrast, the OFB discontinuity is \(2.650\)--\(2.844\) bp, with HAC(21) \(t\)-statistics of \(4.90\)--\(5.28\) and exact \(p\)-values no larger than \(0.0012\).
	
	The OFB jump also changes by only \(0.028\) bp between DGS and DTB for SPX and \(0.068\) bp for RUT. Thus, the evidence identifies a stable discontinuity in the option--benchmark relative price but does not support a stable positive Treasury-side jump as its source. A stronger benchmark test therefore replaces Treasuries altogether.
	
	\subsection{Robustness to a Non-Treasury Benchmark}
	\label{subsec:alt_ois_benchmark}
	
	\begin{figure}[!t]
		\centering
		\includegraphics[width=\textwidth]{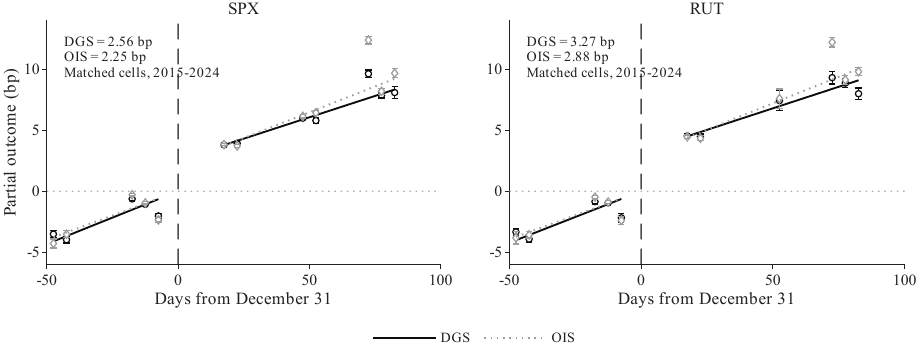}
		\caption{DGS and OIS Local Year-End Discontinuities on the Common Three-Benchmark Sample}
		\label{fig:ois_common_rd}
		
		\vspace{0.4em}
		
		\begin{minipage}{0.96\textwidth}
			\footnotesize
			\textit{Notes:} The figure applies the baseline strict same-date local design to market--trade-date--maturity cells exactly common to DGS, DTB, and OIS over 2015--2024. Solid and dotted lines use DGS and OIS, respectively, with identical option observations, controls, weights, and local-linear specifications within each market. Reported coefficients are in unannualized basis points.
		\end{minipage}
	\end{figure}
	
	The decomposition in \Cref{subsec:alt_option_benchmark_decomposition} weakens explanations based on a particular Treasury-curve construction but does not eliminate Treasury-specific pricing effects. I therefore replace the Treasury benchmark with an EFFR-referenced OIS discount curve.
	
	The comparison uses complete year-end episodes from 2015--2024. The local design uses identical \(30\)--\(365\)-day cells common to DGS, DTB, and OIS, while the full-panel design uses exact DGS--OIS common support through \(1{,}095\) days. Within each comparison, the option observations and all other design features are held fixed.
	
	The positive boundary effect survives under OIS (\Cref{fig:ois_common_rd,tab:ois_common_results}). In the local design, the DGS and OIS estimates are \(2.563\) and \(2.245\) bp for SPX and \(3.268\) and \(2.883\) bp for RUT, with exact \(p\)-values no larger than \(0.0039\). In the full-panel design, the corresponding estimates are \(3.087\) and \(3.205\) bp for SPX and \(2.962\) and \(3.005\) bp for RUT. Replacing the Treasury benchmark with OIS therefore leaves both the sign and economic magnitude of the estimated boundary effect largely unchanged.
	
	\begin{table}[!t]
		\centering
		\singlespacing
		\caption{Year-End Boundary Effects Using a Non-Treasury OIS Benchmark}
		\label{tab:ois_common_results}
		\begin{threeparttable}
			\footnotesize
			\setlength{\tabcolsep}{4.0pt}
			\renewcommand{\arraystretch}{1.12}
			\begin{tabular*}{0.96\textwidth}{@{\extracolsep{\fill}}lllrrrrr@{}}
				\toprule
				Design
				& Market
				& Benchmark
				& \(\widehat\kappa\)
				& HAC \(t\)
				& Exact \(p\)
				& Observations
				& Trade dates \\
				\midrule
				Strict same-date local
				& SPX
				& DGS
				& 2.563
				& 7.982
				& 0.0020
				& 1,418
				& 338 \\
				Strict same-date local
				& SPX
				& OIS
				& 2.245
				& 5.844
				& 0.0020
				& 1,418
				& 338 \\
				Strict same-date local
				& RUT
				& DGS
				& 3.268
				& 7.237
				& 0.0020
				& 1,107
				& 338 \\
				Strict same-date local
				& RUT
				& OIS
				& 2.883
				& 6.223
				& 0.0039
				& 1,107
				& 338 \\
				\addlinespace
				Full-panel
				& SPX
				& DGS
				& 3.087
				& 7.296
				& 0.0039
				& 30,779
				& 2,481 \\
				Full-panel
				& SPX
				& OIS
				& 3.205
				& 7.714
				& 0.0039
				& 30,779
				& 2,481 \\
				Full-panel
				& RUT
				& DGS
				& 2.962
				& 6.799
				& 0.0020
				& 21,325
				& 2,481 \\
				Full-panel
				& RUT
				& OIS
				& 3.005
				& 7.334
				& 0.0020
				& 21,325
				& 2,481 \\
				\bottomrule
			\end{tabular*}
		\end{threeparttable}
		
		\vspace{0.4em}
		
		\begin{minipage}{0.96\textwidth}
			\footnotesize
			\textit{Notes:} The table compares DGS and OIS over 2015--2024 on identical option samples. The local design uses \(30\)--\(365\)-day cells exactly common to DGS, DTB, and OIS and follows the specification in \Cref{eq:data_local_yearend}; the baseline monthly implementation is reported in \Cref{tab:monthly_local_yearend_rd}. The full-panel design uses exact DGS--OIS common support through \(1{,}095\) days with the baseline controls, including the same-day \(\operatorname{ABCP90\!-\!DTB3}_{t}\) spread. HAC \(t\)-statistics use Newey--West HAC(21) trade-date scores. Exact \(p\)-values use two-sided sign flips of 10 year-end episodes in the local design and 10 calendar-year blocks in the full panel. The table does not provide direct inference on DGS--OIS coefficient differences. \(\widehat\kappa\) is in unannualized basis points.
		\end{minipage}
	\end{table}
	
	The benchmark replacement does not systematically attenuate the boundary effect. OIS estimates are somewhat smaller than DGS in the local design but slightly larger in the full-panel design, and the DGS--OIS coefficient differences are not directly tested. OIS may also contain implementation wedges of its own. The relevant conclusion is therefore narrower: a positive year-end discontinuity of similar magnitude survives when Treasuries are replaced by a non-Treasury benchmark on identical option samples.
	
	\subsection{Realized Year-End Turns in Short-Term Funding Markets}
	\label{subsec:alt_money_market_turn}
	
	A separate possibility is that realized year-end increases in secured short-term funding rates mechanically generate the option-implied discount-factor jump through cumulative carry. I evaluate this channel by converting year-end turns in SOFR, TGCR, and BGCR into the same unannualized price units as the OFB discontinuity.
	
	For rate series \(j\in\{\mathrm{SOFR},\mathrm{TGCR},\mathrm{BGCR}\}\), year \(y\), and an \(N\)-calendar-day turn window, define
	\begin{equation}
		\mathcal W_{y,N}
		=
		\left\{
		T_{y}^{\mathrm{YE}}-(N-1),\ldots,T_{y}^{\mathrm{YE}}
		\right\},
		\qquad
		T_{y}^{\mathrm{YE}}
		=
		\operatorname{December~31~of~}y.
		\label{eq:money_market_turn_window}
	\end{equation}
	
	For weekends and holidays, I carry forward the most recent observed rate,
	\begin{equation}
		\widetilde r_{d}^{\,j}
		=
		r_{\max\left\{u\in\mathcal D^{j}:u\leq d\right\}}^{j},
		\label{eq:money_market_rate_carry}
	\end{equation}
	where \(\mathcal D^{j}\) denotes the observation dates for series \(j\). I use two baseline rules: the December 1--15 mean and the December mean through the day preceding the turn window. The cumulative turn effect is
	\begin{equation}
		\mathcal T^{j}_{y,N}(q)
		=
		\frac{100}{360}
		\sum_{d\in\mathcal W_{y,N}}
		\left(
		\widetilde r^{\,j}_{d}
		-
		\overline r^{\,j}_{y}(q)
		\right).
		\label{eq:money_market_turn_pv}
	\end{equation}
	
	Because the rates are in percentage points, \Cref{eq:money_market_turn_pv} converts ACT/360 accruals directly into unannualized basis points. The two baselines use either the December 1--15 mean or the December mean through the day preceding the turn window.
	
	SOFR contributes 12 year-end episodes from 2014--2025, using the historical indicative proxy through 2017 and published SOFR thereafter without mixing series within an episode \citep{SOFR_proxy,FRED_SOFR}. TGCR and BGCR contribute eight episodes from 2018--2025 \citep{NYFedTGCR,NYFedBGCR}. I evaluate \(N\in\{1,2,4,7\}\).
	
	\begin{table}[!t]
		\centering
		\singlespacing
		\caption{Cumulative Price Effects of Year-End Turns in Secured Short-Term Funding Markets}
		\label{tab:money_market_turn_pv}
		\begin{threeparttable}
			\footnotesize
			\setlength{\tabcolsep}{4.2pt}
			\renewcommand{\arraystretch}{1.10}
			
			\begin{tabular*}{0.96\textwidth}{@{\extracolsep{\fill}}lrrrrrrr@{}}
				\toprule
				\multicolumn{8}{l}{\textit{Panel A. December 1--15 baseline rate}} \\
				\addlinespace[0.25em]
				Rate
				& Episodes
				& \makecell{1-day\\mean}
				& \makecell{2-day\\mean}
				& \makecell{4-day\\mean}
				& \makecell{7-day\\mean}
				& \makecell{7-day\\median}
				& \makecell{7-day\\maximum} \\
				\midrule
				SOFR
				& 12
				& 0.044
				& 0.067
				& 0.122
				& 0.188
				& 0.044
				& 0.873 \\
				TGCR
				& 8
				& 0.037
				& 0.047
				& 0.079
				& 0.115
				& 0.004
				& 0.870 \\
				BGCR
				& 8
				& 0.037
				& 0.047
				& 0.079
				& 0.116
				& 0.004
				& 0.872 \\
				\bottomrule
			\end{tabular*}
			
			\vspace{0.75em}
			
			\begin{tabular*}{0.96\textwidth}{@{\extracolsep{\fill}}lrrrrrr@{}}
				\toprule
				\multicolumn{7}{l}{\textit{Panel B. December baseline excluding the turn window}} \\
				\addlinespace[0.25em]
				Rate
				& Episodes
				& \makecell{7-day\\mean}
				& \makecell{7-day\\median}
				& \makecell{7-day\\maximum}
				& \makecell{Mean coverage\\of OFB (\%)}
				& \makecell{Maximum coverage\\of OFB (\%)} \\
				\midrule
				SOFR
				& 12
				& 0.149
				& 0.049
				& 0.563
				& 5.39
				& 20.38 \\
				TGCR
				& 8
				& 0.102
				& 0.004
				& 0.559
				& 3.68
				& 20.26 \\
				BGCR
				& 8
				& 0.102
				& 0.004
				& 0.562
				& 3.68
				& 20.36 \\
				\bottomrule
			\end{tabular*}
		\end{threeparttable}
		
		\vspace{0.4em}
		
		\begin{minipage}{0.96\textwidth}
			\footnotesize
			\textit{Notes:} The table reports the unannualized turn effects in \Cref{eq:money_market_turn_pv}. Panel A uses the December 1--15 mean rate as the baseline; Panel B uses the mean through the day preceding the turn window. Turn windows contain \(N\) calendar days ending December 31, with the most recent observed rate carried over non-observation days and ACT/360 accrual. Coverage divides the 7-day mean or maximum by the \(2.761\) bp benchmark OFB jump, the mean of the SPX and RUT DGS local-RD estimates. All effects are in unannualized basis points.
		\end{minipage}
	\end{table}
	
	The realized secured-funding turns in \Cref{tab:money_market_turn_pv} are quantitatively far smaller than the \(2.761\) bp benchmark OFB jump. Using the December 1--15 baseline, mean 7-day effects are \(0.188\) bp for SOFR and \(0.115\)--\(0.116\) bp for TGCR and BGCR, while even the largest realized effects are about \(0.87\) bp. Under the alternative baseline, mean effects fall to \(0.102\)--\(0.149\) bp, or \(3.7\%\)--\(5.4\%\) of the OFB jump, and the largest effects are about \(0.56\) bp, or roughly \(20\%\).
	
	The time-series variation also runs against simple pass-through. Under both baseline definitions, correlations between 7-day turns and annual DGS local-RD coefficients are negative in all six rate-series--market combinations. The limited number of episodes precludes a causal interpretation, but years with larger OFB discontinuities do not coincide with larger realized secured-funding turns.
	
	Observed overnight and repo year-end carry therefore cannot account for the OFB discontinuity in either magnitude or annual variation. This test does not rule out a broader balance-sheet shadow cost associated with surviving year-end; it rules out reducing the identified boundary wedge to mechanical pass-through of realized secured short-term funding rates.
	
	\subsection{Calendar Boundaries and the Specificity of December 31}
	\label{subsec:alt_december_quarterly}
	
	A natural concern is that the year-end coefficient reflects December option structure rather than exposure to the reporting boundary. Removing all December quarterly expirations from the full-panel sample leaves DGS coefficients of \(3.162\) bp for SPX and \(3.346\) bp for RUT and DTB coefficients of \(3.177\) and \(3.353\) bp, with exact \(p\)-values no larger than \(0.0011\). Interactions between year-end crossing and quarterly-expiration status are also insignificant, with exact \(p\)-values from \(0.188\) to \(0.976\). December quarterly contracts therefore do not account for the baseline effect.
	
	I next ask whether a similar discontinuity appears at calendar quarter-ends generally. For calendar year \(j\), let \(T_j^q\), \(q\in\mathcal Q=\{1,2,3,4\}\), denote March 31, June 30, September 30, and December 31, respectively, and define
	\begin{equation}
		h_{t,T}^{q}
		=
		\operatorname{days}
		\left(
		T-T_{j}^{q}
		\right).
		\label{eq:quarter_boundary_running_variable}
	\end{equation}
	
	The symmetric stacked design uses trade dates within the 92 calendar days preceding each boundary, maturities satisfying \(0<\lvert h_{t,T}^{q}\rvert<90\), and two-sided maturity support within every boundary--trade-date cell. It retains the baseline triangular kernel and bid--ask control, with boundary--trade-date fixed effects and boundary-specific local-linear slopes on each side. The main sample consists of standard-monthly cells exactly matched across DGS and DTB over the complete calendar years 2013--2025, providing symmetric support for all four boundaries and 13 calendar-year blocks for exact inference.
	
	Let \(\kappa_{i,q}^{b}\) denote the jump for market \(i\), benchmark \(b\), and boundary \(q\). I compare December 31 with the equally weighted mean of the other three quarter-ends:
	\begin{align}
		\overline\kappa_{i,\mathrm{nonYE}}^{b}
		&=
		\frac{1}{3}
		\sum_{q=1}^{3}
		\kappa_{i,q}^{b},
		\notag\\
		\Delta_{i}^{\mathrm{YE},b}
		&=
		\kappa_{i,4}^{b}
		-
		\overline\kappa_{i,\mathrm{nonYE}}^{b}.
		\label{eq:quarterly_expiration_contrast}
	\end{align}
	
	\begin{table}[!t]
		\centering
		\singlespacing
		\caption{Symmetric Quarter-End Boundary Comparisons and the Excess Year-End Effect}
		\label{tab:quarterly_expiration_confound}
		\begin{threeparttable}
			\footnotesize
			\setlength{\tabcolsep}{4.8pt}
			\renewcommand{\arraystretch}{1.08}
			\begin{tabular*}{0.96\textwidth}{@{\extracolsep{\fill}}lcccc@{}}
				\toprule
				Boundary or contrast
				& SPX--DGS
				& SPX--DTB
				& RUT--DGS
				& RUT--DTB \\
				\midrule
				
				March 31
				& \makecell{0.068\\(0.239)\\\textnormal{[0.8477]}}
				& \makecell{0.099\\(0.354)\\\textnormal{[0.7615]}}
				& \makecell{\(-1.236\)\\\((-2.196)\)\\\textnormal{[0.0413]}}
				& \makecell{\(-1.277\)\\\((-2.307)\)\\\textnormal{[0.0310]}} \\
				\addlinespace[0.25em]
				
				June 30
				& \makecell{\(-0.254\)\\\((-0.873)\)\\\textnormal{[0.4844]}}
				& \makecell{\(-0.275\)\\\((-0.923)\)\\\textnormal{[0.4585]}}
				& \makecell{\(-0.487\)\\\((-1.319)\)\\\textnormal{[0.2029]}}
				& \makecell{\(-0.516\)\\\((-1.357)\)\\\textnormal{[0.1836]}} \\
				\addlinespace[0.25em]
				
				September 30
				& \makecell{\(-0.136\)\\\((-0.645)\)\\\textnormal{[0.5222]}}
				& \makecell{\(-0.071\)\\\((-0.351)\)\\\textnormal{[0.7346]}}
				& \makecell{0.265\\(0.768)\\\textnormal{[0.4851]}}
				& \makecell{0.322\\(0.951)\\\textnormal{[0.3982]}} \\
				\addlinespace[0.25em]
				
				December 31
				& \makecell{2.568\\(5.318)\\\textnormal{[0.0012]}}
				& \makecell{2.543\\(5.327)\\\textnormal{[0.0012]}}
				& \makecell{3.171\\(5.604)\\\textnormal{[0.0007]}}
				& \makecell{3.101\\(5.586)\\\textnormal{[0.0007]}} \\
				
				\midrule
				
				Mean of non-year-end quarter-ends
				& \makecell{\(-0.107\)\\\((-0.665)\)\\\textnormal{[0.6006]}}
				& \makecell{\(-0.082\)\\\((-0.515)\)\\\textnormal{[0.6724]}}
				& \makecell{\(-0.486\)\\\((-1.993)\)\\\textnormal{[0.1409]}}
				& \makecell{\(-0.491\)\\\((-2.023)\)\\\textnormal{[0.1353]}} \\
				\addlinespace[0.30em]
				
				\textbf{Excess year-end effect}
				& \makecell{\textbf{2.676}\\\textbf{(5.328)}\\\textnormal{\textbf{[0.0007]}}}
				& \makecell{\textbf{2.625}\\\textbf{(5.293)}\\\textnormal{\textbf{[0.0007]}}}
				& \makecell{\textbf{3.658}\\\textbf{(5.953)}\\\textnormal{\textbf{[0.0005]}}}
				& \makecell{\textbf{3.592}\\\textbf{(5.940)}\\\textnormal{\textbf{[0.0005]}}} \\
				\addlinespace[0.30em]
				
				Equality of four boundaries
				& \makecell{\(\chi^{2}(3)=29.684\)\\\textnormal{[0.0015]}}
				& \makecell{\(\chi^{2}(3)=29.402\)\\\textnormal{[0.0015]}}
				& \makecell{\(\chi^{2}(3)=38.931\)\\\textnormal{[0.0007]}}
				& \makecell{\(\chi^{2}(3)=39.151\)\\\textnormal{[0.0010]}} \\
				
				\bottomrule
			\end{tabular*}
		\end{threeparttable}
		
		\vspace{0.4em}
		
		\begin{minipage}{0.96\textwidth}
			\footnotesize
			\textit{Notes:} The table uses standard-monthly market--trade-date--maturity cells exactly matched across DGS and DTB over 2013--2025. Each quarter-end uses trade dates within the preceding 92 calendar days, \(0<\lvert h_{t,T}^{q}\rvert<90\), a triangular kernel, and two-sided maturity support within each boundary--trade-date cell. The stacked regression includes boundary--trade-date fixed effects, boundary-specific left- and right-side local-linear maturity slopes, and one common coefficient on \(\operatorname{BA}_{i,t}(T)\) across the four boundaries. Each boundary and contrast cell reports the coefficient, HAC(21) \(t\)-statistic in parentheses, and two-sided calendar-year score sign-flip exact \(p\)-value in brackets. The excess year-end effect subtracts the equally weighted mean of the March, June, and September jumps from the December jump. The equality row reports the HAC Wald statistic for \(H_{0}:\kappa_{i,1}^{b}=\kappa_{i,2}^{b}=\kappa_{i,3}^{b}=\kappa_{i,4}^{b}\), with the exact max-score \(p\)-value in brackets. The stacked samples contain \(7{,}201\) SPX and \(5{,}634\) RUT observations and use 13 calendar-year score blocks. All coefficients are in unannualized basis points.
		\end{minipage}
	\end{table}
	
	\Cref{tab:quarterly_expiration_confound} shows no common positive discontinuity at the three non-year-end quarter-ends. SPX estimates are close to zero throughout, while RUT exhibits a negative March 31 jump and small or negative estimates at the other two boundaries. December 31 instead produces precisely positive jumps of \(2.568\) and \(3.171\) bp under DGS and \(2.543\) and \(3.101\) bp under DTB for SPX and RUT, respectively.
	
	The direct contrasts sharpen this distinction. The December jump exceeds the mean of the other three quarter-ends by \(2.676\) bp for SPX and \(3.658\) bp for RUT under DGS, with corresponding DTB estimates of \(2.625\) and \(3.592\) bp. Exact \(p\)-values are no larger than \(0.0007\), and the joint null that all four boundary coefficients are equal is rejected under both benchmarks.
	
	The same pattern survives changes in benchmark and contract group. On cells exactly common to DGS, DTB, and OIS over 2015--2024, the excess year-end effect is \(2.786\) bp for SPX and \(3.897\) bp for RUT under OIS, compared with \(2.889\) and \(3.930\) bp under DGS. Regular weeklies produce DGS excess effects of \(1.520\) bp for SPXW and \(2.158\) bp for RUTW; the latter is less precise because RUTW has a shorter sample and weaker right-side maturity support.
	
	I finally broaden the comparison beyond quarter-ends to a prespecified grid of 15 month-end, holiday, and pseudo-calendar cutoffs. Some non-year-end boundaries exhibit seasonal discontinuities, so the data do not support the stronger claim that December 31 is the only priced calendar boundary; in RUT, the largest positive placebo point estimate even exceeds the December estimate. The relevant comparison is therefore multiplicity-adjusted rather than a sequence of individual null tests.
	
	Under a family-wise max-statistic correction across the 15 cutoffs, the December 31 statistic remains significant, with global \(p\)-values of \(0.0055\) for SPX and \(0.0183\) for RUT. Thus, the year-end discontinuity survives both symmetric quarter-end comparisons and a broader search allowing other seasonal calendar effects to exist.
	
	Taken together, the evidence rules out two narrower calendar explanations. The effect is not generated by December quarterly expirations and is not a generic positive quarter-end discontinuity. More broadly, December 31 remains statistically distinctive after accounting for other calendar-boundary effects and multiple testing, consistent with pricing of exposure to the year-end reporting boundary rather than generic calendar mechanics.
	
	\subsection{Robustness to Maturity Structure and Calendar Conventions}
	\label{subsec:alt_maturity_calendar}
	
	I examine whether the year-end coefficient reflects maturity misspecification or generic calendar exposure. I vary the full-panel maturity controls, test prominent holidays as alternative cutoffs, and reconstruct maturity and contract-horizon exposure using business days, holidays, and weekends.
	
	\begin{table}[!t]
		\centering
		\singlespacing
		\caption{Robustness to Alternative Maturity-Structure Controls}
		\label{tab:maturity_structure_robustness}
		\begin{threeparttable}
			\footnotesize
			\setlength{\tabcolsep}{4.5pt}
			\renewcommand{\arraystretch}{1.05}
			\begin{tabular*}{0.96\textwidth}{@{\extracolsep{\fill}}lcccc@{}}
				\toprule
				Specification
				& \makecell{SPX\\DGS}
				& \makecell{RUT\\DGS}
				& \makecell{SPX\\DTB}
				& \makecell{RUT\\DTB} \\
				\midrule
				Centered cubic polynomial
				& \makecell{2.938\\(6.468)}
				& \makecell{3.199\\(5.302)}
				& \makecell{2.869\\(6.747)}
				& \makecell{3.165\\(5.773)} \\
				Fourth-order polynomial
				& \makecell{2.940\\(6.494)}
				& \makecell{3.196\\(5.304)}
				& \makecell{2.866\\(6.746)}
				& \makecell{3.170\\(5.781)} \\
				Fifth-order polynomial
				& \makecell{2.923\\(6.478)}
				& \makecell{3.186\\(5.305)}
				& \makecell{2.865\\(6.751)}
				& \makecell{3.173\\(5.785)} \\
				Restricted cubic spline
				& \makecell{2.927\\(6.473)}
				& \makecell{3.184\\(5.297)}
				& \makecell{2.869\\(6.745)}
				& \makecell{3.172\\(5.778)} \\
				7-day maturity-bin fixed effects
				& \makecell{2.919\\(6.470)}
				& \makecell{3.188\\(5.315)}
				& \makecell{2.867\\(6.752)}
				& \makecell{3.164\\(5.775)} \\
				14-day maturity-bin fixed effects
				& \makecell{2.917\\(6.469)}
				& \makecell{3.186\\(5.315)}
				& \makecell{2.865\\(6.749)}
				& \makecell{3.162\\(5.774)} \\
				Expiration-month fixed effects
				& \makecell{2.799\\(7.585)}
				& \makecell{3.213\\(5.668)}
				& \makecell{2.478\\(8.030)}
				& \makecell{3.157\\(6.082)} \\
				\bottomrule
			\end{tabular*}
		\end{threeparttable}
		
		\vspace{0.4em}
		
		\begin{minipage}{0.96\textwidth}
			\footnotesize
			\textit{Notes:} Each cell reports the coefficient on \(D_{t,T}^{\mathrm{YC}}/\tau_{t,T}\), with the HAC(21) \(t\)-statistic in parentheses. All specifications retain the baseline same-day samples, calendar-year fixed effects, \(\operatorname{BA}_{i,t}(T)/\tau_{t,T}\), the NFCI, and \(\Delta_{t}^{\mathrm{ABCP-TB}}=100(\operatorname{ABCP90}_{t}-\operatorname{DTB3}_{t})\) from \Cref{tab:yearend_regression_spx,tab:yearend_regression_rut}; only the maturity specification changes. The restricted cubic spline uses five knots. The maturity-bin specifications replace the polynomial with residual-maturity-bin fixed effects, and the final row adds expiration-month fixed effects. All two-sided calendar-year score sign-flip \(p\)-values are at most \(0.0012\). Coefficients are in unannualized basis points.
		\end{minipage}
	\end{table}
	
	The coefficient is nearly invariant to maturity specification in \Cref{tab:maturity_structure_robustness}. Higher-order polynomials, a restricted cubic spline, and maturity-bin fixed effects produce estimates of about \(2.9\) bp for SPX and \(3.2\) bp for RUT. Expiration-month fixed effects lower the SPX estimates to \(2.478\)--\(2.799\) bp but leave all four coefficients precisely positive. The effect therefore does not depend on a particular maturity polynomial or seasonal expiration composition.
	
	A related concern is that crossing a prominent holiday, rather than December 31 specifically, may generate a maturity discontinuity. I apply the strict same-date design to Thanksgiving, Independence Day, and a November 15 pseudo-cutoff using DGS observations on cells exactly common to DGS and DTB. Each cutoff uses the same 92-day pre-cutoff window, 90-day bandwidth, triangular kernel, trade-date fixed effects, side-specific local-linear slopes, bid--ask control, and strict two-sided maturity support as the year-end design.
	
	\begin{table}[!t]
		\centering
		\singlespacing
		\caption{Direct Holiday-Cutoff Placebos}
		\label{tab:holiday_cutoff_placebos}
		\begin{threeparttable}
			\footnotesize
			\setlength{\tabcolsep}{5.0pt}
			\renewcommand{\arraystretch}{1.06}
			\begin{tabular*}{0.96\textwidth}{@{\extracolsep{\fill}}llrrrrrr@{}}
				\toprule
				Market
				& Cutoff
				& Jump
				& HAC(21) \(t\)
				& Exact \(p\)
				& Observations
				& Trade dates
				& Episodes \\
				\midrule
				SPX
				& Thanksgiving
				& \(-1.294\)
				& \(-5.952\)
				& 0.0016
				& 2,023
				& 517
				& 14 \\
				SPX
				& Independence Day
				& \(-0.290\)
				& \(-0.862\)
				& 0.4177
				& 1,705
				& 413
				& 13 \\
				SPX
				& November 15
				& 0.812
				& 0.916
				& 0.4563
				& 1,243
				& 315
				& 14 \\
				SPX
				& December 31
				& 2.678
				& 5.264
				& 0.0004
				& 1,993
				& 479
				& 14 \\
				\addlinespace[0.4em]
				RUT
				& Thanksgiving
				& \(-1.846\)
				& \(-4.887\)
				& 0.0002
				& 1,520
				& 517
				& 14 \\
				RUT
				& Independence Day
				& \(-0.303\)
				& \(-0.782\)
				& 0.3772
				& 1,372
				& 413
				& 13 \\
				RUT
				& November 15
				& 0.419
				& 0.597
				& 0.5093
				& 1,065
				& 315
				& 14 \\
				RUT
				& December 31
				& 2.844
				& 4.921
				& 0.0012
				& 1,548
				& 479
				& 14 \\
				\bottomrule
			\end{tabular*}
		\end{threeparttable}
		
		\vspace{0.4em}
		
		\begin{minipage}{0.96\textwidth}
			\footnotesize
			\textit{Notes:} The table applies the strict same-date local design to alternative calendar cutoffs using DGS observations on market--trade-date--maturity cells exactly common to DGS and DTB. Each cutoff uses trade dates within the preceding 92 calendar days, \(0<\lvert h\rvert<90\), a triangular kernel, trade-date fixed effects, separate local-linear maturity slopes, the maturity-level median bid--ask control, and two-sided support on every retained trade date. Thanksgiving is the fourth Thursday of November, Independence Day is July 4, and November 15 is a nonholiday pseudo-cutoff. December 31 reproduces the baseline year-end design. HAC(21) \(t\)-statistics use trade-date scores, and exact \(p\)-values use two-sided calendar-year episode score sign flips. Coefficients are in unannualized basis points.
		\end{minipage}
	\end{table}
	
	The placebo cutoffs in \Cref{tab:holiday_cutoff_placebos} do not reproduce the positive year-end pattern. Thanksgiving instead generates precisely negative jumps, while Independence Day and November 15 are insignificant in both markets. Prominent holidays therefore do not generically generate the positive maturity-crossing effect observed at December 31.
	
	I next change the maturity coordinate and contract-horizon calendar controls. Full-panel regressions replace calendar-day maturity with business-day maturity or control for holidays and weekends spanned by each contract. The local design analogously replaces the running variable with business-day distance or adds horizon-exposure controls.
	
	\begin{table}[!t]
		\centering
		\singlespacing
		\caption{Robustness to Day-Count and Calendar Conventions}
		\label{tab:calendar_convention_robustness}
		\begin{threeparttable}
			\footnotesize
			\setlength{\tabcolsep}{4.5pt}
			\renewcommand{\arraystretch}{1.05}
			\begin{tabular*}{0.96\textwidth}{@{\extracolsep{\fill}}lcccc@{}}
				\toprule
				Specification
				& \makecell{SPX\\DGS}
				& \makecell{RUT\\DGS}
				& \makecell{SPX\\DTB}
				& \makecell{RUT\\DTB} \\
				\midrule
				\multicolumn{5}{l}{\textit{Panel A. Full-panel design}} \\
				\addlinespace[0.15em]
				Calendar-day basis
				& \makecell{2.938\\(6.468)}
				& \makecell{3.199\\(5.302)}
				& \makecell{2.869\\(6.747)}
				& \makecell{3.165\\(5.773)} \\
				Business-day basis
				& \makecell{3.215\\(7.058)}
				& \makecell{3.479\\(5.818)}
				& \makecell{3.126\\(7.344)}
				& \makecell{3.432\\(6.332)} \\
				Christmas-exposure control
				& \makecell{3.101\\(4.073)}
				& \makecell{2.508\\(3.246)}
				& \makecell{2.940\\(3.765)}
				& \makecell{2.391\\(3.267)} \\
				Holiday and weekend exposure controls
				& \makecell{2.966\\(6.089)}
				& \makecell{3.256\\(4.369)}
				& \makecell{2.981\\(6.355)}
				& \makecell{3.451\\(4.960)} \\
				Holiday-exposure control
				& \makecell{2.985\\(5.955)}
				& \makecell{3.297\\(4.359)}
				& \makecell{2.984\\(6.272)}
				& \makecell{3.482\\(4.940)} \\
				\addlinespace[0.4em]
				\multicolumn{5}{l}{\textit{Panel B. Strict same-date local design}} \\
				\addlinespace[0.15em]
				Calendar-day running variable
				& \makecell{2.678\\(5.264)}
				& \makecell{2.844\\(4.921)}
				& \makecell{2.650\\(5.276)}
				& \makecell{2.776\\(4.904)} \\
				Business-day running variable
				& \makecell{2.857\\(5.651)}
				& \makecell{3.046\\(5.468)}
				& \makecell{2.828\\(5.665)}
				& \makecell{2.978\\(5.456)} \\
				Holiday and weekend exposure controls
				& \makecell{2.882\\(3.993)}
				& \makecell{3.469\\(2.208)}
				& \makecell{2.824\\(3.953)}
				& \makecell{3.341\\(2.158)} \\
				Holiday-exposure control
				& \makecell{2.903\\(4.105)}
				& \makecell{3.662\\(2.480)}
				& \makecell{2.846\\(4.064)}
				& \makecell{3.538\\(2.433)} \\
				\bottomrule
			\end{tabular*}
		\end{threeparttable}
		
		\vspace{0.4em}
		
		\begin{minipage}{0.96\textwidth}
			\footnotesize
			\textit{Notes:} Each cell reports the unannualized year-end coefficient, with the HAC(21) \(t\)-statistic in parentheses. Panel A starts from the same-day full-panel \(\mathcal M_{2}\) specifications in \Cref{tab:yearend_regression_spx,tab:yearend_regression_rut}, including \(\operatorname{ABCP90\!-\!DTB3}\). The business-day specification replaces calendar-day residual maturity with its business-day counterpart; the remaining rows add contract-horizon calendar controls. Panel B starts from the strict same-date local design in \Cref{eq:data_local_yearend,tab:monthly_local_yearend_rd}. The business-day specification replaces calendar-day distance with business-day distance from December 31. ``Holiday and weekend exposure controls'' include counts of holidays and weekend days over the contract horizon, while ``Holiday-exposure control'' includes only the holiday count. Among identified full-panel alternatives, two-sided calendar-year score sign-flip \(p\)-values are at most \(0.0178\). Exact \(p\)-values for the local business-day specifications are at most \(0.0006\). For local holiday-control specifications, exact \(p\)-values are at most \(0.0103\) for SPX and range from \(0.0985\) to \(0.1740\) for RUT. Coefficients are in unannualized basis points.
		\end{minipage}
	\end{table}
	
	Changing the maturity coordinate does not attenuate the effect. In \Cref{tab:calendar_convention_robustness}, business-day full-panel estimates are \(3.126\)--\(3.479\) bp, while local estimates are \(2.828\)--\(2.857\) bp for SPX and \(2.978\)--\(3.046\) bp for RUT.
	
	Contract-horizon controls give the same conclusion. Identified full-panel estimates remain between \(2.391\) and \(3.482\) bp, and local estimates between \(2.824\) and \(3.662\) bp. Local holiday controls reduce precision for RUT because horizon exposure is highly correlated with crossing, but the point estimates do not attenuate.
	
	Some calendar controls are not separately identified. In the local sample, indicators for spanning Christmas or New Year's Day are perfectly collinear with year-end crossing. New Year exposure is also nearly collinear with the crossing term in the full-panel DTB sample, so I do not interpret those specifications.
	
	Overall, alternative maturity controls, business-day coordinates, direct holiday cutoffs, and contract-horizon controls do not explain the December 31 discontinuity. Together with the broader calendar-boundary tests in \Cref{subsec:alt_december_quarterly}, the evidence makes maturity curvature and generic calendar mechanics unlikely explanations for the year-end wedge.
	
	\subsection{Quoted Liquidity and Data Quality}
	\label{subsec:alt_quote_quality}
	
	A remaining concern is that the year-end discontinuity reflects deterioration in option-quote quality rather than a pricing wedge. I examine this possibility by testing for a discontinuity in quoted bid--ask spreads, removing the bid--ask control from the OFB regression, and pruning cells with the weakest intraday support.
	
	\begin{table}[!t]
		\centering
		\singlespacing
		\caption{Quoted Liquidity and Data-Quality Robustness of the Local Year-End Discontinuity}
		\label{tab:quote_quality_robustness}
		\begin{threeparttable}
			\footnotesize
			\setlength{\tabcolsep}{4.2pt}
			\renewcommand{\arraystretch}{1.10}
			\begin{tabular*}{0.96\textwidth}{@{\extracolsep{\fill}}lrrrrrr@{}}
				\toprule
				&
				\multicolumn{3}{c}{SPX}
				&
				\multicolumn{3}{c}{RUT} \\
				\cmidrule(lr){2-4}
				\cmidrule(lr){5-7}
				Specification
				& DGS
				& DTB
				& \(N\)
				& DGS
				& DTB
				& \(N\) \\
				\midrule
				
				\multicolumn{7}{l}{\textit{Panel A. OFB outcome}} \\
				\addlinespace[0.2em]
				
				Baseline
				& \makecell{2.678 (5.264)\\\textnormal{[0.0004]}}
				& \makecell{2.650 (5.276)\\\textnormal{[0.0004]}}
				& 1,993
				& \makecell{2.844 (4.921)\\\textnormal{[0.0012]}}
				& \makecell{2.776 (4.904)\\\textnormal{[0.0012]}}
				& 1,548 \\
				
				No bid--ask control
				& \makecell{2.553 (5.120)\\\textnormal{[0.0010]}}
				& \makecell{2.522 (5.127)\\\textnormal{[0.0009]}}
				& 1,993
				& \makecell{2.937 (4.945)\\\textnormal{[0.0011]}}
				& \makecell{2.877 (4.909)\\\textnormal{[0.0013]}}
				& 1,548 \\
				
				\makecell[l]{Drop bottom-decile\\minute support}
				& \makecell{2.614 (4.014)\\\textnormal{[0.0039]}}
				& \makecell{2.599 (4.050)\\\textnormal{[0.0039]}}
				& 1,543
				& \makecell{2.327 (2.485)\\\textnormal{[0.0547]}}
				& \makecell{2.334 (2.521)\\\textnormal{[0.0547]}}
				& 858 \\
				
				\midrule
				
				\multicolumn{7}{l}{\textit{Panel B. Quoted bid--ask spread as the outcome}} \\
				\addlinespace[0.2em]
				
				Bid--ask discontinuity
				& \multicolumn{2}{c}{\makecell{0.322 (1.546)\\\textnormal{[0.2496]}}}
				& 1,993
				& \multicolumn{2}{c}{\makecell{0.684 (2.264)\\\textnormal{[0.0836]}}}
				& 1,548 \\
				
				\bottomrule
			\end{tabular*}
		\end{threeparttable}
		
		\vspace{0.4em}
		
		\begin{minipage}{0.96\textwidth}
			\footnotesize
			\textit{Notes:} Panel A reports the unannualized OFB jump from the strict same-date local design. Each cell reports the coefficient followed by the HAC(21) \(t\)-statistic in parentheses; brackets report the two-sided calendar-year episode sign-flip exact \(p\)-value. The baseline is identical to \Cref{tab:monthly_local_yearend_rd}. ``No bid--ask control'' removes \(\operatorname{BA}_{i,t}(T)\) while retaining the same sample and local specification. ``Drop bottom-decile minute support'' removes market--trade-date--maturity cells in the bottom decile of intraday minute support before reimposing strict same-date two-sided support. Panel B uses the maturity-level median quoted bid--ask spread as the dependent variable. Because this measure does not depend on the benchmark curve, DGS and DTB columns are combined. \(N\) denotes final local-sample observations. All jump coefficients are in basis points.
		\end{minipage}
	\end{table}
	
	Quoted liquidity itself does not exhibit a discontinuity comparable to the OFB jump. In Panel B of \Cref{tab:quote_quality_robustness}, the bid--ask jump is \(0.322\) bp for SPX with exact \(p=0.250\) and \(0.684\) bp for RUT with exact \(p=0.084\). Quoted liquidity may therefore vary modestly at year-end, particularly in RUT, but it does not reproduce the common \(2\)--\(3\) bp OFB discontinuity.
	
	The OFB result also does not depend on conditioning on bid--ask spreads. Removing the bid--ask control leaves estimates of \(2.522\)--\(2.553\) bp for SPX and \(2.877\)--\(2.937\) bp for RUT, with all four coefficients precisely positive.
	
	Pruning weakly supported cells changes precision more than economic magnitude. Dropping the bottom decile of intraday minute support leaves SPX estimates of \(2.599\)--\(2.614\) bp and RUT estimates of \(2.327\)--\(2.334\) bp. Reimposing strict two-sided support after the screen reduces the RUT sample from \(1{,}548\) to \(858\) observations and from \(479\) to \(275\) trade dates, raising its exact \(p\)-values to \(0.0547\). The point estimates nevertheless remain positive and economically close to the baseline. Screening on raw quote-row support gives the same result.
	
	Further tightening intraday coverage requirements from \(90\%\) to \(95\%\) or \(97.5\%\) changes neither the local sample nor the estimates because baseline observations already satisfy these thresholds. Restrictions targeting December 24--31, half-day sessions, or the final three trading days likewise remove no observations from the monthly local sample and therefore provide no additional identifying variation.
	
	Overall, neither quoted-liquidity behavior nor weak intraday support explains the year-end discontinuity. The OFB jump remains economically stable without the bid--ask control and after substantial pruning of low-support cells. Together with the synthetic-forward fit documented in \Cref{app:ab_fit,tab:app_ab_fit}, the evidence makes a quote-quality or local-liquidity artifact unlikely.
	
	\subsection{Separating Dividend Contributions from the Option-Implied Discount Factor}
	\label{subsec:alt_dividend}
	
	Dividends are separated from the option-implied discount factor by construction in \Cref{eq:data_ab_regression,eq:data_ab_discount}: prepaid-forward and dividend components enter the intercept, while \(\widehat B_{i,t,s}^{\mathrm{AB}}(T)\) is identified from the strike slope. I nevertheless use external dividend data to test whether finite strike support or quote composition allows the intercept component to leak into the estimated slope.
	
	Let \(PR_{i,t}\) and \(TR_{i,t}\) denote the price and total-return indices for market \(i\). I measure the realized cumulative dividend contribution from trade date to maturity as
	\begin{equation}
		\operatorname{DIV}_{i,t}(T)
		=
		10^{4}
		\left[
		\log
		\left(
		\frac{TR_{i,T}}{TR_{i,t}}
		\right)
		-
		\log
		\left(
		\frac{PR_{i,T}}{PR_{i,t}}
		\right)
		\right].
		\label{eq:realized_dividend_contribution}
	\end{equation}
	
	If the index is unavailable on expiration, I use the most recent preceding trading-day observation. \(\operatorname{DIV}_{i,t}(T)\) is constructed independently of option prices and benchmark discount curves and is expressed in basis points.
	
	I first use \(\operatorname{DIV}_{i,t}(T)\) as the outcome in the strict same-date local design. I then augment the OFB regression with the dividend level, dividend--maturity and dividend--crossing interactions, and all three jointly. All other local-design features are unchanged.
	
	\begin{table}[!t]
		\centering
		\singlespacing
		\caption{Year-End Dividend Contributions and the OFB after Dividend Controls}
		\label{tab:dividend_robustness}
		\begin{threeparttable}
			\footnotesize
			\setlength{\tabcolsep}{5.0pt}
			\renewcommand{\arraystretch}{1.06}
			\begin{tabular*}{0.96\textwidth}{@{\extracolsep{\fill}}lcccc@{}}
				\toprule
				&
				\multicolumn{2}{c}{SPX}
				&
				\multicolumn{2}{c}{RUT} \\
				\cmidrule(lr){2-3}
				\cmidrule(lr){4-5}
				Outcome and specification
				& DGS
				& DTB
				& DGS
				& DTB \\
				\midrule
				\multicolumn{5}{l}{\textit{Panel A. Realized dividend contribution}} \\
				\addlinespace[0.2em]
				Local jump in dividend contribution
				& \multicolumn{2}{c}{\makecell{\(-4.701\)\\\((-11.632)\)\\\([0.0002]\)}}
				& \multicolumn{2}{c}{\makecell{\(-2.039\)\\\((-3.892)\)\\\([0.0033]\)}} \\
				\addlinespace[0.45em]
				\multicolumn{5}{l}{\textit{Panel B. OFB after dividend controls}} \\
				\addlinespace[0.2em]
				Baseline
				& \makecell{2.678\\(5.264)\\\([0.0004]\)}
				& \makecell{2.650\\(5.276)\\\([0.0004]\)}
				& \makecell{2.844\\(4.921)\\\([0.0012]\)}
				& \makecell{2.776\\(4.904)\\\([0.0012]\)} \\
				Dividend-level control
				& \makecell{2.467\\(4.118)\\\([0.0054]\)}
				& \makecell{2.450\\(4.123)\\\([0.0055]\)}
				& \makecell{2.489\\(4.164)\\\([0.0100]\)}
				& \makecell{2.420\\(4.158)\\\([0.0107]\)} \\
				Dividend--maturity interaction
				& \makecell{2.448\\(4.382)\\\([0.0040]\)}
				& \makecell{2.455\\(4.398)\\\([0.0038]\)}
				& \makecell{2.490\\(4.163)\\\([0.0100]\)}
				& \makecell{2.420\\(4.157)\\\([0.0107]\)} \\
				Dividend--crossing interaction
				& \makecell{3.128\\(2.245)\\\([0.0629]\)}
				& \makecell{2.949\\(2.195)\\\([0.0709]\)}
				& \makecell{2.466\\(4.086)\\\([0.0123]\)}
				& \makecell{2.397\\(4.080)\\\([0.0128]\)} \\
				All dividend controls
				& \makecell{3.186\\(2.694)\\\([0.0347]\)}
				& \makecell{3.263\\(2.801)\\\([0.0322]\)}
				& \makecell{2.495\\(4.115)\\\([0.0127]\)}
				& \makecell{2.424\\(4.107)\\\([0.0135]\)} \\
				\bottomrule
			\end{tabular*}
		\end{threeparttable}
		
		\vspace{0.4em}
		
		\begin{minipage}{0.96\textwidth}
			\footnotesize
			\textit{Notes:} Panel A reports the year-end jump from the strict same-date local design using the realized dividend contribution in \Cref{eq:realized_dividend_contribution} as the outcome. Because dividend contribution is benchmark-independent, DGS and DTB columns are combined within each market. Panel B uses the unannualized OFB \(G_{i,t}^{b}(T)\). Each cell reports the year-end coefficient, HAC(21) \(t\)-statistic in parentheses, and two-sided calendar-year episode sign-flip exact \(p\)-value in brackets. The baseline sample contains \(1{,}993\) SPX and \(1{,}548\) RUT observations; specifications requiring dividend data contain \(1{,}888\) and \(1{,}489\), respectively. Differences from the full-sample baseline therefore also reflect sample composition. All coefficients are in unannualized basis points.
		\end{minipage}
	\end{table}
	
	Panel A of \Cref{tab:dividend_robustness} shows realized dividend jumps of \(-4.701\) bp for SPX and \(-2.039\) bp for RUT. Their signs are opposite to the positive OFB discontinuity, so realized dividend seasonality does not reproduce the year-end pricing pattern.
	
	The OFB coefficient also remains economically stable after controlling for dividends. Dividend-level and dividend--maturity controls leave estimates near \(2.4\)--\(2.5\) bp in both markets. Allowing dividend--crossing interactions reduces precision for SPX but does not attenuate the coefficient, and the joint specification leaves all four market--benchmark estimates positive at \(2.424\)--\(3.263\) bp.
	
	The result is similar under a non-Treasury benchmark. On the 2015--2024 OIS common sample, realized dividend jumps are \(-5.338\) bp for SPX and \(-2.256\) bp for RUT, while the OIS-based OFB coefficients with all dividend controls remain positive at \(3.689\) and \(2.811\) bp.
	
	Realized dividends do not directly observe the expectations embedded in option prices at the trade date, so these tests cannot eliminate every dividend-expectation error. Nevertheless, dividends are separated from the discount-factor slope by construction, the external dividend discontinuity has the opposite sign, and flexible dividend controls do not remove the OFB jump. Ordinary dividend seasonality or leakage from the synthetic-forward intercept is therefore unlikely to explain the year-end discontinuity.
	
	\section{External Evidence on the Year-End Boundary and Regime Shift}
	\label{sec:external_evidence}
	
	I use two external datasets to test whether the year-end pattern depends on my option-processing pipeline or reflects a broader price of carrying balance-sheet exposure across reporting dates. Government-bond CIP provides a distinct relative-value market in which intermediary balance-sheet effects are well established \citep{DTV18,CDW21,COZ21}, while the option-implied-rate panel of \citet{DVT26,DVT26Data} provides an independently constructed option sample. Because these tests differ from the main strict same-date design, I use them as external validation rather than as alternative estimators of the same structural object.
	
	\subsection{Government-Bond CIP}
	\label{subsec:cip_boundary_effect}
	
	For currency \(c\), define
	\begin{equation}
		Y_{c,t}
		=
		\operatorname{CIP}^{\mathrm{govt}}_{c,t}(3M)
		-
		\operatorname{CIP}^{\mathrm{govt}}_{c,t}(1Y).
		\label{eq:cip_3m_1y_spread}
	\end{equation}
	
	Around September 30, the three-month trade begins to span December 31, whereas the one-year trade spans the boundary on both sides of the cutoff. A discontinuity in \(Y_{c,t}\) therefore isolates the relative price change associated with adding one year-end boundary to the short-dated position.
	
	For magnitude comparisons, I map the annualized three-month discontinuity into the fixed-price representation identified in the OFB data:
	\begin{equation}
		\widehat\kappa^{\mathrm{CIP}}
		=
		\frac{\widehat\delta^{\mathrm{CIP}}}{4}.
		\label{eq:cip_toll_mapping}
	\end{equation}
	
	This mapping is a normalization rather than a separately identified CIP functional form. The CIP data contain only the three-month and one-year tenors needed for this comparison, so the sign and timing of the discontinuity are more informative than the mapped level.
	
	Using complete 2015--2024 episodes for 10 currencies against the U.S.\ dollar, I estimate local discontinuities around September 30 with a 90-day bandwidth, triangular weights, and separate linear trends on the two sides.
	
	\begin{table}[!t]
		\centering
		\singlespacing
		\caption{Year-End Reporting-Boundary Effects in Government-Bond CIP}
		\label{tab:cip_yearend_regime}
		\begin{threeparttable}
			\footnotesize
			\setlength{\tabcolsep}{6pt}
			\renewcommand{\arraystretch}{1.05}
			\begin{tabular*}{0.96\textwidth}{@{\extracolsep{\fill}}lccc@{}}
				\toprule
				&
				\makecell{Daily equal-weighted\\mean}
				&
				\makecell{Daily cross-sectional\\median}
				&
				\makecell{Currency--trade-date\\pooled panel} \\
				\midrule
				\(1/\tau\)-mapped boundary charge
				& 2.60
				& 2.26
				& 2.57 \\
				HAC 95\% confidence interval
				& [1.69, 3.50]
				& [1.32, 3.20]
				& [1.68, 3.47] \\
				Exact \(p\)-value
				& 0.002
				& 0.006
				& 0.002 \\
				Positive currency-level estimates
				& 10/10
				& 10/10
				& 10/10 \\
				\bottomrule
			\end{tabular*}
		\end{threeparttable}
		
		\vspace{0.25em}
		
		\begin{minipage}{0.96\textwidth}
			\footnotesize
			\textit{Notes:} The boundary charge applies the \(1/\tau\) mapping in \Cref{eq:cip_toll_mapping} and is expressed in unannualized basis points. The 2015--2024 sample includes AUD, CAD, CHF, DKK, EUR, GBP, JPY, NOK, NZD, and SEK against the U.S.\ dollar. The dependent variable is the same-date difference between three-month and one-year government-bond CIP deviations. Regressions use September 30 as the cutoff, a 90-day bandwidth, triangular weights, and separate linear trends on the two sides. Exact \(p\)-values are from two-sided sign-flip tests over 10 calendar-year score blocks.
		\end{minipage}
	\end{table}
	
	As reported in \Cref{tab:cip_yearend_regime}, the mapped charges range from \(2.26\) to \(2.60\) bp, exact \(p\)-values range from \(0.002\) to \(0.006\), and all 10 currency-level estimates are positive. These magnitudes are close to the \(2\)--\(3\) bp OFB discontinuity, although the comparison is descriptive because only the option data directly identify the \(1/\tau\) functional form.
	
	\subsection{A Common Mid-2010s Regime Shift}
	\label{subsec:cip_regime_shift}
	
	The time variation in government-bond CIP provides a stronger external comparison. I estimate annual boundary effects over 2000--2024 and apply the same first-post-year convention used for the OFB regime analysis in \Cref{subsec:yearly_regime_shift}. The primary series is the daily equal-weighted mean across currencies, while daily cross-sectional medians and a currency--trade-date pooled panel provide robustness checks.
	
	Candidate first-post years range from 2010 through 2020. The 2007--2009 global financial crisis is retained as nuisance annual variation but excluded from the PRE--POST contrast. Scan-adjusted inference repeats the complete break search under null-imposed calendar-year sign flips.
	
	\begin{table}[!t]
		\centering
		\singlespacing
		\caption{Regime Change in the Government-Bond CIP Boundary Effect}
		\label{tab:cip_regime_break}
		\begin{threeparttable}
			\footnotesize
			\setlength{\tabcolsep}{5.2pt}
			\renewcommand{\arraystretch}{1.07}
			\begin{tabular*}{0.96\textwidth}{@{\extracolsep{\fill}}lccc@{}}
				\toprule
				&
				\makecell{Daily equal-weighted\\mean}
				&
				\makecell{Daily cross-sectional\\median}
				&
				\makecell{Currency--trade-date\\pooled panel} \\
				\midrule
				Selected first-post year
				& 2018
				& 2018
				& 2018 \\
				Post minus pre, bp
				& 2.474
				& 2.039
				& 2.547 \\
				Scan-adjusted \(p\)-value
				& 0.0169
				& 0.0239
				& 0.0097 \\
				95\% compatible break years
				& 2014--2019
				& 2014--2019
				& 2014--2019 \\
				\addlinespace[0.25em]
				PRE boundary charge, bp
				& 0.784
				& 0.817
				& 0.633 \\
				PRE exact \(p\)-value
				& 0.0274
				& 0.0128
				& 0.0505 \\
				POST boundary charge, bp
				& 3.256
				& 2.855
				& 3.219 \\
				POST exact \(p\)-value
				& 0.0156
				& 0.0156
				& 0.0156 \\
				\bottomrule
			\end{tabular*}
		\end{threeparttable}
		
		\vspace{0.4em}
		
		\begin{minipage}{0.96\textwidth}
			\footnotesize
			\textit{Notes:} The table uses annual September 30 discontinuities in \(\operatorname{CIP}^{\mathrm{govt}}(3M)-\operatorname{CIP}^{\mathrm{govt}}(1Y)\) over 2000--2024, normalized by four as in \Cref{eq:cip_toll_mapping}. The break year is the first calendar year in the POST regime. Candidate first-post years range from 2010 through 2020, with at least five non-GFC years on each side. The 2007--2009 GFC years are retained as nuisance states but excluded from the PRE--POST contrast and identifying sign blocks. Scan-adjusted \(p\)-values account for searching over candidate dates using null-imposed calendar-year sign flips. The 95\% compatibility sets invert candidate-date tests conditional on a one-break specification. All boundary charges are in unannualized basis points.
		\end{minipage}
	\end{table}
	
	All three specifications select 2018 as the first post-regime year (\Cref{tab:cip_regime_break}). The estimated increase ranges from \(2.04\) to \(2.55\) bp, scan-adjusted \(p\)-values range from \(0.0097\) to \(0.0239\), and all three 95\% compatibility sets span 2014--2019. Government-bond CIP therefore provides stronger unknown-break evidence than the OFB scan itself while independently locating the transition in essentially the same mid-2010s window.
	
	The pre-break CIP charge is already positive at \(0.63\)--\(0.82\) bp, so the regime change represents an increase rather than the appearance of a previously absent effect. Post-break estimates rise to \(2.86\)--\(3.26\) bp. The selected break is also insensitive to individual episodes: leave-one-year-out scans continue to select 2018 except when 2017 is omitted, in which case the optimum shifts to 2016.
	
	The common timing is more informative than the similarity in mapped magnitudes. The OFB and government-bond CIP designs use different contracts, prices, participants, and estimation procedures, yet both identify a substantial increase in the price of year-end exposure within overlapping transition windows (\Cref{subsec:yearly_regime_shift,tab:cip_regime_break}). This timing overlaps the post-crisis tightening of balance-sheet regulation. I do not interpret the parallel breaks as identifying the causal effect of any particular regulation; rather, they provide cross-market evidence consistent with a higher shadow price of carrying intermediary balance-sheet exposure across year-end.
	
	\subsection{Replication in an Independent Option-Implied-Rate Panel}
	\label{subsec:dvt_rep}
	
	The international panel of \citet{DVT26,DVT26Data} provides a second external check using option-implied rates constructed from different option sources and an independent estimation pipeline. Because the published series are interpolated to fixed tenors, this exercise is less local than the strict maturity design in \Cref{subsec:data_local_design} and is therefore treated as a replication rather than an equivalent estimator.
	
	A three-month position begins to span December 31 around September 30, while a six-month position begins to do so around June 30. Comparing each tenor with the same-date one-year rate yields annualized boundary effects \(\Delta_{3M}\) and \(\Delta_{6M}\). Under a common fixed price wedge,
	\begin{equation}
		0.25\,\Delta_{3M}
		=
		0.50\,\Delta_{6M}
		=
		\kappa.
		\label{eq:dvt_tau_restriction}
	\end{equation}
	
	In the U.S.\ box-rate panel, a 60-day local bandwidth produces annualized discontinuities of approximately \(4.69\) bp at three months and \(1.90\) bp at six months, corresponding to unannualized effects of \(1.17\) and \(0.95\) bp. The cross-tenor restriction in \Cref{eq:dvt_tau_restriction} is not rejected (\(p=0.893\)), and the restricted common wedge is approximately \(1.13\) bp.
	
	Inference on the common wedge is weak, with an exact \(p\)-value of \(0.567\). The DVT evidence is therefore not a separate high-powered rejection of zero. Its contribution is construction independence: a positive U.S.\ crossing pattern and cross-tenor magnitudes consistent with the fixed-price representation also appear in data produced by other researchers.
	
	The international evidence is more heterogeneous. Europe shows little unconditional three- or six-month discontinuity, although synchronized break searches select 2015 for both the box-rate and convenience-yield series. The corresponding post-minus-pre increases are approximately \(2.48\) and \(2.63\) bp, with scan-adjusted \(p\)-values of \(0.136\) and \(0.102\). These results are suggestive rather than statistically sharp. They nevertheless show that the U.S.\ pattern is not a mechanical consequence of applying the same boundary construction to interpolated option rates.
	
	One possible source of this international heterogeneity is an interaction between balance-sheet constraints and the interest-rate environment, since \citet{DVT26} document a strong relation between convenience yields and nominal rates. I do not pursue that mechanism here. Fixed-tenor interpolation, the short post-regulation history, and simultaneous cross-country differences in interest rates and intermediary structure make a clean decomposition difficult.
	
	Taken together, the two external exercises sharpen the interpretation of the main OFB result. Government-bond CIP independently exhibits both a year-end boundary charge and a closely aligned mid-2010s strengthening, while the Diamond--Van Tassel panel reproduces the positive U.S.\ option-implied crossing pattern and cross-tenor scaling consistent with a fixed price wedge. These findings make explanations based solely on my option-processing pipeline, fixed option-expiration rules, or generic calendar seasonality less plausible and are consistent with a time-varying price of year-end balance-sheet capacity.
	
	\section{What Do Option-Implied Rates Measure?}
	\label{sec:measurement}
	
	The empirical distinction in this paper is between statistical precision and economic purity. The synthetic-forward regressions fit extremely well (\Cref{app:ab_fit,tab:app_ab_fit}), yet the resulting option--benchmark spread contains a reporting-boundary discontinuity and covaries with asset-backed funding conditions (\Cref{fig:ofbrd,tab:monthly_local_yearend_rd,tab:yearend_regression_spx,tab:yearend_regression_rut}). Precise recovery of the price of a riskless payoff therefore does not by itself establish a frictionless risk-free rate.
	
	\subsection{Relative Purity and Economic Interpretation}
	\label{subsec:same_spread_identification}
	
	With a Treasury benchmark, the OFB is algebraically identical to the option-implied-rate minus Treasury-rate spread used as a convenience-yield measure \citep{BDG22,DVT26}. A completed box has a fixed terminal payoff and provides no Treasury convenience services, but maintaining the position may still require cash, collateral, margin, internal limits, or balance-sheet capacity. A riskless payoff and frictionless implementation are therefore distinct conditions.
	
	By \Cref{eq:theory_ofb_decomposition}, the observed spread satisfies
	\begin{equation}
		\operatorname{OFB}_{i,t}^{b}(T)
		=
		CY_{t}^{b}(T)
		+
		\phi_{i,t}^{\mathrm{opt}}(T)
		-
		\phi_{t}^{b}(T).
		\label{eq:measurement_pure_convenience_restriction}
	\end{equation}
	
	Identifying the entire spread with the benchmark convenience yield consequently requires the relative-purity condition \(\phi_{i,t}^{\mathrm{opt}}(T)=\phi_{t}^{b}(T)\) in \Cref{eq:theory_relative_purity}. Neither put--call parity nor payoff certainty imposes this cancellation.
	
	The empirical results show why this distinction matters. The unannualized OFB rises by about \(2\)--\(3\) bp when maturity first crosses December 31 among contracts priced on the same trade date (\Cref{fig:ofbrd,tab:monthly_local_yearend_rd}). The jump is nearly unchanged across DGS and DTB and remains positive against OIS on identical option cells, while the benchmark-side component does not exhibit a stable positive discontinuity (\Cref{tab:option_benchmark_jump_decomposition,fig:ois_common_rd,tab:ois_common_results}). Separately, the full panel loads positively on the same-day \(\operatorname{ABCP90\!-\!DTB3}\) spread, while the year-end coefficient remains positive after controlling for funding conditions, quoted liquidity, the NFCI, and maturity structure (\Cref{tab:yearend_regression_spx,tab:yearend_regression_rut}).
	
	The functional-form and external evidence reinforce this interpretation. The fixed-charge specification is favored over a duration-proportional premium (\Cref{tab:yearend_functional_form}). Government-bond CIP also exhibits a positive boundary effect in an economically distinct relative-value market (\Cref{subsec:cip_boundary_effect,tab:cip_yearend_regime}). These results do not imply that the OFB contains no convenience yield; they show instead that parity and payoff certainty alone cannot identify convenience as the entire spread.
	
	This creates an asymmetry in identification. One economically meaningful implementation wedge is sufficient to reject frictionlessness as a mechanical consequence of put--call parity, whereas establishing economic purity requires ruling out relevant wedges or showing that option- and benchmark-side wedges offset. The identified year-end toll is therefore an existence proof rather than a complete decomposition.
	
	\subsection{Economic Magnitude and the Difficulty of Correction}
	\label{subsec:measurement_magnitude}
	
	The local \(2\)--\(3\) bp estimate is an unannualized price effect. Under the fixed-price specification favored in \Cref{tab:yearend_functional_form}, a boundary charge \(\kappa\) contributes approximately
	\begin{equation}
		\Delta \operatorname{OFB}^{\mathrm{YC}}(\tau)
		\simeq
		\frac{\kappa}{\tau}
		\label{eq:measurement_annualized_boundary}
	\end{equation}
	to the annualized option--benchmark spread. A representative \(2.5\) bp charge therefore corresponds to about \(5\) annualized bp at six months, \(10\) bp at three months, and \(30\) bp at one month.
	
	The identified wedge cannot be removed by subtracting a constant. It appears only when maturity spans the reporting boundary, scales nonlinearly with \(1/\tau\) in annualized units, and varies substantially over time (\Cref{fig:yearly_method_comparison,fig:yearly_regime_break}). Moreover, December 31 provides unusually sharp identification, whereas funding, collateral, netting, inventory, or balance-sheet costs that vary smoothly may leave no comparable cutoff. Removing the identified year-end component would therefore not establish that the residual option-implied rate is frictionless.
	
	\subsection{Implications for International Decompositions}
	\label{subsec:international_decompositions}
	
	The same identification issue applies when option-implied rates are used to decompose international arbitrage deviations. Allowing for option-side implementation wedges in foreign-currency \(f\) and U.S.\ dollar \(\$\) markets gives
	\begin{equation}
		\widehat{\operatorname{CIP}}_{f,t}^{\mathrm{opt}}(T)
		=
		\operatorname{CIP}_{f,t}^{0}(T)
		+
		\phi_{\$,t}^{\mathrm{opt}}(T)
		-
		\phi_{f,t}^{\mathrm{opt}}(T),
		\label{eq:international_option_cip}
	\end{equation}
	where \(\operatorname{CIP}_{f,t}^{0}(T)\) denotes the international-arbitrage deviation stripped of option-side implementation wedges.
	
	Equating the option-based measure with \(\operatorname{CIP}_{f,t}^{0}(T)\) additionally requires
	\begin{equation}
		\phi_{f,t}^{\mathrm{opt}}(T)
		=
		\phi_{\$,t}^{\mathrm{opt}}(T).
		\label{eq:international_option_wedge_cancellation}
	\end{equation}
	
	Put--call parity identifies the discount slope within each option market but does not require implementation wedges to coincide across collateral regimes, netting arrangements, intermediary structures, or reporting regimes. The accounting decomposition remains valid when such wedges exist, but the economic attribution of its components changes.
	
	The U.S.\ evidence does not identify foreign option-side wedges and therefore does not directly test the cancellation condition in \Cref{eq:international_option_wedge_cancellation}. The independent international panel examined in \Cref{subsec:dvt_rep} also exhibits substantial cross-country heterogeneity rather than mechanical equality of year-end effects. Cross-country cancellation of option-side wedges must therefore be justified or tested separately rather than inferred from put--call parity.
	
	\subsection{Regulatory Interpretation and Scope}
	\label{subsec:measurement_regulatory_regime}
	
	The regime evidence also informs the interpretation of the boundary wedge. The OFB transition is broadly localized to 2015--2019 (\Cref{subsec:yearly_regime_shift}), while government-bond CIP independently yields a closely overlapping 2014--2019 window (\Cref{subsec:cip_regime_shift}). The Diamond--Van Tassel panel provides weaker but directionally similar evidence of a mid-2010s transition (\Cref{subsec:dvt_rep}). Taken together, these results favor a strengthening of year-end pricing during the post-crisis regulatory transition rather than a stable option-market seasonal.
	
	This timing overlaps the tightening of balance-sheet regulation for large U.S.\ intermediaries. The G-SIB capital surcharge was finalized in 2015, phased in from 2016, and became fully effective in 2019 \citep{FedGSIB15}. The framework is particularly relevant to year-end incentives because systemic indicators used in G-SIB measurement have generated documented balance-sheet adjustments around reporting dates \citep{BMPW22}. The enhanced supplementary leverage ratio, which tightened a non-risk-weighted capital constraint for the largest U.S.\ banking organizations, became effective in 2018 \citep{FedESLR14}.
	
	The timing is corroborating rather than causal evidence. The regulations overlap with one another and with other post-crisis changes, while the statistical break date is itself estimated with uncertainty. The designs therefore do not identify the causal effect of the G-SIB surcharge, the enhanced supplementary leverage ratio, or any other particular rule. The relevant observation is narrower: independently constructed OFB and government-bond CIP measures both exhibit a materially stronger price of year-end exposure during essentially the same transition period (\Cref{subsec:yearly_regime_shift,subsec:cip_regime_shift}).
	
	More generally, the paper identifies a reporting-boundary component in the relative price of option-implied and benchmark discount factors, not the level of \(\phi_{i,t}^{\mathrm{opt}}(T)\). The positive relation with \(\operatorname{ABCP90\!-\!DTB3}\) likewise identifies funding-related covariation rather than the borrowing rate or internal capital cost of the marginal box trader. The benchmark decomposition and OIS tests reduce the plausibility of a benchmark-side explanation, while the external evidence shows that related patterns appear outside my primary option-processing pipeline (\Cref{tab:option_benchmark_jump_decomposition,tab:ois_common_results,sec:external_evidence}). None of these tests separately identifies funding, collateral, margin, internal capital, or regulatory channels.
	
	The annualized magnitudes in \Cref{subsec:measurement_magnitude} are therefore mechanical price-equivalent mappings of the estimated fixed wedge rather than structural funding-cost estimates. The narrower conclusion is sufficient for the paper: a precisely estimated discount factor and a riskless terminal payoff do not establish a frictionless discount rate. Interpreting the entire option--benchmark spread as a convenience yield requires the relative-purity condition in \Cref{eq:theory_relative_purity} to be justified or tested separately.
	
	\section{Conclusion}
	\label{sec:conclusion}
	
	Put--call parity can identify option-implied discount factors with high statistical precision without making the resulting rates frictionless. A completed box delivers a fixed terminal payoff, yet maintaining it may require cash, collateral, margin, internal limits, and balance-sheet capacity. High synthetic-forward fit therefore establishes measurement precision rather than economic purity (\Cref{app:ab_fit,tab:app_ab_fit}).
	
	The clearest evidence is the price discontinuity when maturity first crosses a future December 31. Strict same-date comparisons produce an average unannualized jump of about \(2\)--\(3\) bp, while full-panel estimates are of similar magnitude (\Cref{fig:ofbrd,tab:monthly_local_yearend_rd,tab:yearend_regression_spx,tab:yearend_regression_rut}). Functional-form tests favor a fixed price wedge that enters annualized rates through \(1/\tau\), so a representative \(2.5\) bp charge corresponds to about \(10\) annualized bp at three months (\Cref{tab:yearend_functional_form}). The effect survives alternative Treasury and OIS benchmarks, regular-weekly contracts, and calendar-placebo tests, while the OFB separately covaries positively with benchmark-relative asset-backed funding conditions (\Cref{fig:ois_common_rd,tab:ois_common_results,fig:rd_weekly,tab:weekly_local_yearend_rd,tab:quarterly_expiration_confound,tab:yearend_regression_spx,tab:yearend_regression_rut}).
	
	The boundary charge is also strongly time varying. Annual estimates from the full-panel and strict local designs comove closely, and the option evidence places a substantial strengthening of the year-end effect in the mid-2010s (\Cref{fig:yearly_method_comparison,fig:yearly_regime_break}). Government-bond CIP independently exhibits both a positive year-end boundary charge and a closely aligned regime shift despite using different securities, prices, participants, and estimation procedures (\Cref{subsec:cip_boundary_effect,subsec:cip_regime_shift,tab:cip_yearend_regime,tab:cip_regime_break}). An independently constructed international option-implied-rate panel also reproduces the positive U.S.\ crossing pattern and cross-tenor magnitudes consistent with the fixed-price representation (\Cref{subsec:dvt_rep}). Together, these results are consistent with a higher shadow price of carrying intermediary balance-sheet exposure across year-end during the post-crisis regulatory transition, although they do not identify the causal effect of any particular regulation.
	
	These findings do not imply that the entire OFB is an option-side implementation cost or that safe-asset convenience yields are absent. They establish the narrower result that at least one economically meaningful implementation wedge enters the relative price of option-implied and benchmark discount factors. Because the identified component is discontinuous, maturity dependent, and time varying, it cannot be removed by a constant adjustment. Nor would removing it establish that smoother funding, collateral, margin, or balance-sheet wedges are absent (\Cref{sec:measurement}).
	
	Interpreting the entire option--benchmark spread as a convenience yield therefore requires the additional relative-purity condition that option- and benchmark-side implementation wedges offset. Put--call parity identifies the discount rate precisely; it does not identify its economic purity.
	
	\newpage
	
	\newpage
	\singlespacing

	\newpage
	\thispagestyle{plain}
	
	\begin{center}
		{\Large\bfseries Internet Appendix to}\\[0.5em]
		{\large\bfseries The Year-End Toll}
	\end{center}
	
	\vspace{1em}
	
	
	\counterwithout{table}{section}
	\counterwithout{figure}{section}
	
	\setcounter{section}{0}
	\setcounter{subsection}{0}
	\setcounter{subsubsection}{0}
	\setcounter{equation}{0}
	\setcounter{table}{0}
	\setcounter{figure}{0}
	
	\renewcommand{\thesection}{IA.\arabic{section}}
	\renewcommand{\thesubsection}{\thesection.\arabic{subsection}}
	\renewcommand{\thesubsubsection}{\thesubsection.\arabic{subsubsection}}
	
	\renewcommand{\theequation}{IA.\arabic{equation}}
	\renewcommand{\thetable}{IA.\arabic{table}}
	\renewcommand{\thefigure}{IA.\arabic{figure}}
	
	\renewcommand*{\theHsection}{IA.\arabic{section}}
	\renewcommand*{\theHsubsection}{IA.\arabic{section}.\arabic{subsection}}
	\renewcommand*{\theHsubsubsection}{IA.\arabic{section}.\arabic{subsection}.\arabic{subsubsection}}
	
	\renewcommand*{\theHequation}{IA.\arabic{equation}}
	\renewcommand*{\theHtable}{IA.\arabic{table}}
	\renewcommand*{\theHfigure}{IA.\arabic{figure}}
	
	\onehalfspacing
	
	\section{Data Sources and Sample Coverage}
	\label{app:data_sources}
	
	This section summarizes the sources, frequencies, sample periods, and uses of the raw data. Option-implied and benchmark discount factors are constructed independently and merged only afterward. The NFCI and ABCP--Treasury-bill spread enter only the full-panel regressions; secured overnight rates, equity indices, and government-bond CIP data are used only in robustness or external-validity tests. None enters the construction of the option-implied discount factor or baseline OFB.
	
	Option data are one-minute National Best Bid and Offer quotes for SPX, RUT, SPXW, and RUTW from Theta Data \citep{ThetaData}, with contract, settlement, and expiration conventions following Cboe specifications \citep{CBOESPX26,CBOERUT26}. SPX and RUT form the standard-monthly sample used in the full-panel and baseline local analyses, while SPXW and RUTW form the separate regular-weekly replication sample.
	
	The primary Treasury benchmark is the DGS curve constructed from FRED constant-maturity Treasury yields at 1, 3, and 6 months and 1, 2, and 3 years \citep{FRED_DGS}, supporting maturities through three years. DTB uses FRED Treasury-bill discount yields at 4 weeks, 3 months, 6 months, and 1 year and supports maturities through one year \citep{FRED_DTB}. The two curves use different source quotations and discount-factor conversions, providing a direct robustness check on Treasury curve construction.
	
	The non-Treasury benchmark is a U.S.\ dollar OIS curve referenced to the EFFR and bootstrapped from market quotes. The source data span January 2015 through October 2025, while the analysis ends in December 2024 to retain complete calendar-year episodes. OIS is used only in matched-sample tests of whether the year-end result reflects Treasury-specific convenience demand, collateral value, or curve construction.
	
	Full-panel regressions control for the weekly NFCI from FRED \citep{FRED_NFCI}, matched to the latest observation available on or before each option trade date with a maximum carry of 14 calendar days. NFCI captures broad leverage, credit, liquidity, and risk-bearing conditions rather than the funding cost of a particular arbitrageur.
	
	Asset-backed short-term funding conditions are measured by the daily spread
	\[
	\Delta_{t}^{\mathrm{ABCP-TB}}
	=
	100
	\left[
	\operatorname{ABCP90}_{t}
	-
	\operatorname{DTB3}_{t}
	\right],
	\]
	where \(\operatorname{ABCP90}_{t}\) is the 90-day AA asset-backed commercial-paper rate and \(\operatorname{DTB3}_{t}\) is the three-month Treasury-bill secondary-market rate from FRED \citep{FRED_CP,FRED_DTB}. The baseline full-panel sample requires both components on the option trade date and therefore uses an exact same-day spread. A robustness specification carries only a previously observed completed spread for at most three calendar days; future observations are never used and the component rates are never filled separately. The spread enters only the full-panel regressions and affects neither option-implied nor benchmark discount factors.
	
	Secured year-end turn tests use daily SOFR, TGCR, and BGCR. For 2014--2017, before official SOFR publication, I use the Federal Reserve Bank of New York's historical indicative SOFR series; for 2018--2025, I use published SOFR from FRED \citep{SOFR_proxy,FRED_SOFR}. The two SOFR series are linked across years but never combined within a single year-end episode. TGCR and BGCR come from the New York Fed \citep{NYFedTGCR,NYFedBGCR}. These series are used only to measure realized cumulative price effects around year-end.
	
	Dividend robustness tests use price and total-return indices for the S\&P 500 and Russell 2000 \citep{compustat}. Realized dividend contributions from trade date to maturity are measured as the difference between total-return and price-return log growth. This variable is an ex post robustness outcome and is not interpreted as a real-time measure of expected dividends.
	
	The external-market test uses the government-bond covered-interest-parity data of \citet{DKSData25}, covering multiple currencies and tenors against the U.S.\ dollar from January 2000 through June 2025. These data share no price inputs with the option, Treasury, or OIS samples and provide independent evidence on the year-end boundary effect.
	
	Option data and all benchmark curves therefore have no common price inputs. Each source is cleaned and converted independently, and option and benchmark observations are merged only when market, trade date, and maturity coincide and the target maturity lies within benchmark support. The OFB is consequently the relative price of independently estimated option-implied and external benchmark discount factors; auxiliary series affect only the regressions or robustness tests in which they are explicitly introduced.
	
	\section{Construction of Option-Implied and Benchmark Discount Curves}
	\label{app:pipeline_construction}
	
	This section describes the construction of option-implied and benchmark discount factors and their combination into the analysis panels. Option discount factors use only contemporaneous call and put quotes, while DGS, DTB, and OIS curves are constructed independently with no future information or extrapolation beyond observed support. The two sides are merged only after construction at exactly matched market--trade-date--maturity cells. Standard monthlies and regular weeklies use the same synthetic-forward estimator and quality screens but separate contract selection, maturity conventions, and daily aggregation; they are never pooled. \Cref{app:ab_fit} reports synthetic-forward fit and intraday precision.
	
	\subsection{Option Quotes and Contract Selection}
	\label{app:option_pipeline}
	
	The option data are one-minute NBBO quotes for European-style SPX, RUT, SPXW, and RUTW index options \citep{ThetaData}. Standard-monthly SPX and RUT and regular-weekly SPXW run from June 2012 through December 2025; RUTW begins in January 2016.
	
	The standard-monthly sample retains only regular monthly expirations under the SPX and RUT roots and excludes weekly, daily, end-of-month, and other nonstandard contracts before daily aggregation. These contracts are AM-settled, with ACT/365 maturity measured from each quote timestamp to 9:30 a.m.\ Eastern Time on expiration.
	
	The regular-weekly sample retains only SPXW and RUTW expirations classified as regular weeklies by the exchange calendar: Friday, or the immediately preceding NYSE trading day when Friday is closed. Daily, end-of-month, standard-monthly, and other irregular expirations are excluded before aggregation. Regular weeklies are PM-settled, with ACT/365 maturity measured to 4:00 p.m.\ Eastern Time on expiration, or 1:00 p.m.\ on early-close expiration dates. This expiration-specific adjustment is applied before minute-level maturity is computed and therefore propagates through daily aggregation and benchmark matching.
	
	Trade dates are restricted to NYSE business days using the default NYSE calendar in the MATLAB Financial Toolbox. Weekends and full-market holidays are excluded, while early-close sessions remain valid trading days; for affected PM-settled weeklies, only the terminal timestamp changes. Contracts with expiration before the trade date or nonpositive or nonfinite maturity are removed.
	
	I retain noncrossed quotes with positive bids and asks no lower than bids and use bid--ask midpoints. Duplicate observations within market, timestamp, maturity, strike, and option side are collapsed by medians of the price fields. Calls and puts are matched only when market, timestamp, maturity, and strike coincide exactly.
	
	Each minute--maturity cell retains strikes observed on both option sides. Cells failing minimum-strike or discount-factor validity requirements are excluded. Only AB estimates with \(R^{2}\geq0.999\) enter daily discount-factor aggregation, while the bid--ask measure is screened independently.
	
	Residual maturity is restricted to \(30\leq\tau_{t,T}\leq1{,}095\) calendar days for standard monthlies and \(7\leq\tau_{t,T}\leq365\) days for regular weeklies. These restrictions are imposed before benchmark matching and verified afterward.
	
	Minute-level discount factors and liquidity measures are aggregated by intraday medians within market--trade-date--maturity cells, with market identity preserved even when multiple markets share a file. Discount-factor estimation is detailed in \Cref{app:option_discount_estimators}, and liquidity construction and aggregation in \Cref{app:liquidity_intraday_aggregation}.
	
	\subsection{Option-Implied Discount-Factor Estimators}
	\label{app:option_discount_estimators}
	
	For option market \(i\), trade date \(t\), minute \(s\), maturity \(T\), and strike \(K\), the baseline AB estimator uses the synthetic-forward relation of \citet{AB21,BDG22}:
	\begin{align}
		C_{i,t,s}(K,T)-P_{i,t,s}(K,T)
		&=
		B_{i,t,s}(T)
		\left[
		F_{i,t,s}(T)-K
		\right]
		\notag\\
		&=
		a_{i,t,s}(T)
		+
		\xi_{i,t,s}(T)K
		+
		\varepsilon_{i,t,s}^{\mathrm{AB}}(K,T).
		\label{eq:app_ab_discount_identification}
	\end{align}
	
	The intercept \(a_{i,t,s}(T)=B_{i,t,s}(T)F_{i,t,s}(T)\) absorbs the prepaid-forward level and dividend component. Since \(\xi_{i,t,s}(T)=-B_{i,t,s}(T)\), the strike slope identifies the discount factor without separately estimating dividends:
	\begin{equation}
		\widehat B_{i,t,s}^{\mathrm{AB}}(T)
		=
		-\widehat\xi_{i,t,s}(T).
		\label{eq:app_ab_discount_factor}
	\end{equation}
	
	Each minute-level regression requires at least three matched strikes. Before intraday aggregation, I retain only finite estimates satisfying
	\begin{equation}
		0.5
		\leq
		\widehat B_{i,t,s}^{\mathrm{AB}}(T)
		\leq
		1.5,
		\qquad
		R_{i,t,s}^{2}(T)
		\geq
		0.999.
		\label{eq:app_ab_validity}
	\end{equation}
	
	The baseline daily discount factor \(\widehat B_{i,t}(T)\) is the intraday median of valid \(\widehat B_{i,t,s}^{\mathrm{AB}}(T)\) estimates within each market--trade-date--maturity cell. AB uses the full strike cross section rather than individual-strike parity residuals and is the baseline estimator throughout. \Cref{app:ab_fit} reports its fit and intraday precision.
	
	For standard monthlies, I also construct an adjacent-strike Box estimator:
	\begin{equation}
		\widehat B_{i,t,s,g}^{\mathrm{Box}}(T)
		=
		\frac{
			[C-P]_{i,t,s}(K_g,T)
			-
			[C-P]_{i,t,s}(K_{g+1},T)
		}{
			K_{g+1}-K_g
		}.
		\label{eq:app_box_discount_identification}
	\end{equation}
	
	Here, \(g\) indexes adjacent strike pairs. I interpolate the strike at which \(C-P=0\) to locate forward ATM and retain up to six adjacent boxes around it. Boxes are weighted by quoted width and distance from forward ATM, and the fitted value at forward ATM is obtained from a weighted local-linear regression, with a weighted mean used when the local slope is unidentified. Because Box uses only a narrow strike neighborhood, it is not subject to the AB \(R^{2}\) screen.
	
	Regular weeklies use AB only; the weekly pipeline neither constructs smoothed Box estimates nor imports monthly Box values.
	
	\subsection{Liquidity Measurement and Intraday Aggregation}
	\label{app:liquidity_intraday_aggregation}
	
	Quoted liquidity is measured at the strike nearest the AB-implied prepaid forward. I take the median of the call and put dollar bid--ask spreads, divide by the AB-implied forward, and multiply by \(10^{4}\) to express the measure in basis points.
	
	AB discount factors and bid--ask measures are aggregated separately by intraday medians within market--trade-date--maturity cells; monthly Box discount factors are aggregated at the same level. Final AB cells require at least \(95\%\) valid minute-level coverage for both the discount factor and bid--ask measure. The same coverage rule applies to monthly Box discount factors, and direct AB--Box comparisons require both sets of criteria. The \(R^{2}\) screen applies only to AB discount factors and therefore does not mechanically select the liquidity measure.
	
	\subsection{Benchmark Discount-Curve Construction}
	\label{app:benchmark_curve_construction}
	
	DGS and DTB curves are anchored at \(D_t(0)=1\), while OIS is normalized at the contract's two-business-day spot date. Target discount factors are obtained by linear interpolation in log discount-factor space using each curve's native maturity coordinate, with no extrapolation outside observed support.
	
	Benchmarks are matched to option trade dates on an as-of basis. Same-day curves are used when available; otherwise, the latest prior curve is carried forward for at most 5 calendar days for DGS, 3 for DTB, and 4 for OIS. Cells without an admissible prior curve are dropped, and future curves are never used. This trade-date matching is distinct from missing-tenor treatment within a curve date: DGS and DTB require complete source tenors, whereas OIS permits the limited tenor-level carry described below.
	
	\subsubsection{DGS Treasury Curve}
	\label{app:dgs_pipeline}
	
	The 1- and 3-month DGS yields are converted directly into zero-coupon nodes. Published yields from 6 months through 3 years are treated as par yields under a bond-equivalent semiannual convention. The 6-month quote supplies the first semiannual par node; subsequent par yields are linearly interpolated to the semiannual coupon grid and discount factors are bootstrapped recursively.
	
	Maturity coordinates use unadjusted constant-maturity dates under ACT/365, and target option maturities are obtained by log-linear interpolation between nodes. Curves are constructed only when all required source tenors are observed; missing tenors are neither interpolated across dates nor carried forward during curve construction. The final curve spans June 2012 through December 2025 and supports \(30\)--\(1{,}095\)-day target maturities.
	
	\subsubsection{DTB Treasury-Bill Curve}
	\label{app:dtb_pipeline}
	
	The DTB curve uses fixed nodes at 28, 91, 182, and 365 days. Let \(q_{t,n}^{\mathrm{DTB}}\) denote the published bank-discount yield and \(d_n\) node maturity. The discount factor is
	\begin{equation}
		D_{t}^{\mathrm{DTB}}(d_n)
		=
		1-
		\left(
		\frac{q_{t,n}^{\mathrm{DTB}}}{100}
		\right)
		\frac{d_n}{360}.
		\label{eq:app_dtb_discount_factor}
	\end{equation}
	
	With \(D_t^{\mathrm{DTB}}(0)=1\), define cumulative log discount as
	\begin{equation}
		A_t^{\mathrm{DTB}}(d)
		=
		-\log D_t^{\mathrm{DTB}}(d).
		\label{eq:app_dtb_log_discount}
	\end{equation}
	
	Target discount factors from 1 to 365 days are obtained by linearly interpolating \(A_t^{\mathrm{DTB}}(d)\) between nodes and transforming back to discount factors. DTB therefore requires neither coupon bootstrapping nor par-yield interpolation. Curves require all four source quotes. The final sample spans June 2012 through December 2025, with \(30\)--\(365\)-day target maturities used in the analysis.
	
	\subsubsection{OIS Curve}
	\label{app:ois_pipeline}
	
	The OIS curve is bootstrapped directly from EFFR-referenced U.S.\ dollar OIS quotes. Raw quotes run from January 2, 2015 through October 31, 2025, while the empirical sample ends on December 31, 2024 to retain ten complete year-end episodes.
	
	Curve construction uses a two-business-day spot lag, ACT/360 accrual, Modified Following adjustment, and zero payment lag. Tenors of one year or less are treated as single-period contracts; longer maturities use annual fixed-leg payments, with front stubs for the 15- and 18-month contracts. Intermediate payment-date discount factors are log-linearly interpolated between previously constructed nodes. The 21-month quote is excluded because it appears only from September 2023 and would change the interpolation grid over the sample.
	
	Isolated missing quotes for retained tenors may be carried forward for at most five business days. A node is retained only when every intermediate discount factor required by its cash-flow schedule can be constructed. Target option maturities are converted to ACT/360 from the spot date, and only internal curve support corresponding to option maturities of \(30\)--\(1{,}095\) calendar days is used.
	
	\subsection{Maturity Matching and OFB Panels}
	\label{app:ofb_pipeline}
	
	Let \(e\) index the option discount-factor estimator. For standard monthlies, \(\mathcal E^{\mathrm M}=\{\mathrm{AB},\mathrm{Box}\}\); for regular weeklies, \(\mathcal E^{\mathrm W}=\{\mathrm{AB}\}\). Benchmarks are \(b\in\mathcal B=\{\mathrm{DGS},\mathrm{DTB},\mathrm{OIS}\}\).
	
	The daily option discount factor \(\widehat B_{i,t}^{e}(T)\) is the market--trade-date--maturity aggregate of estimator \(e\): the intraday median of valid minute-level estimates for AB, and the daily aggregation of locally combined adjacent-strike estimates for Box.
	
	Option maturity \(\tau_{t,T}\) is measured on ACT/365. For benchmark interpolation, the same target maturity is mapped into the DGS ACT/365 coordinate, DTB calendar-day coordinate, or OIS spot-date ACT/360 coordinate. These mappings affect only benchmark interpolation; OFB annualization always uses the common option maturity \(\tau_{t,T}\). Cells outside native benchmark support are excluded.
	
	For each valid cell, estimator \(e\), and benchmark \(b\),
	\begin{equation}
		\operatorname{OFB}_{i,t}^{e,b}(T)
		=
		\frac{10^{4}}{\tau_{t,T}}
		\log\!\left(
		\frac{
			D_{t}^{b}(T)
		}{
			\widehat B_{i,t}^{e}(T)
		}
		\right).
		\label{eq:app_ofb_definition}
	\end{equation}
	
	The baseline uses AB, so I suppress the estimator superscript:
	\begin{equation}
		\operatorname{OFB}_{i,t}^{b}(T)
		\equiv
		\operatorname{OFB}_{i,t}^{\mathrm{AB},b}(T),
		\qquad
		\widehat B_{i,t}(T)
		\equiv
		\widehat B_{i,t}^{\mathrm{AB}}(T).
		\label{eq:app_ofb_baseline_notation}
	\end{equation}
	
	A positive OFB means that the option-implied continuously compounded rate exceeds the maturity-matched benchmark rate. It is a relative discount-rate measure identified from the common strike slope of call--put spreads, not an individual-strike parity residual.
	
	For standard monthlies, the natural DGS panel spans June 2012 through December 2025 with \(30\)--\(1{,}095\)-day maturities, while DTB covers \(30\)--\(365\) days over the same period. The partial 2012 sample is retained in the baseline full panel, annual available-sample estimates, and the 2012 local episode; complete-calendar-year results for 2013--2025 are reported separately as robustness.
	
	The baseline strict same-date monthly design uses only \(30\)--\(365\)-day market--trade-date--maturity cells observed under both DGS and DTB, so the two Treasury estimates use identical option observations and maturity support.
	
	OIS analyses cover January 2015 through December 2024. DGS--OIS full-panel comparisons up to three years use exactly matched \(30\)--\(1{,}095\)-day cells. Direct DGS--DTB--OIS comparisons up to one year and the OIS strict same-date local design use cells jointly observed under all three benchmarks within \(30\)--\(365\) days.
	
	Regular-weekly DGS and DTB panels use \(7\)--\(365\)-day maturities. SPXW spans June 2012 through December 2025 and RUTW January 2016 through December 2025. The baseline weekly local design uses cells observed under both DGS and DTB within each market.
	
	The weekly OIS replication uses only cells jointly observed under DGS, DTB, and OIS, covering 2015--2024 for SPXW and 2016--2024 for RUTW. It serves only as a non-Treasury common-support robustness test.
	
	DGS is the primary benchmark because it spans the full sample and long-maturity cross section. DTB provides a short-maturity Treasury replication based on directly converted bill discount yields, while OIS provides a matched non-Treasury robustness benchmark. Standard monthlies are used for full-panel, functional-form, and repeated-crossing analyses; regular weeklies are used only as a separate contract-group replication of the local year-end discontinuity.
	
	\section{Fit and Precision of the Synthetic-Forward Regressions}
	\label{app:ab_fit}
	
	This section evaluates how precisely the synthetic-forward regressions identify the common strike slope in standard monthlies and regular weeklies. Both contract groups use the same quote-validation, call--put-matching, minimum-strike, discount-factor, and \(R^{2}\) screens, while contract selection and maturity ranges are constructed separately. Diagnostics are computed after contract selection and before benchmark matching or the final OFB restrictions.
	
	For market \(i\), trade date \(t\), minute \(s\), and maturity \(T\), define the daily cross-sectional fit error as
	\begin{equation}
		\operatorname{FitError}_{i,t}
		=
		10^{6}
		\left[
		1-
		\operatorname{median}_{s,T}
		\left(
		R^{2}_{i,t,s,T}
		\right)
		\right].
		\label{eq:app_ab_fit_ppm}
	\end{equation}
	The measure is expressed in parts per million, so 1 ppm corresponds to unexplained cross-sectional variation of \(10^{-6}\) in the typical synthetic-forward regression for that trade date.
	
	To assess intraday sampling precision, let \(n_{i,t,T}\), \(\widetilde B_{i,t}(T)\), and \(\operatorname{MAD}_{i,t}(T)\) denote the number, median, and median absolute deviation of valid intraday AB estimates. I approximate the standard error of the maturity-specific intraday median by
	\begin{equation}
		\begin{aligned}
			\widehat{\sigma}^{\log B}_{i,t,T}
			&=
			1.4826
			\frac{
				\operatorname{MAD}_{i,t}(T)
			}{
				\widetilde B_{i,t}(T)
			},
			\\
			\widehat{\operatorname{se}}_{i,t,T}
			&=
			10^{4}
			\frac{
				1.2533\,
				\widehat{\sigma}^{\log B}_{i,t,T}
			}{
				\sqrt{n_{i,t,T}}
			}.
		\end{aligned}
		\label{eq:app_ab_median_precision}
	\end{equation}
	
	\subsection{Standard Monthlies}
	\label{app:ab_fit_monthly}
	
	\begin{table}[!t]
		\centering
		\singlespacing
		\caption{Fit and Precision of Synthetic-Forward Regressions: Standard Monthlies}
		\label{tab:app_ab_fit}
		\begin{threeparttable}
			\footnotesize
			\setlength{\tabcolsep}{6pt}
			\renewcommand{\arraystretch}{1.02}
			\begin{tabular*}{0.96\textwidth}{@{\extracolsep{\fill}}lrrr@{}}
				\toprule
				& SPX & RUT & Pooled \\
				\midrule
				\multicolumn{4}{l}{\textit{Panel A. Cross-sectional fit before quality screening}} \\
				\addlinespace[0.2em]
				Minute--maturity regressions
				& 16,170,396 & 11,397,715 & 27,568,111 \\
				Market--date--maturity cells
				& 41,538 & 29,282 & 70,820 \\
				Median daily fit error, ppm
				& 0.129 & 0.823 & 0.276 \\
				95th percentile
				& 0.515 & 3.670 & 2.948 \\
				Maximum
				& 4.431 & 15.618 & 15.618 \\
				\midrule
				\multicolumn{4}{l}{\textit{Panel B. Minute-level quality screening}} \\
				\addlinespace[0.2em]
				Candidate AB estimates
				& 16,170,201 & 11,397,141 & 27,567,342 \\
				Retained AB estimates
				& 16,163,595 & 11,395,145 & 27,558,740 \\
				Exclusion rate
				& 0.0409\% & 0.0175\% & 0.0312\% \\
				\midrule
				\multicolumn{4}{l}{\textit{Panel C. Precision of intraday median, unannualized bp}} \\
				\addlinespace[0.2em]
				Median
				& 0.078 & 0.167 & 0.110 \\
				95th percentile
				& 0.222 & 0.484 & 0.386 \\
				\bottomrule
			\end{tabular*}
		\end{threeparttable}
		
		\vspace{0.4em}
		
		\begin{minipage}{0.96\textwidth}
			\footnotesize
			\textit{Notes:} The sample contains standard-monthly SPX and RUT options from June 2012 through December 2025 with \(30\leq\tau_{t,T}\leq1{,}095\) calendar days. Weekly, daily, end-of-month, and other nonstandard expirations are excluded. Panel A reports the market--trade-date distribution of the fit-error measure defined in \Cref{eq:app_ab_fit_ppm}. Panel B applies the minimum-strike requirement, \(0.5\leq\widehat B^{\mathrm{AB}}\leq1.5\), and \(R^{2}\geq0.999\) before intraday aggregation. Panel C first takes the median maturity-specific standard error defined in \Cref{eq:app_ab_median_precision} within each market--trade date and then reports its time-series distribution. Pooled distribution statistics weight market--trade dates equally.
		\end{minipage}
	\end{table}
	
	\Cref{tab:app_ab_fit} shows extremely tight cross-sectional fit. Median daily fit errors are \(0.129\) ppm for SPX and \(0.823\) ppm for RUT, while the pooled maximum of \(15.618\) ppm still corresponds to a typical daily \(R^{2}\) of approximately \(0.999984\). The screens remove only 8,602 of 27,567,342 candidate AB estimates, or \(0.0312\%\), and are applied before and independently of benchmark rates, the OFB, and year-end crossing status. Median intraday sampling errors are \(0.078\) bp for SPX and \(0.167\) bp for RUT, with 95th percentiles of \(0.222\) and \(0.484\) bp. These magnitudes are well below the \(2\)--\(3\) bp year-end discontinuities reported in \Cref{tab:monthly_local_yearend_rd}.
	
	These diagnostics support the numerical precision of the option-implied discount factors. They do not, however, rule out economically meaningful price components common across strikes or imply equality between option-implied and external benchmark discount factors.
	
	\subsection{Regular Weeklies}
	\label{app:ab_fit_weekly}
	
	\begin{table}[!t]
		\centering
		\singlespacing
		\caption{Fit and Precision of Synthetic-Forward Regressions: Regular Weeklies}
		\label{tab:app_ab_fit_weekly}
		\begin{threeparttable}
			\footnotesize
			\setlength{\tabcolsep}{6pt}
			\renewcommand{\arraystretch}{1.02}
			\begin{tabular*}{0.96\textwidth}{@{\extracolsep{\fill}}lrrr@{}}
				\toprule
				& SPXW & RUTW & Pooled \\
				\midrule
				\multicolumn{4}{l}{\textit{Panel A. Cross-sectional fit before quality screening}} \\
				\addlinespace[0.2em]
				Minute--maturity regressions
				& 15,281,604 & 7,546,543 & 22,828,147 \\
				Market--date--maturity cells
				& 39,186 & 19,363 & 58,549 \\
				Median daily fit error, ppm
				& 0.420 & 0.423 & 0.420 \\
				95th percentile
				& 1.731 & 6.592 & 4.645 \\
				Maximum
				& 9.484 & 21.533 & 21.533 \\
				\midrule
				\multicolumn{4}{l}{\textit{Panel B. Minute-level quality screening}} \\
				\addlinespace[0.2em]
				Candidate AB estimates
				& 15,281,600 & 7,546,543 & 22,828,143 \\
				Retained AB estimates
				& 15,280,726 & 7,543,796 & 22,824,522 \\
				Exclusion rate
				& 0.0057\% & 0.0364\% & 0.0159\% \\
				\midrule
				\multicolumn{4}{l}{\textit{Panel C. Precision of intraday median, unannualized bp}} \\
				\addlinespace[0.2em]
				Median
				& 0.157 & 0.134 & 0.148 \\
				95th percentile
				& 0.458 & 0.996 & 0.668 \\
				\bottomrule
			\end{tabular*}
		\end{threeparttable}
		
		\vspace{0.4em}
		
		\begin{minipage}{0.96\textwidth}
			\footnotesize
			\textit{Notes:} The sample contains regular SPXW options from June 2012 through December 2025 and RUTW options from January 2016 through December 2025, with \(7\leq\tau_{t,T}\leq365\) calendar days. A regular weekly expiration is Friday, or the preceding NYSE business day when that Friday is a market holiday. Daily, end-of-month, standard-monthly, and other irregular expirations are excluded. Panel A reports the market--trade-date distribution of the fit-error measure defined in \Cref{eq:app_ab_fit_ppm}. Panel B applies the minimum-strike requirement, \(0.5\leq\widehat B^{\mathrm{AB}}\leq1.5\), and \(R^{2}\geq0.999\) before intraday aggregation. Panel C first takes the median maturity-specific standard error defined in \Cref{eq:app_ab_median_precision} within each market--trade date and then reports its time-series distribution. Pooled distribution statistics weight market--trade dates equally.
		\end{minipage}
	\end{table}
	
	\Cref{tab:app_ab_fit_weekly} shows similarly tight cross-sectional fit for regular weeklies. Median daily fit errors are \(0.420\) ppm for SPXW and \(0.423\) ppm for RUTW, implying a typical \(R^{2}\) of about \(0.9999996\). Although RUTW has a heavier right tail, the pooled maximum of \(21.533\) ppm still corresponds to an \(R^{2}\) of approximately \(0.999978\). Only 3,621 of 22,828,143 candidate estimates are removed, for a pooled exclusion rate of \(0.0159\%\). No candidate has a nonfinite \(R^{2}\), and the independently reconstructed quality audit produces no disagreement with the final \texttt{ab\_pass} indicator.
	
	Median intraday standard errors are \(0.157\) bp for SPXW and \(0.134\) bp for RUTW, with 95th percentiles of \(0.458\) and \(0.996\) bp. These errors remain small relative to the approximately \(1.5\) bp SPXW and \(2.1\)--\(2.2\) bp RUTW year-end discontinuities reported in \Cref{tab:weekly_local_yearend_rd}. The complete weekly pipeline also processes all 3,415 SPXW and 2,513 RUTW Stage-0 files without failure.
	
	Taken together, the monthly and weekly diagnostics show that the common strike slope is estimated with high numerical precision. The empirical issue is therefore not noisy recovery of the option-implied discount factor, but the economic content of a precisely measured implied rate.
	
	\FloatBarrier
	
	\section{Robustness of the Year-End Local-Discontinuity Design}
	\label{app:local_yearend_design}
	
	This section tests the sensitivity of the strict same-date local design in \Cref{subsec:data_local_design} to bandwidth, kernel, trade-date window, bid--ask controls, and individual year-end episodes. Every specification uses DGS--DTB matched market--trade-date--maturity cells, requires both sides of the cutoff within each trade date, uses the unannualized OFB \(G_{i,t}^{b}(T)\), and retains trade-date fixed effects and separate maturity slopes on the two sides.
	
	The baseline uses fourth-quarter trade dates, \(0<\lvert h_{t,T}\rvert<90\), a triangular kernel, and the maturity-level median bid--ask control. Core alternatives change one element at a time: 60- or 120-day bandwidths, uniform or Epanechnikov kernels, a September--December trade window, no bid--ask control, exclusion of the 2018 or 2025 episode, and a 10-day donut. I additionally report a November--December window and local quadratic as diagnostics because they sharply reduce local support or impose curvature that the discrete maturity grid may not identify reliably.
	
	\begin{table}[!t]
		\centering
		\singlespacing
		\caption{Robustness of Year-End Local-Discontinuity Estimates}
		\label{tab:local_rd_robustness}
		\begin{threeparttable}
			\scriptsize
			\setlength{\tabcolsep}{3.3pt}
			\renewcommand{\arraystretch}{1.04}
			\begin{tabular*}{0.96\textwidth}{@{\extracolsep{\fill}}lrrrrrr@{}}
				\toprule
				&
				\multicolumn{3}{c}{SPX}
				&
				\multicolumn{3}{c}{RUT} \\
				\cmidrule(lr){2-4}
				\cmidrule(lr){5-7}
				Specification
				& DGS
				& DTB
				& \(N\)
				& DGS
				& DTB
				& \(N\) \\
				\midrule
				
				\multicolumn{7}{l}{\textit{Panel A. Standard monthlies}} \\
				\addlinespace[0.2em]
				
				Baseline
				& \(2.678^{***}\) (0.509)
				& \(2.650^{***}\) (0.502)
				& 1,993
				& \(2.844^{***}\) (0.578)
				& \(2.776^{***}\) (0.566)
				& 1,548 \\
				
				60-day bandwidth
				& \(2.720^{***}\) (0.504)
				& \(2.723^{***}\) (0.501)
				& 1,514
				& \(4.287^{**}\) (1.344)
				& \(4.334^{**}\) (1.338)
				& 921 \\
				
				120-day bandwidth
				& \(2.700^{***}\) (0.519)
				& \(2.678^{***}\) (0.514)
				& 2,207
				& \(2.854^{***}\) (0.577)
				& \(2.783^{***}\) (0.565)
				& 1,548 \\
				
				Uniform kernel
				& \(2.651^{***}\) (0.529)
				& \(2.610^{***}\) (0.521)
				& 1,993
				& \(2.861^{***}\) (0.581)
				& \(2.786^{***}\) (0.569)
				& 1,548 \\
				
				Epanechnikov kernel
				& \(2.670^{***}\) (0.509)
				& \(2.638^{***}\) (0.502)
				& 1,993
				& \(2.845^{***}\) (0.573)
				& \(2.774^{***}\) (0.561)
				& 1,548 \\
				
				Sep.--Dec. trade dates
				& \(2.702^{***}\) (0.360)
				& \(2.696^{***}\) (0.357)
				& 3,433
				& \(2.916^{***}\) (0.422)
				& \(2.896^{***}\) (0.417)
				& 2,701 \\
				
				No bid--ask control
				& \(2.553^{***}\) (0.499)
				& \(2.522^{***}\) (0.492)
				& 1,993
				& \(2.937^{***}\) (0.594)
				& \(2.877^{***}\) (0.586)
				& 1,548 \\
				
				Exclude 2018
				& \(2.335^{***}\) (0.497)
				& \(2.327^{***}\) (0.497)
				& 1,848
				& \(2.426^{***}\) (0.548)
				& \(2.382^{***}\) (0.544)
				& 1,423 \\
				
				Exclude 2025
				& \(2.275^{***}\) (0.441)
				& \(2.242^{***}\) (0.431)
				& 1,838
				& \(2.420^{***}\) (0.500)
				& \(2.354^{***}\) (0.486)
				& 1,439 \\
				
				10-day donut
				& \(2.988^{***}\) (0.453)
				& \(2.971^{***}\) (0.454)
				& 1,675
				& \(3.225^{***}\) (0.523)
				& \(3.160^{***}\) (0.521)
				& 1,291 \\
				
				\addlinespace[0.2em]
				\multicolumn{7}{l}{\textit{Additional diagnostics}} \\
				\addlinespace[0.2em]
				
				Nov.--Dec. trade dates
				& 3.519 (4.373)
				& 3.363 (4.255)
				& 676
				& 4.498 (5.376)
				& 4.329 (5.262)
				& 519 \\
				
				Local quadratic
				& 1.122 (2.670)
				& 1.102 (2.647)
				& 1,993
				& 4.903 (3.570)
				& 4.886 (3.525)
				& 1,548 \\
				
				\midrule
				
				&
				\multicolumn{3}{c}{SPXW}
				&
				\multicolumn{3}{c}{RUTW} \\
				\cmidrule(lr){2-4}
				\cmidrule(lr){5-7}
				Specification
				& DGS
				& DTB
				& \(N\)
				& DGS
				& DTB
				& \(N\) \\
				\midrule
				
				\multicolumn{7}{l}{\textit{Panel B. Regular weeklies}} \\
				\addlinespace[0.2em]
				
				Baseline
				& \(1.530^{**}\) (0.313)
				& \(1.456^{**}\) (0.312)
				& 7,256
				& \(2.183^{*}\) (0.814)
				& 2.093 (0.814)
				& 2,974 \\
				
				60-day bandwidth
				& \(1.230^{**}\) (0.333)
				& \(1.157^{**}\) (0.332)
				& 6,219
				& 1.415 (0.750)
				& 1.320 (0.752)
				& 2,078 \\
				
				120-day bandwidth
				& \(1.629^{***}\) (0.311)
				& \(1.550^{***}\) (0.309)
				& 7,818
				& \(2.787^{***}\) (0.902)
				& \(2.686^{***}\) (0.912)
				& 3,661 \\
				
				Uniform kernel
				& \(1.794^{***}\) (0.305)
				& \(1.699^{***}\) (0.304)
				& 7,256
				& \(3.073^{**}\) (0.992)
				& \(2.928^{**}\) (1.005)
				& 2,974 \\
				
				Epanechnikov kernel
				& \(1.626^{***}\) (0.309)
				& \(1.549^{***}\) (0.309)
				& 7,256
				& \(2.458^{*}\) (0.855)
				& \(2.360^{*}\) (0.855)
				& 2,974 \\
				
				Sep.--Dec. trade dates
				& \(1.582^{**}\) (0.297)
				& \(1.507^{**}\) (0.297)
				& 9,617
				& 2.242 (0.787)
				& 2.169 (0.788)
				& 3,844 \\
				
				No bid--ask control
				& \(1.528^{***}\) (0.304)
				& \(1.448^{**}\) (0.303)
				& 7,256
				& \(2.352^{*}\) (0.821)
				& \(2.259^{*}\) (0.822)
				& 2,974 \\
				
				Exclude 2018
				& \(1.432^{**}\) (0.337)
				& \(1.368^{**}\) (0.337)
				& 6,581
				& 2.264 (0.883)
				& 2.172 (0.882)
				& 2,643 \\
				
				Exclude 2025
				& \(1.250^{**}\) (0.271)
				& \(1.175^{**}\) (0.268)
				& 6,560
				& 1.317 (0.537)
				& 1.226 (0.538)
				& 2,495 \\
				
				10-day donut
				& \(2.953^{***}\) (0.392)
				& \(2.898^{***}\) (0.397)
				& 5,222
				& \(7.446^{**}\) (2.435)
				& \(7.410^{**}\) (2.436)
				& 1,967 \\
				
				\addlinespace[0.2em]
				\multicolumn{7}{l}{\textit{Additional diagnostics}} \\
				\addlinespace[0.2em]
				
				Nov.--Dec. trade dates
				& \(1.258^{**}\) (0.335)
				& \(1.157^{**}\) (0.332)
				& 4,411
				& 1.979 (0.919)
				& 1.864 (0.918)
				& 1,860 \\
				
				Local quadratic
				& 0.699 (0.347)
				& 0.621 (0.348)
				& 7,256
				& 0.321 (0.593)
				& 0.212 (0.597)
				& 2,974 \\
				
				\bottomrule
			\end{tabular*}
		\end{threeparttable}
		
		\vspace{0.4em}
		
		\begin{minipage}{0.96\textwidth}
			\footnotesize
			\textit{Notes:} Each cell reports the unannualized year-end discontinuity \(\widehat\kappa\) in basis points, with Newey--West HAC(21) standard errors based on trade-date scores in parentheses. Significance stars use two-sided calendar-year episode sign-flip exact \(p\)-values: \({}^{***}\), \({}^{**}\), and \({}^{*}\) denote \(p<0.01\), \(p<0.05\), and \(p<0.10\), respectively. All specifications use DGS--DTB matched cells with strict same-date two-sided support; \(N\) is the final number of market--trade-date--maturity observations. Standard monthlies and SPXW contain 14 episodes, while RUTW contains 10; the November--December and local-quadratic specifications are reported as diagnostics because of limited local support.
		\end{minipage}
	\end{table}
	
	\Cref{tab:local_rd_robustness} shows that the monthly discontinuity remains positive and significant across all core specifications. SPX estimates are generally \(2.2\)--\(3.0\) bp and RUT estimates \(2.4\)--\(3.2\) bp, apart from the larger RUT estimate under the 60-day bandwidth. DGS--DTB differences remain below \(0.08\) bp, making the result effectively invariant to the two Treasury curve constructions.
	
	Regular-weekly estimates are also uniformly positive. SPXW remains significant at the 5\% level throughout the core specifications. RUTW inference is less stable because it contains only ten episodes and has thinner right-side maturity support on some trade dates, as documented in \Cref{tab:local_support_comparison}.
	
	The 10-day donut preserves the positive sign but increases the weekly estimates, especially for RUTW, to about \(7.4\) bp. Removing near-cutoff contracts leaves a sparser maturity grid and greater sensitivity to the fitted slopes, so this specification does not support invariance of the magnitude.
	
	The additional diagnostics expose the same support limitation. Restricting standard monthlies to November--December raises HAC standard errors above \(4\)--\(5\) bp, while the local quadratic is poorly conditioned for standard monthlies, with condition numbers of about \(376\) for SPX and \(716\) for RUT, and substantially attenuates the weekly estimates. The discrete maturity grid therefore does not reliably identify separate second-order curvature near the cutoff.
	
	Overall, the positive year-end discontinuity is robust to conventional changes in bandwidth, kernel, trade-date window, bid--ask controls, and individual episodes. Magnitudes and precision become unstable only when substantial near-cutoff support is removed or additional local curvature is imposed.
	
	\FloatBarrier
	
	\section{Full-Panel Robustness and Influence Diagnostics}
	\label{app:full_panel_robustness}
	
	This section examines whether the full-panel year-end coefficient in \Cref{eq:data_full_panel_m2,tab:yearend_regression_spx,tab:yearend_regression_rut} is sensitive to sample construction, cross-benchmark maturity support, missing funding-spread observations, the option discount-factor estimator, HAC bandwidth, or individual years. The coefficient of interest is \(\kappa\) on \(D_{t,T}^{\mathrm{YC}}/\tau_{t,T}\), reported in unannualized basis points.
	
	\subsection{Sample Construction, Missing Data, and Discount-Factor Measurement}
	\label{app:full_panel_sample_measurement}
	
	The baseline uses the available June 2012--December 2025 sample and requires an exact same-day \(\operatorname{ABCP90\!-\!DTB3}\) observation. I reestimate the model on complete calendar years from 2013--2025 and on market--trade-date--maturity cells of at most 365 days exactly shared by DGS and DTB, separating partial-year and maturity-support effects from benchmark construction.
	
	I also relax same-day funding availability by carrying only the most recent completed \(\operatorname{ABCP90\!-\!DTB3}\) spread for at most three calendar days. Future observations are never used, component rates are never filled separately, and an indicator identifies carried observations.
	
	Finally, I replace the baseline AB discount factor with the adjacent-strike completed-box measure while retaining the same sample and regression controls.
	
	\begin{table}[!t]
		\centering
		\singlespacing
		\caption{Sample and Measurement Robustness of Full-Panel Year-End Estimates}
		\label{tab:full_panel_sample_measurement_robustness}
		\begin{threeparttable}
			\footnotesize
			\setlength{\tabcolsep}{3.0pt}
			\renewcommand{\arraystretch}{1.03}
			\begin{tabular*}{0.96\textwidth}{@{\extracolsep{\fill}}lrrrrrrrr@{}}
				\toprule
				&
				\multicolumn{4}{c}{DGS}
				&
				\multicolumn{4}{c}{DTB} \\
				\cmidrule(lr){2-5}
				\cmidrule(lr){6-9}
				Specification
				& \(\widehat\kappa\)
				& HAC \(t\)
				& Exact \(p\)
				& \(N\)
				& \(\widehat\kappa\)
				& HAC \(t\)
				& Exact \(p\)
				& \(N\) \\
				\midrule
				
				\multicolumn{9}{l}{\textit{Panel A. SPX}} \\
				\addlinespace[0.1em]
				
				\makecell[l]{AB, available sample\\same-day funding}
				& 2.938
				& 6.468
				& 0.0011
				& 40,855
				& 2.869
				& 6.747
				& 0.0011
				& 27,626 \\
				
				\makecell[l]{AB, complete years\\2013--2025}
				& 3.135
				& 6.694
				& 0.0010
				& 39,458
				& 3.077
				& 7.070
				& 0.0007
				& 26,774 \\
				
				\makecell[l]{AB, DGS--DTB\\1-year common support}
				& 2.885
				& 6.900
				& 0.0011
				& 27,626
				& 2.869
				& 6.747
				& 0.0011
				& 27,626 \\
				
				\makecell[l]{AB, limited\\prior-day carry}
				& 2.917
				& 6.476
				& 0.0011
				& 41,315
				& 2.855
				& 6.779
				& 0.0011
				& 27,932 \\
				
				\makecell[l]{Box, available sample\\same-day funding}
				& 2.840
				& 4.853
				& 0.0104
				& 40,855
				& 2.750
				& 4.851
				& 0.0117
				& 27,626 \\
				
				\midrule
				
				\multicolumn{9}{l}{\textit{Panel B. RUT}} \\
				\addlinespace[0.1em]
				
				\makecell[l]{AB, available sample\\same-day funding}
				& 3.199
				& 5.302
				& 0.0005
				& 28,841
				& 3.165
				& 5.773
				& 0.0005
				& 18,767 \\
				
				\makecell[l]{AB, complete years\\2013--2025}
				& 3.373
				& 5.352
				& 0.0007
				& 27,843
				& 3.350
				& 5.882
				& 0.0005
				& 18,057 \\
				
				\makecell[l]{AB, DGS--DTB\\1-year common support}
				& 3.172
				& 5.846
				& 0.0005
				& 18,767
				& 3.165
				& 5.773
				& 0.0005
				& 18,767 \\
				
				\makecell[l]{AB, limited\\prior-day carry}
				& 3.179
				& 5.320
				& 0.0005
				& 29,169
				& 3.149
				& 5.803
				& 0.0005
				& 18,984 \\
				
				\makecell[l]{Box, available sample\\same-day funding}
				& 4.250
				& 4.021
				& 0.0033
				& 28,841
				& 4.308
				& 4.210
				& 0.0035
				& 18,767 \\
				
				\bottomrule
			\end{tabular*}
		\end{threeparttable}
		
		\vspace{0.35em}
		
		\begin{minipage}{0.96\textwidth}
			\footnotesize
			\textit{Notes:} \(\widehat\kappa\) is the coefficient on \(D_{t,T}^{\mathrm{YC}}/\tau_{t,T}\). HAC \(t\)-statistics use Newey--West HAC(21) inference based on trade-date scores; exact \(p\)-values are two-sided calendar-year score sign-flip \(p\)-values. The available AB sample spans June 2012--December 2025, with DGS maturities of \(30\)--\(1{,}095\) days and DTB maturities of \(30\)--\(365\) days. The baseline requires same-day \(\Delta_{t}^{\mathrm{ABCP-TB}}=100(\operatorname{ABCP90}_{t}-\operatorname{DTB3}_{t})\). The complete-year specification excludes partial-year 2012. The 1-year common-support specification retains cells of at most 365 days observed under both DGS and DTB. The limited-carry specification uses only a previously observed completed funding spread for at most three calendar days and includes a carry indicator. The Box specification replaces the AB discount factor with the adjacent-strike completed-box measure. All regressions include an intercept, calendar-year fixed effects, a centered and standardized cubic in maturity, \(\operatorname{BA}/\tau\), the NFCI, and the funding-spread control. Coefficients are in unannualized basis points.
		\end{minipage}
	\end{table}
	
	Across the AB specifications in \Cref{tab:full_panel_sample_measurement_robustness}, \(\widehat\kappa\) ranges from \(2.855\) to \(3.373\) bp and remains precisely positive. Excluding partial-year 2012 raises the estimates by about \(0.18\)--\(0.21\) bp. On exact one-year common support, the DGS--DTB differences are only \(0.015\) bp for SPX and \(0.007\) bp for RUT, while allowing limited prior-day funding carry changes the baseline coefficient by at most \(0.021\) bp.
	
	Replacing AB with completed boxes also preserves a positive and significant boundary effect. SPX Box estimates are \(2.750\)--\(2.840\) bp, close to the AB estimates, whereas RUT Box estimates rise to \(4.250\)--\(4.308\) bp. Discount-factor measurement therefore affects the magnitude more for RUT, but not the sign or presence of the year-end wedge.
	
	\subsection{Inference Bandwidth and Year-Level Influence}
	\label{app:full_panel_inference_influence}
	
	Baseline inference uses Newey--West HAC(21) standard errors based on trade-date scores. I vary the bandwidth from 5 to 63 trading days and retain calendar-year score sign flips as a bandwidth-independent inference check.
	
	I also reestimate the full-panel model after removing each trade year, rebuilding the calendar-year fixed effects and centered maturity basis within each fold. The local counterpart removes each year-end episode and reestimates the strict same-date design.
	
	\begin{table}[!t]
		\centering
		\singlespacing
		\caption{Inference Bandwidth and Year-Level Influence Diagnostics}
		\label{tab:full_panel_inference_influence}
		\begin{threeparttable}
			\footnotesize
			\setlength{\tabcolsep}{4.5pt}
			\renewcommand{\arraystretch}{1.06}
			
			\begin{tabular*}{0.96\textwidth}{@{\extracolsep{\fill}}llrrrrr@{}}
				\toprule
				\multicolumn{7}{l}{\textit{Panel A. Full-panel HAC bandwidth}} \\
				\addlinespace[0.2em]
				Market
				& Benchmark
				& \(\widehat\kappa\)
				& HAC(5) \(t\)
				& HAC(21) \(t\)
				& HAC(63) \(t\)
				& Exact \(p\) \\
				\midrule
				SPX & DGS & 2.938 & 10.731 & 6.468 & 5.205 & 0.0011 \\
				SPX & DTB & 2.869 & 11.151 & 6.747 & 5.458 & 0.0011 \\
				RUT & DGS & 3.199 & 8.500 & 5.302 & 4.304 & 0.0005 \\
				RUT & DTB & 3.165 & 9.149 & 5.773 & 4.747 & 0.0005 \\
				\bottomrule
			\end{tabular*}
			
			\vspace{0.8em}
			
			\begin{tabular*}{0.96\textwidth}{@{\extracolsep{\fill}}lllrrrcr@{}}
				\toprule
				\multicolumn{8}{l}{\textit{Panel B. Leave-one-year-end-out influence diagnostics}} \\
				\addlinespace[0.2em]
				Design
				& Market
				& Benchmark
				& Full sample
				& Minimum
				& Maximum
				& \makecell{All folds\\positive}
				& \makecell{Share with\\exact \(p<0.05\)} \\
				\midrule
				Full panel & SPX & DGS & 2.938 & 2.489 & 3.227 & Yes & 100\% \\
				Full panel & SPX & DTB & 2.869 & 2.438 & 3.164 & Yes & 100\% \\
				Full panel & RUT & DGS & 3.199 & 2.486 & 3.543 & Yes & 100\% \\
				Full panel & RUT & DTB & 3.165 & 2.457 & 3.492 & Yes & 100\% \\
				\addlinespace[0.2em]
				Local RD & SPX & DGS & 2.678 & 2.275 & 2.981 & Yes & 100\% \\
				Local RD & SPX & DTB & 2.650 & 2.242 & 2.949 & Yes & 100\% \\
				Local RD & RUT & DGS & 2.844 & 2.420 & 3.185 & Yes & 100\% \\
				Local RD & RUT & DTB & 2.776 & 2.354 & 3.111 & Yes & 100\% \\
				\bottomrule
			\end{tabular*}
		\end{threeparttable}
		
		\vspace{0.4em}
		
		\begin{minipage}{0.96\textwidth}
			\footnotesize
			\textit{Notes:} Panel A varies only the HAC bandwidth in the available-sample same-day \(\mathcal M_{2}\) specifications from \Cref{tab:yearend_regression_spx,tab:yearend_regression_rut}, including \(\operatorname{ABCP90\!-\!DTB3}\). Exact \(p\)-values are two-sided sign-flip \(p\)-values based on 14 calendar-year score blocks. Panel B reestimates the full-panel \(\mathcal M_{2}\) after omitting each trade year and reconstructing calendar-year fixed effects and the centered maturity polynomial within each fold. Local RD rows similarly omit each year-end episode from the baseline design in \Cref{tab:monthly_local_yearend_rd}. Minimum and maximum report the range of \(\widehat\kappa\) across folds. The final column reports the share of folds with two-sided calendar-year or episode-level exact \(p<0.05\). Coefficients are in unannualized basis points.
		\end{minipage}
	\end{table}
	
	As shown in \Cref{tab:full_panel_inference_influence}, full-panel HAC \(t\)-statistics remain between \(4.304\) and \(11.151\) across 5--63-day bandwidths, consistent with the exact sign-flip inference. Full-panel leave-one-year-out estimates remain between \(2.438\) and \(3.543\) bp, while local leave-one-episode-out estimates remain between \(2.242\) and \(3.185\) bp. Every fold is positive and has exact \(p<0.05\), so neither result is driven by a single year or year-end episode.
	
	\subsection{First and Additional Year-End Crossings}
	\label{app:additional_yearend_crossings}
	
	The baseline indicator records only whether a contract spans at least one year-end. For longer maturities, define
	\[
	N_{t,T}^{\mathrm{YC}}
	=
	\max\{\operatorname{year}(T)-\operatorname{year}(t),0\}.
	\]
	I use the natural DGS panel and estimate separate marginal effects for the first three crossings with trade-date fixed effects.
	
	The regression jointly includes
	\begin{equation}
		\frac{\mathbbm 1\{N_{t,T}^{\mathrm{YC}}\geq j\}}{\tau_{t,T}},
		\qquad
		j=1,2,3.
		\label{eq:additional_yearend_crossings}
	\end{equation}
	The coefficient for \(j=1\) measures the effect of moving from zero to one crossing, while the later coefficients measure incremental effects conditional on the preceding crossings.
	
	\begin{table}[!t]
		\centering
		\singlespacing
		\caption{Marginal Price Effects of First and Additional Year-End Crossings}
		\label{tab:additional_yearend_crossings}
		\begin{threeparttable}
			\footnotesize
			\setlength{\tabcolsep}{5.5pt}
			\renewcommand{\arraystretch}{1.08}
			\begin{tabular*}{0.96\textwidth}{@{\extracolsep{\fill}}lrrrrrr@{}}
				\toprule
				&
				\multicolumn{3}{c}{SPX}
				&
				\multicolumn{3}{c}{RUT} \\
				\cmidrule(lr){2-4}
				\cmidrule(lr){5-7}
				Marginal year-end crossing
				& Coefficient
				& HAC SE
				& Exact \(p\)
				& Coefficient
				& HAC SE
				& Exact \(p\) \\
				\midrule
				
				First:
				\(\mathbbm 1\{N^{\mathrm{YC}}\geq1\}/\tau\)
				& 2.888
				& 0.355
				& 0.0001
				& 3.173
				& 0.520
				& 0.0001 \\
				
				Second:
				\(\mathbbm 1\{N^{\mathrm{YC}}\geq2\}/\tau\)
				& \(-0.295\)
				& 1.444
				& 0.9281
				& \(-0.327\)
				& 2.810
				& 0.9585 \\
				
				Third:
				\(\mathbbm 1\{N^{\mathrm{YC}}\geq3\}/\tau\)
				& 2.941
				& 12.987
				& 0.8573
				& \(-3.046\)
				& 13.771
				& 0.8699 \\
				
				\midrule
				
				Observations
				& \multicolumn{3}{c}{41,404}
				& \multicolumn{3}{c}{29,244} \\
				
				Trade dates
				& \multicolumn{3}{c}{3,416}
				& \multicolumn{3}{c}{3,415} \\
				
				Calendar-year episodes
				& \multicolumn{3}{c}{14}
				& \multicolumn{3}{c}{14} \\
				
				\bottomrule
			\end{tabular*}
		\end{threeparttable}
		
		\vspace{0.4em}
		
		\begin{minipage}{0.96\textwidth}
			\footnotesize
			\textit{Notes:} The sample is the available \(30\)--\(1{,}095\)-day DGS panel. The regression jointly includes \(\mathbbm 1\{N_{t,T}^{\mathrm{YC}}\geq j\}/\tau_{t,T}\) for \(j=1,2,3\), so each coefficient measures the marginal unannualized effect of adding one year-end crossing conditional on the preceding crossings. The specification uses trade-date fixed effects, a centered and standardized cubic in maturity, and \(\operatorname{BA}/\tau\). HAC standard errors are Newey--West HAC(21) estimates based on trade-date scores, and exact \(p\)-values are two-sided calendar-year score sign-flip \(p\)-values. Coefficients are in unannualized basis points.
		\end{minipage}
	\end{table}
	
	As shown in \Cref{tab:additional_yearend_crossings}, the first crossing is precisely estimated at \(2.888\) bp for SPX and \(3.173\) bp for RUT, closely matching the approximately \(3\) bp full-panel boundary effect despite the different sample and fixed-effect structure.
	
	Later crossings are weakly identified. Second-crossing estimates are near zero but have standard errors of \(1.444\) and \(2.810\) bp, while third-crossing standard errors rise to about \(13\) bp. The data therefore identify a positive first-crossing effect but cannot determine whether additional reporting boundaries impose further charges.
	
	\FloatBarrier
	
\end{document}